\documentclass[submission, PhysLectNotes]{SciPost}
\hypersetup{
    colorlinks,
    linkcolor={red!50!black},
    citecolor={blue!50!black},
    urlcolor={blue!80!black}
}

\usepackage[bitstream-charter]{mathdesign}
\DeclareSymbolFont{usualmathcal}{OMS}{cmsy}{m}{n}
\DeclareSymbolFontAlphabet{\mathcal}{usualmathcal}

\usepackage[utf8]{inputenc}
\usepackage{physics}
\usepackage{tikz}
\usepackage{pgfplots}
\usetikzlibrary{positioning}
\usepackage{graphicx}
\usepackage{float}
\usepackage{mathtools}
\usepackage{xcolor}
\usepackage{hyperref}
\usepackage{latexsym}
\usepackage{booktabs}
\usepackage{color}
\usepackage{authblk}
\usepackage{slashed}
\usepackage{bbm}
\usepackage{amsmath}
\usepackage{comment}
\usepackage[normalem]{ulem}

\def \del{\partial}
\newcommand{\nn}{\nonumber}

\def \pt{\partial}

\newcommand{\mn}{{\mu\nu}}
\newcommand{\ee}{\end{equation}}
\newcommand{\be}{\begin{equation}}

\newcommand{\ba}{\begin{align}}
\newcommand{\ena}{\end{align}}

\newcommand{\bea}{\begin{eqnarray}}
\newcommand{\eea}{\end{eqnarray}}

\newcommand{\I}[1]{\vcenter{\hbox to 2.5mm{%
\begin{tikzpicture}[#1]%
\draw[] (-0.05,-0.03)--(0.05,0.03) node at (0,0) {I};%
\end{tikzpicture}%
}}}

\newcommand{\Idot}[1]{\vbox to 4.6mm{\hbox to 2.8mm{%
\begin{tikzpicture}[#1]%
\draw[] (-0.05,-0.03)--(0.05,0.03) node at (0,0) {I};
\node at (0,0.2) {..};%
\end{tikzpicture}%
}}}

\numberwithin{equation}{section}
\begin{document}

\begin{flushright}
\bf MS-TP-22-57
\end{flushright}

% TODO: write your article's title here.
% The article title is centered, in Large boldface, and should fit in two lines
\begin{center}{\Large \emph{
Gravitational waves from the early universe\\
}}\end{center}

% TODO: write the author list here. Use first name (+ other initials) + surname format.
% Separate subsequent authors by a comma, omit the comma and use "and" for the last author.
% Mark the corresponding author with a superscript star.
\begin{center}
Rafael R. Lino dos Santos\textsuperscript{1,2,3,4,æ*} and
Linda M. van Manen \textsuperscript{5,ą}
\end{center}

% TODO: write all affiliations here.
% Format: institute, city, country
\begin{center}

{\bf 1} University  of  Southern  Denmark,  Campusvej  55,  DK-5230  Odense  M,  Denmark
\\
{\bf 2} University of Münster, Institute for Theoretical Physics, 48149 Münster, Germany
\\
{\bf 3} National Centre for Nuclear Research, Pasteura 7, 02-093 Warsaw, Poland
\\
{\bf 4} Instituto de Física, Universidade Federal Fluminense, 24210-346 Niterói, RJ, Brazil
\\
{\bf 5} Friedrich-Schiller-Universität, Institute for Theoretical Physics, 07743 Jena, Germany 
\\

\textsuperscript{æ}  {\href{mailto:rrlsantos@id.uff.br}{rrlsantos@id.uff.br}}
\textsuperscript{ą} {\href{mailto:linda.van.manen@uni-jena.de}{linda.van.manen@uni-jena.de}}

\end{center}

\begin{center}
\today
\end{center}

% For convenience during refereeing (optional),
% you can turn on line numbers by uncommenting the next line:
%\linenumbers
% You should run LaTeX twice in order for the line numbers to appear.

\section*{Abstract}
{\bf

Even though different models could already be constrained using CMB data, these models remained unconstrained at higher frequency scales, and, therefore, the knowledge of new physics at these scales remained limited, relying on theoretical assumptions and extrapolations. Recently, however, we experienced the advent of gravitational-wave and multi-messenger astronomy, including the outstanding detections by the LIGO-Virgo collaboration and the searches for Hellings-Downs correlations in pulsar timing data. 

Over the next decades, ongoing and future collaborations will explore different frequency ranges of the gravitational wave spectrum, granting access to regimes inaccessible by the CMB and collider experiments. These include ground-based, space-borne, and pulsar timing array searches for gravitational waves. By comparing their data with model-dependent gravitational-wave predictions, we can unprecedentedly constrain or eliminate different models or even discover new physics.

In these lecture notes, as an entrance door for graduate students, we focus on how we can search for a cosmological background of gravitational waves and how this can be used to probe cosmology and beyond-Standard-Model physics. For this purpose, we review the formalism of gravitational waves in General Relativity, introduce stochastic background gravitational waves, and derive the Hellings-Downs correlation for pulsar timing array searches. We comment on detection efforts and present some of the most important cosmological sources that could produce such a background.
\\

These lecture notes were inspired by the course ``Gravitational Waves from the Early Universe'' given at the 27th W.E. Heraeus ``Saalburg'' Summer School 2021 by Valerie Domcke. 
}

\vspace{10pt}
\noindent\rule{\textwidth}{1pt}
\tableofcontents\thispagestyle{fancy}
\noindent\rule{\textwidth}{1pt}
\vspace{10pt}

\section{Introduction and motivation}

Observations of the Universe allow us to look back in time to epochs when the cosmic temperature was of order $T \approx eV$. Standard cosmology then claims that, at this stage, free electrons combined with protons to form neutral hydrogen for the first time, in an event known as recombination.\footnote{The prefix \emph{re} in recombination might be confusing about the number of times the event of protons and electrons combining has occurred. This event has happened only once.} After recombination, photons were able to propagate freely through space. Some of these photons have traveled essentially unimpeded ever since and are observed today as the Cosmic Microwave Background (CMB). The moment when the first photons could freely travel through the universe is known as photon decoupling.

Prior to this epoch, the Universe was filled with a hot, dense plasma of electrons, protons, and photons. Frequent scattering of photons off charged particles rendered the Universe opaque, effectively obscuring any electromagnetic information from earlier times. As a result, direct observations of the Universe before photon decoupling using light are fundamentally limited.

However, a variety of early-Universe phenomena could have created gravitational waves (GWs). They may have been produced as early as cosmic inflation, a hypothetical rapid (accelerated) expansion of the early Universe (see Fig. \ref{fig:diagram}), creating a relic background of gravitational waves, that appears as noise and is similar to the CMB. This relic GW background is known as the gravitational wave background or the stochastic gravitational wave background (SGWB). Unlike photons before decoupling, gravitational waves interact extremely weakly with matter and traveled through the early universe largely unperturbed. Hence, no fundamental physical barrier prevents from observing early GWs carrying information from epochs far earlier than photon decoupling. Although there are no fundamental obstacles, there are plenty of experimental challenges. 

Similarly to the CMB, the GW background is expected to appear as an approximately isotropic background, manifesting as weak noise coming from all directions. Extracting this signal requires disentangling it from instrumental noise and from astrophysical foregrounds. However, different collaborations expect to detect the GW background in the coming years.\footnote{Some collaborations are already active. They rely on ground-based detectors (LIGO, Virgo, KAGRA) or are pulsar timing array (PTA) collaborations (NANOGrav, EPTA, PPTA, CPTA, InPTA, IPTA). Future collaborations include space-based detectors (LISA, DECIGO, Taiji, and TianQin) as well as next-generation ground-based observatories (Cosmic Explorer and the Einstein Telescope).} Since each collaboration probes a different spectrum range of frequencies, their combined observations will provide access to different physical processes and cosmic epochs. 

In this sense, gravitational-wave observations can be viewed as a \emph{gravitational-wave orchestra}, in which different detectors act as complementary instruments, each probing a distinct frequency band and, correspondingly, a different stage of cosmic history\footnote{Continuing the musical analogy, in a string quintet we can go from the double bass (low) to the cello, and then to the viola and the violins (high). Likewise, in the gravitational wave orchestra of the early universe, we can go from matter domination to radiation domination era, then reheating and inflation.} in a given cosmological interpretation. For a visual presentation of such stages in the standard model of cosmology, the so-called $\Lambda$CDM model, see Fig. \ref{fig:diagram}.

Lastly, different early-universe sources can produce GWs. Examples are inflation and Beyond Standard Model (BSM) physics phenomena, such as cosmic strings and first-order phase transitions. (These examples will be addressed in chapters 6 and 7). Remarkably, BSM depends on energy scales far beyond the reach of electromagnetic probes such as the CMB and accelerators on Earth. These scales, however, can be probed by GW experiments. Therefore, the gravitational-wave background can also be seen as a laboratory to probe new physics, further complementing our knowledge of cosmology and particle physics by going beyond other probes. 
\begin{figure}[H]
\centering
\begin{tikzpicture}
\draw[<-](-1,0)--(6,0) node[anchor=west]{$T$};
\draw[] (-0.8,0.2)--(-0.8,-0.2) node[anchor=north]{?};
\draw[xshift=-17] (0.47,-0.15)--(0.63,0.15);
\draw[xshift=-17] (0.57,-0.15)--(0.73,0.15);
\draw[] (1,0.2)--(1,-0.2) node[anchor=north]{MeV};
\draw[] (3,0.2)--(3,-0.2) node[anchor=north]{eV};
\draw[] (5,0.2)--(5,-0.2) node[anchor=north]{meV};
\draw[] (-0.9,2)..controls (-0.2,2) and (-0.4,1)..(1.1,1);
\draw[] (1.1,1)..controls (5,0.8)..(5.5,0.5);
\draw[] (-0.9,2)..controls (-0.2,2) and (-0.4,3)..(1.1,3);
\draw[] (1.1,3)..controls (5,3.2)..(5.5,3.5);
\draw[dashed, blue] (5,0.2)--(5,4) node[anchor=west]{$\Lambda$};
\draw[dashed, blue] (3,0.2)--(3,4) node[anchor=west]{Matter};
\draw[dashed, blue] (0.6,0.2)--(0.6,4) node[anchor=west]{Radiation} node[anchor=east]{Cosmic inflation};
\draw[orange,->] (3,2.2)--(6,2.2) node[anchor=west] {$\gamma$};
\draw[orange,->] (-0.3,1.8)--(6,1.8) node[anchor=west] {GW};
\node[blue] at (1.75,2.35) {BBN};
\node[blue] at (3.5,2.5) {CMB};
\end{tikzpicture}
\caption{\textit{A visualization of cosmic history within the standard model of cosmology ($\Lambda$CDM). In this interpretation, starting from the Big Bang, the Universe underwent a brief period of cosmic inflation, followed by a radiation-dominated era, a matter-dominated era, and finally the present epoch dominated by dark energy, modeled as a cosmological constant $\Lambda$, which drives the accelerated expansion of the Universe. The horizontal axis shows cosmic time, equivalently represented by the decreasing temperature $T$ and the corresponding characteristic energy scales. The earliest freely propagating photons observable today were released at photon decoupling and form the cosmic microwave background (CMB), indicated by $\gamma$. In contrast, gravitational waves (GWs) from the early Universe could originate from much earlier times, potentially as early as inflation, propagate largely unimpeded, and may be detected in the coming decades, offering a unique probe of inflation and physics beyond the Standard Model.}}
\label{fig:diagram}
\end{figure}

\subsubsection*{Organization of lecture notes}
The lectures are organized in the following sequence. In Sec.~\ref{sec:Linearized Einstein equations}, we obtain GWs as vacuum solutions of the linearized Einstein equations and study the effects of GWs on test masses. Next, in Sec.~\ref{sec:Emission of gravitational waves}, we study the sourced emission of GWs and their energy-momentum tensor, we derive Einstein's quadrupole formula for the power emitted by a source, and write expressions for the power spectrum of tensor perturbations, which are used extensively in the next sections. Then, in Sec.~\ref{sec:The stochastic gravitational wave background}, we focus on the background of stochastic gravitational waves, and we briefly comment on sources and detection efforts, while in Sec.~\ref{sec:searching} we elaborate more on the main properties of interferometers and pulsar timing array searches concerning GW detection efforts. In Sec.~\ref{sec:Primordial gravitational waves}, we give a very short review on cosmic inflation, explore properties of GWs in the expanding Friedmann-Robertson-Lemaître-Walker universe, and show how data can be used to constrain BSM physics. Finally, in Sec.~\ref{sec:Probing cosmology and BSM physics with the SGWB}, we non-exhaustively discuss some cosmological sources of GWs (cosmic gravitational microwave background, single-field slow-roll inflation, axion-inflation, scalar-induced gravitational waves, first-order phase transitions, and cosmic strings). We conclude in Sec.~\ref{sec:conclusions}.

\newpage
\subsection{Aim and scope of these lecture notes}
The purpose of these lecture notes is to serve as an entrance door to the subject of cosmic gravitational waves for graduate students. The emphasis is on building physical intuition, introducing the key theoretical ideas, and providing a coherent overview of the main mechanisms for the production, propagation, and observation of gravitational waves in a cosmological context. Wherever possible, the discussion prioritizes clarity and conceptual understanding over technical completeness.

These notes are therefore not intended to be a comprehensive or definitive reference on the subject. Readers seeking a more rigorous and detailed treatment are strongly encouraged to read the literature provided below.
These references should be regarded as complementary to the present notes and as essential resources for readers wishing to develop a deeper and more complete understanding of the field.

Furthermore, these notes do not intend to educate the reader on the basics of general relativity and cosmology, but rather focus on the advanced topic of gravitational waves. Hence, we assume the reader has a solid understanding of general relativity, and a basic understanding of gauge theory and cosmology, throughout these lecture notes.

\subsection{Reference list}
\subsubsection*{Theory: Linearized gravity and emission of GW:}
\begin{enumerate}
    \item[] \cite{maggiore} Gravitational Waves: Volume 1: Theory and Experiments by M. Maggiore 
\item[] \cite{carroll} Spacetime and Geometry. An Introduction to General Relativity by S. Carroll 
\item[] \cite{MisnerThorneWheeler1973} Gravitation by C. W.  Misner,  K. S. Thorne, and J. A. Wheeler
\end{enumerate}

\subsubsection*{Stochastic gravitational wave background (SGWB):}
\begin{enumerate}
    \item[] \cite{Maggiore:2018sht} Gravitational Waves: Volume 2: Astrophysics and Cosmology by M. Maggiore 
\item[] \cite{Regimbau:2011rp} The astrophysical gravitational wave stochastic background by T. Regimbau
\item[] \cite{Caprini:2018mtu}  Cosmological Backgrounds of Gravitational Waves by C. Caprini and D. G. Figueroa

\end{enumerate}

\subsubsection*{Pulsar timing arrays and Hellings Downs:}
\begin{enumerate}
\item[] \cite{1983ApJ...265L..39H} Upper limits on the isotropic gravitational radiation background from pulsar timing analysis by R. Hellings and G. Downs
\item[] \cite{Taylor:2021yjx} The Nanohertz Gravitational Wave Astronomer by S. R. Taylor
\item[] \cite{Taylor:2014isa} Exploring the Cosmos with Gravitational Waves by S. R. Taylor
\item[]\cite{Romano:2023zhb}  Answers to frequently asked questions about the pulsar timing array Hellings and Downs curve by J. D. Romano and B. Allen
\end{enumerate}

\subsubsection*{Remarkable events:}
\begin{enumerate}
\item[] \cite{1968Natur.217..709H} Observation of a Rapidly Pulsating Radio Source by J. Bell et al
\item[] \cite{1975ApJ...195L..51H} Discovery of a pulsar in a binary system by R. Hulse and J. Taylor
\item[] \cite{LIGOScientific:2016aoc} Observation of Gravitational Waves from a Binary Black Hole Merger by B. P. Abbott et al
\item[] \cite{LIGOScientific:2017vwq} Observation of Gravitational Waves from a Binary Neutron Star Inspiral by B. P. Abbott et al
\item[] \cite{EventHorizonTelescope:2019dse}  First M87 Event Horizon Telescope Results. I. The Shadow of the Supermassive Black Hole by Akiyama et al for the EHT collaboration
\item[] \cite{EventHorizonTelescope:2022wkp}  First Sagittarius A* Event Horizon Telescope Results. I. The Shadow of the Supermassive Black Hole in the Center of the Milky Way by Akiyama et al for the EHT collaboration

\end{enumerate}

\subsubsection*{Evidence for Hellings-Downs correlation in PTA datasets:}
\begin{enumerate}
\item[] \cite{NANOGrav:2023gor} The NANOGrav 15 yr Data Set: Evidence for a Gravitational-wave Background by Agazie et al for the NANOGrav collaboration
\item[] \cite{EPTA:2023fyk} The second data release from the European Pulsar Timing Array III. Search for gravitational wave signals by J. Antoniadis et al for the EPTA collaboration
\item[] \cite{Reardon:2023gzh} Search for an isotropic gravitational-wave background with the Parkes Pulsar Timing Array by D. Reardon et al for the PPTA collaboration
\item[] \cite{Xu:2023wog} Searching for the Nano-Hertz Stochastic Gravitational Wave Background with the Chinese Pulsar Timing Array Data Release I by H. Xu et al for the CPTA collaboration
\end{enumerate}

\subsubsection*{Probing cosmology and BSM physics with the SGWB:}
\begin{enumerate}
\item[]  \cite{Baumann:2009ds} Inflation by D. Baumann
\item[] \cite{Domcke:2013pma} Matter, Dark Matter and Gravitational Waves from a GUT-Scale $U(1)$ Phase Transition by V. Domcke
\item[] \cite{LISACosmologyWorkingGroup:2022jok} Cosmology with the Laser Interferometer Space Antenna by P. Auclair et al
\item[] \cite{NANOGrav:2023hvm} The NANOGrav 15 yr Data Set: Search for Signals from New Physics by A. Afzal et al for the NANOGrav collaboration
\end{enumerate}

\newpage
\section{Linearized Einstein equations}
\label{sec:Linearized Einstein equations}

In this section, we start by evaluating linearized general relativity (GR), which describes the dynamics of a slightly perturbed gravitational field. After all, we can think about GWs as \emph{small} ripples in flat spacetime. Hence, we consider a metric tensor decomposed into the Minkowski metric and a small perturbation,
 \be
 g_\mn = \eta_\mn + h_\mn(x), \quad \text{with} \;|h_\mn| <<1,
 \ee
 where higher order in $h$ can be omitted due to the smallness of $h$. Furthermore, we use the $(-,+,+,+)$ sign notation for $\eta_\mn$ and the indices are raised with $\eta_\mn$ after the perturbative expansion, i.e.,
 $g^{\mn} = \eta^{\mn} - h^{\mn}.$ Afterward, we will look into the number of degrees of freedom the metric perturbation contains and discuss the most used gauge for fixing the unphysical degrees of freedom. Lastly, we will solve the Einstein equation for test masses far from the source of GWs. All the material in the first two chapters is based on the book ``Gravitational Waves: Volume 1: Theory and Experiments'' by Michele Maggiore \cite{maggiore} and ``Spacetime and Geometry. An ``Introduction to General Relativity'' by Sean Carroll \cite{carroll}. Another noteworthy reference is the book ``Gravitation'' by Charles W. Misner, Kip S Thorne, and John Archibald Wheeler \cite{MisnerThorneWheeler1973}. We recommend these references for an elaborate and detailed explanation of linearized general relativity and GWs.\\
 
\subsection{The linearized Einstein equations}

The familiar Einstein's equations are given by,
\be
G_{\mu\nu} \equiv R_\mn - \frac{1}{2}g_\mn R = \frac{8 \pi G}{c^4}T_\mn,
\ee
which relates the spacetime geometry, encoded in the metric $g_{\mn}$, to matter described by the energy-momentum tensor $T_{\mn}$.
The Ricci tensor $R_{\mn}$ and Ricci scalar $R$ for the linearized theory are computed following the usual scheme, starting from the Christoffel symbol. One can easily check that the linearized Christoffel symbol is given by
 \begin{align}
\Gamma_\mn^\rho =& \frac{1}{2}g^{\rho \sigma} [\partial_\mu g_{\nu\sigma}+\partial_\nu g_{\mu\sigma}-\partial_\sigma g_\mn]\nonumber\\
 =& \frac{1}{2}\eta^{\rho \sigma} [ \partial_\mu h_{\nu\sigma} + \partial_\nu h_{\mu\sigma} -\partial_\sigma h_\mn] + \order{h^2},
 \label{gamma}
 \end{align}
 which leads to the following Riemann curvature tensor,
 \begin{align}
 R^{\mu}_{\nu \sigma \rho} =& \partial_{\sigma} \Gamma^{\mu}_{\nu \rho} - \partial_{\rho} \Gamma^{\mu}_{\nu \sigma} + \Gamma^{\mu}_{\sigma \lambda}\Gamma^{\lambda}_{\nu \rho} - \Gamma^{\mu}_{\rho \lambda}\Gamma^{\lambda}_{\nu \sigma}\nonumber\\
 =& \frac{1}{2}[\eta^{\mu \lambda}(\partial_{\sigma}\partial_\nu h_{\rho \lambda} - \partial_{\sigma}\partial_\rho h_{\nu \lambda} - \partial_{\sigma}\partial_\lambda h_{\nu \rho}  - (\sigma \leftrightarrow \rho)] +\order{h^2}.
 \label{riemann}
 \end{align}
Note that the $\Gamma^2$ terms are higher-order terms in $h$ and will not contribute to the first-order Einstein equations. With a bit of algebra, one can find the Ricci tensor, 
\ba
R_\mn = R^{\rho}_{\mu \rho \nu} = \frac{1}{2}(\partial_{\rho}\partial_{\mu}h_{\nu}^{\rho}+\partial_{\rho}\partial_{\nu}h_{\mu}^{\rho}-\partial_{\mu}\partial_{\nu}h-\Box h_{\mn})+\order{h^2},
\end{align} and the Ricci scalar
\begin{align}
R=g^{\mn} R_\mn = \partial_{\mu}\partial_{\nu} h^{\mn}-\Box h +\order{h^2},
\end{align} with $h=h^{\mu}_{\;\mu}$ the trace and $\Box=\partial_{\mu}\partial^{\mu}$. Combining all the results gives us the linearized Einstein tensor
\begin{align}
G_\mn = \frac{-1}{2}[\Box h_\mn + \eta_\mn \partial^\rho \partial^\sigma h_{\rho \sigma}- \eta_\mn \Box h - \partial^\rho \partial_\nu h_{\mu \rho} - \partial_\rho \partial_\mu h_{\nu}^{\rho}+ \partial_\nu \partial_\mu h] +\order{h^2}.
\end{align}
This is a rather lengthy equation; hence it is usually preferred to define the trace-reversed quantity $\Bar{h}_\mn \equiv h_\mn - \frac{1}{2}\eta_\mn h$. This simplifies the equation a little to
\begin{align}
G_\mn = \frac{-1}{2}[\Box \Bar{h}_\mn + \eta_\mn \partial^\rho \partial^\sigma \Bar{h}_{\rho \sigma} - \partial^\rho \partial_\nu \Bar{h}_{\mu \rho}- \partial^\rho \partial_\mu \Bar{h}_{\nu \rho}] +\order{h^2}.
\end{align}

\subsection{Gauge conditions}

General relativity is invariant under all coordinate transformations $x^{\mu} \rightarrow x'^{\mu} (x)$, where $x'^{\mu}(x)$ is an arbitrary function of $x^{\mu}$. The metric will transform under this coordinate transformation as 
\begin{equation}
    g_{\mn} \rightarrow g_{\mn}(x') = \frac{\partial x^{\alpha}}{\partial x'^{\mu}}\frac{ \partial x^{\beta}}{\partial x'^{\nu}} g_{\alpha \beta}(x),
    \label{metrictrans}
\end{equation}
which is known as diffeomorphism invariance.\\

In the linearized theory, we expand around a fixed Minkowski background, $g_{\mn} = \eta_{\mn} + h_{\mn}$. The linearized equations are invariant only under infinitesimal coordinate transformations, $x^{\mu} \rightarrow x^{\mu} + \xi^{\mu}(x)$, which induce the linearized gauge transformation $h_{\mn} \rightarrow h_{\mn} - \partial_{\mu} \xi_{\nu} - \partial_{\nu} \xi_{\mu}$, and under finite, global Poincar\'{e} transformations that leave the Minkowski background $\eta_{\mn}$ unchanged.

Since the Minkowski metric $\eta_{\mn}$ is fixed, we choose small coordinate transformations under which $h_{\mn}$ slightly changes, although leaves $\eta_{\mn}$ unchanged. It is convenient to choose a fixed inertial coordinate system on the Minkowski background, in which $\eta_{\mn}$ explicitly exhibits its full set of symmetries: three spatial translations, three spatial rotations, one time translation, and three Lorentz boosts \cite{carroll}. Working in this frame allows us to decompose the perturbation $h_{\mn}$ based on its transformation under \emph{spatial} rotations on a hypersurface. Under these spatial rotations, the metric perturbation can be decomposed into scalars, vectors, and tensors, which transform independently from each other. This allows us to write the Einstein equations for the \emph{linearized theory} as a set of uncoupled ordinary differential equations.\\

\subsubsection{Scalar-vector-tensor decomposition}
\label{sec:svt}

Before discussing the gauge transformation, it is educational to have a closer look at the scalar-vector-tensor (SVT) decomposition. The metric perturbation $h_{\mn}$ is a $(0,2)$ tensor with a spatial $SO(3)$ symmetry. Under these rotations, the $h_{00}$ component is a scalar, $h_{0i}$ is a three-vector, and $h_{ij}$ is a spatial rank $2$ symmetric tensor \cite{carroll}. This tensor can further be decomposed into a trace and a trace-free part. In group theory language, these are the irreducible representations of the spatial rotation group. 

The components of the metric $g_{\mn}$ can then be written as \cite{Lifshitz:1945du,Mukhanov:1990me}
\ba
g_{00} = & -(1+2 \Phi), \nonumber\\
g_{i0} = & g_{0i} = 2(\partial_i B-S_i),\\
g_{ij} = & [(1-2\Psi)\delta_{ij} + 2 \partial_{ij} F + (\partial_i T_j + \partial_j T_i) + t_{ij}].\nonumber
\end{align}
Here, $g_{00}=-1$, and $g_{ij}= \delta_{ij}$ are components of the background (Minkowski) metric. The remaining terms are part of the perturbation $h_{\mn}$ consisting of $4$ scalars ($\Phi, B, \Psi,F)$, $2$ vectors ($S_i, T_i$), and $1$ tensor $(t_{ij})$, with
\be
\partial_i T^i = 0, \quad \partial_i S^i = 0, \quad
t^i_{\,i}=0, \quad \text{and} \quad \partial_i t^i_{j}=0.
\label{constraint}
\ee
We find $10$ independent functions in the decomposed metric. Namely, $4$ of the scalars, $4$ vector components, and $2$ tensor components of the $3 \times 3$ symmetric tensor $t_{ij}$. 
 
\subsection{Gauge transformation and gauge fixing}

Now, before we can start solving the linearized Einstein equations, we need to address the ambiguous definition of the perturbation $h_{\mn}$. The metric perturbation may have different forms depending on the choice of coordinate system. Indeed, if we consider an infinitesimal coordinate transformation 
\be 
x'^{\mu} = x^{\mu} + \xi^{\mu},
\ee 
then Eq.~\ref{metrictrans} tells us that the perturbation transforms as
\ba
h_{\mn} \rightarrow h'_{\mn} =  h_{\mn}- \partial_{\mu} \xi_{\nu} - \partial_{\mu} \xi_{\nu},
\label{coord-transformation-perturb}
\end{align} 
where $\xi$ is small such that the conditions for a linearized theory, $|h_{\mn}| << 1$, are preserved. Here, equation (\ref{coord-transformation-perturb}) corresponds to the Lie derivative along $\xi$. Since $h'_{\mn}$ is also a solution to the linearized Einstein equation, this is known as the \emph{gauge transformation} of the linearized theory and is entirely analogous to the gauge transformation in electrodynamics. That is, we know from electrodynamics that if the vector potentials $A_{\mu}$ solves the Maxwell equation, then $A'_{\mu} = A_{\mu} - \partial_{\mu} \psi$, with $\psi$ any scalar field, is also a solution. Hence, multiple potentials give rise to the same field strength and, thus, the same physical effects. Similarly, the Riemann tensor can be constructed from different metric tensors.\\

In general, the vector $\xi_\mu$ can be written in terms of two scalars and one transverse three-dimensional vector so that $\xi^{\mu} = (\xi^0, \partial_i f+f_i)$, with $\partial_i f^i=0$. Consequently, the scalar, vector, and tensor parts transform as
\begin{align}
    \Phi &\rightarrow \Phi + \partial_0 \xi_0 \nonumber\\
    B &\rightarrow B-\xi_0-\partial_0 f \nonumber\\
    \Psi &\rightarrow \Psi + \frac{1}{3} \nabla^2 f \nonumber\\
    F &\rightarrow F-2f\\
    S_i &\rightarrow S_i-\partial_0 f_i \nonumber\\
    T_i &\rightarrow T_i - f_i \nonumber\\
    t_{ij} &\rightarrow t_{ij} \nonumber,
    \label{eq:decomposition_svt}
\end{align}
which shows that the only gauge-invariant quantity is the transverse and traceless tensor $t_{ij}$, which contains only two degrees of freedom. One can show that the other vector and scalar parts do not dynamically propagate, i.e., do not satisfy a wave equation and can be gauged away \cite{carroll,MisnerThorneWheeler1973}.

A convenient choice of gauge, which is commonly used to fix $h_{\mu\nu}$, is the Lorenz gauge,\footnote{As a historical note, the Lorenz gauge makes reference to the seminal work of Ludwig Lorenz in 1867 on electrodynamics, where the gauge condition is $\partial_\mu A^\mu = 0$. His name is often confused with the name of Hendrik Lorentz, correctly associated with the Lorentz transformations.} $\partial_{\mu} \Bar{h}^{\mn} = 0$ \cite{maggiore}. Note that we are using the trace-reversed metric here. The Lorenz gauge is always applicable in Minkowski spacetime, as we show next. Assume an arbitrary perturbation for which $\partial^{\mu}\Bar{h}_{\mn} \neq 0$. Then under an infinitesimal coordinate transformation, this transforms as 
\begin{align}
 \partial^{'\mu}\Bar{h}'_{\mn}(x')=\partial^{\mu}\Bar{h}_{\mn}(x)-\Box\xi_{\nu}.   
\end{align}
By simply choosing $\Box \xi_{\nu}=\partial^{\mu}\Bar{h}_{\mn}$ the term on the left side will become zero, i.e., $\partial^{'\mu}\Bar{h}'_{\mn}(x')=0$. A solution can always be found since the d'Alembertian operator is invertible. Hence, by choosing the appropriate  $\xi^{\mu}$, any metric perturbation that initially does not obey the Lorenz gauge $\partial^{\mu}\Bar{h}_{\mn} \neq 0$, can always be written in the Lorenz gauge. Now that $\partial^{'\mu}\Bar{h}'_{\mn}(x')=0$, after a new gauge transformation we get,
\ba
\partial^{''\mu}\Bar{h}''_{\mn}(x'')&=\partial^{'\mu}\Bar{h}'_{\mn}(x')-\Box\xi_{\nu}\\
&=\partial^{\mu}\Bar{h}_{\mn}(x)-2 \Box\xi_{\nu}\\
&= -\partial^{\mu}\Bar{h}_{\mn}(x).
\end{align}
To remain in the Lorenz gauge, the vector $\xi_\mu$ needs to satisfy $\Box \xi_\mu =0$ (harmonic function). Hence, from the 10 independent components that we started with, the Lorenz gauge $\partial^{'\mu}\Bar{h}'_{\mn}(x')=0$ removes $4$ components; however, it still leaves some residual freedom for gauge transformations with $\Box \xi_\mu =0$. Here, $\Box \xi_\mu$ depends on 4 independent arbitrary functions $\xi_{\mu}$, as described above, and will therefore remove 4 more components. \\

We choose $\xi_0$ such that the trace of the trace-reversed metric is zero, i.e., $\Bar{h}^{\mu}_{\;\mu}=0$, and the functions $\xi_i$ are chosen such that $\Bar{h}_{0i} = 0$. Note that by making the metric traceless we get $\Bar{h}_{\mn}=h_{\mn}$. Furthermore, it follows that
$$\partial^{\mu} \Bar{h}_{\mu \nu} = \partial^0 \Bar{h}_{00}+\partial^i \Bar{h}_{0i} = 0,$$ where we fixed $\Bar{h}_{0i}=0$. Thus we have $\partial^0 \Bar{h}_{00} = 0$, i.e. $\Bar{h}_{00}$ is a constant in time. This static part of the metric is the Newtonian potential of the source. The GW is given by the time-dependent part of the metric and, as long as we are interested in GWs, we set $\partial^0 \Bar{h}_{00} =0$, which means $\Bar{h}_{00}=0$.

To summarize, we have set 
\ba
h^{TT}_{0\mu}=0, \quad h^{i \, TT}_{\,i}=0, \quad \partial^j h^{TT}_{ij}=0.
\label{TTgauge}
\end{align}
This is known as the transverse traceless gauge or TT gauge. After applying the TT gauge, we have reduced the symmetric metric with 10 degrees of freedom to only 2 degrees of freedom. These are the 2 degrees of freedom that describe the two polarizations of a GW. It is important to notice that the TT gauge described above in (\ref{TTgauge}) is valid for the \emph{vacuum solution}. In general, the linearized Einstein equations in terms of the trace-reversed metric will reduce to a simple expression:
 \ba
 \Box h^{TT}_{\mn} = \frac{-16 \pi G}{c^4} \Lambda_{\mn,\rho \sigma} T_{\rho \sigma}, 
 \end{align} where the lambda tensor $\Lambda_{\mn,\rho \sigma}$ is the TT projection operator that projects any rank-2 tensor to its symmetric, transverse and traceless subspace \cite{maggiore}. To define this operation, we can go to the Fourier space and introduce a transverse projector $P_{\mu \nu} = \delta_{\mu \nu} - \omega_{\mu\nu}$, with $\omega_{\mu\nu}=\frac{k_\mu k_\nu}{k^2}$, which is used to construct
\be\Lambda_{\mu\nu,\alpha\beta} \equiv \frac{1}{2}(P_{\mu\alpha} P_{\nu\beta} + P_{\mu\beta} P_{\nu\alpha}) - \frac{1}{D-1} P_{\mu\nu} P_{\alpha\beta}. \label{eq:tt_proj}
\ee
Here the Greek indices run from 0 to $D$, where $D$ is the spacetime dimension. We can use the same expression in Euclidean signatures. Note that $\Lambda_{\mu\nu,\alpha\beta}$ is transverse in \emph{all} indices, i.e, $k^\mu \Lambda_{\mu\nu,\alpha\beta} = 0$, and it projects out the trace $g^{\mu\nu}\Lambda_{\mu\nu,\alpha\beta} = g^{\alpha\beta}\Lambda_{\mu\nu,\alpha\beta} = 0$. 

The fact that we have only two propagating degrees of freedom here is consistent with the propagation of a massless particle. The fact that we were left only with a TT projection operator is consistent with the propagation of a spin-2 particle. That is why we can conclude that the vacuum solution is associated with the propagation of a free, \emph{massless spin-2 particle}, known as the \emph{graviton}.  See, for instance, more details on the projection operators in the App. C of \cite{LinodosSantos:2020pbu}, as well as generalizations for massive particles and other spins in \cite{Fierz:1939ix,Wigner:1939cj,Bargmann:1948ck,Weinberg:1995mt,Silva:2023pfc}.

\subsection{Vacuum solutions}
Let us think about a GW detector far away from any GW source. Hence, in a vacuum  where $T_{\mn}=0$, such that the Einstein equation reduces to
\ba
\Box h_{\mn}^{TT}=0.
\end{align} This has a plane wave solution,
\ba
h_{\mn}^{TT}(x) = A_{\mn}^{TT}(k)\sin{k^{\alpha}x_{\alpha}},
\end{align} where $k^{\alpha} = (\omega/c, \vec{k})$ is the wave vector, and $A_{\mn}(k)$ is the polarization tensor, which contains information about the amplitude of the GW and the polarization properties. Due to the restrictions imposed by the TT gauge, we can conclude that the amplitude is traceless and purely spatial. What is left to ensure that we are in the \emph{transverse} traceless gauge is to check if the perturbation is transverse. In other words, 
\ba
\partial^{\mu} h_{\mu\nu}^{TT}= k^{\mu} A_{\mn}(k) \sin{k^{\alpha}x_{\alpha}}=0.
\end{align}
This relation is true if the wave vector is orthogonal to the polarization tensor, $k^{\mu}A_{\mn}=0$. For example \cite{maggiore}, if the wave is propagating in the $\hat{z}$- direction, then $A_{z\nu}=0$. Considering that $A_{0\nu} = A_{\;i}^{i} = 0$ and  $A_{\mn}$ is also symmetric.

In this example, the projector $P$ in  Eq.~\ref{eq:tt_proj} has the form
\be 
P = \begin{pmatrix}
1 & 0 & 0\\
0 & 1 & 0\\
0 & 0 & 0
\end{pmatrix},\ee
and any arbitrary symmetric matrix takes the form
\begin{center}
\begin{tikzpicture}
\node at (0,0) {$A_{ij}^{TT}=\Lambda_{ij,kl} \underbrace{A^{kl}}_\text{\parbox{1.7cm}{Arbitrary symmetric $3\times3$ matrix}} =\begin{pmatrix}
\frac{1}{2} (A_{xx}-A_{yy}) & A_{xy} & 0\\
A_{yx} & -\frac{1}{2}(A_{xx}-A_{yy}) & 0\\
0 & 0 & 0
\end{pmatrix}.$};

\draw[->] (0,1.2) to [out=90, in=180] (0.5,1.7) node[anchor=west]{$h_+$};
\draw[->] (2.5,1.2) to [out=90, in=180] (3,1.7) node[anchor=west]{$h_{\times}$};
\end{tikzpicture}
\end{center}
Therefore, we can generally write
\begin{align}
\frac{1}{2}(A_{xx} - A_{yy}) &\equiv h_+,\nonumber\\
A_{xy} = A_{yx} &\equiv h_{\times}, 
\end{align}
and the rest is zero. Thus
\be
A_{\mn}^{TT}(k)=
\begin{pmatrix}
0 & 0 & 0 & 0\\
0 & h_+ & h_{\times} & 0\\
0 & h_{\times} & -h_+ & 0\\
0 & 0 & 0 & 0\\
\end{pmatrix}.
\ee
To check if this is a solution, we can plug it into the equation of motion:
\ba
\Box \Bar{h}_{\mn}^{TT} = k^{\alpha}k_{\alpha} A_{\mn}^{TT}(k) \sin{k^{\beta}x_{\beta}} = 0.
\end{align}
Note that not all components of $A_{\mn}$ are zero, which means that
\ba
k^{\alpha}k_{\alpha} =0\; \rightarrow \;\omega^2=c^2|\vec{k}|^2.
\end{align}
This is a rough proof that GWs travel at the speed of light in general relativity.

\subsection{Effects of gravitational waves on test masses}

To examine the effect of GWs on mass, let us first consider a single particle, with a geodesic trajectory parametrized by $x^{\mu}(\tau)$. The geodesic equation is given by
\be
\frac{d^2 x^{\mu}}{d\tau^2}+\Gamma^{\mu}_{\rho \nu}(x) \frac{dx^{\nu}}{d\tau}\frac{dx^{\rho}}{d\tau}=0.
\ee
We assume that the particle is approximately static, such that $\frac{d x^{\mu}}{d \tau} \approx (1,0,0,0)$, and we can assume that $\tau \approx x^0$. With these assumptions, the geodesic equation will be
\be
\frac{d^2 x^{\mu}}{d\tau^2} = -(\Dot{h}_{\mu 0}- \frac{1}{2}\partial_{\mu} h_{00}), \ee where the dot stands for the time derivative. After applying the TT gauge, we can see that in this gauge and at \emph{linear order}, single particles are not affected by GWs.\\

Now consider a second particle with a geodesic parametrized by $x^{\mu}(\tau)+\xi^{\mu}(\tau)$. This particle will satisfy the geodesic equation

\be
\frac{d^2 (x^{\mu}+\xi^{\mu})}{d\tau^2}+\Gamma^{\mu}_{\rho \nu}(x+\xi) \frac{d(x^{\nu}+\xi^{\nu})}{d\tau}\frac{d(x^{\rho}+\xi^{\rho})}{d\tau}=0.
\ee
It is assumed that $\xi$ is much smaller than the length scale of the GWs, such that we can expand in $\xi$ to the first order. Then, by taking the difference between the two geodesic equations, we obtain the geodesic deviation equation,

\be
\frac{d^2 \xi^{\mu}}{d\tau^2}+2\Gamma^{\mu}_{\rho \nu}(x) \frac{dx^{\nu}}{d\tau}\frac{\xi^{\rho}}{d\tau} +\xi^{\sigma}\partial_{\sigma}\Gamma^{\mu}_{\nu \rho}(x)\frac{dx^{\nu}}{d\tau}\frac{x^{\rho}}{d\tau}=0,
\label{geodev}
\ee
 describing the motion of the test particles relative to each other. We choose coordinates such that the Christoffel symbol vanishes ($ \Gamma_{\mu \rho}^{\nu}(x)=0$) at the spacetime position of the first point particle. The derivative of the Christoffel symbol, however, will not vanish. This choice is always possible \cite{maggiore}. 
 
For the covariant derivative of $\xi^{\mu}$ it is convenient to introduce basis vectors by $\hat{e}$, which are normalized to satisfy $\hat{e}_{\mu}\hat{e}^{\mu}=1$. Writing the vector as \(\boldsymbol{\xi}= \xi^{\nu} \hat{e}_{\nu}\), the covariant derivative is given by:
\begin{align}
\frac{D^2 \boldsymbol{\xi} }{d\tau^2} =& \, \frac{d^2 \xi^{\nu}}{d\tau^2} \,\hat{e}_{\nu} + 2  \Gamma_{\mu \rho}^{\nu} \frac{d \xi^{\mu}}{d\tau} u^{\rho}\, \hat{e}_{\nu} + \xi^{\mu} \partial_{\sigma} \Gamma_{\mu \rho}^{\nu} u^{\rho} u^{\sigma} \hat{e}_{\nu}\nonumber \\[4pt]
&+ \xi^{\mu} \Gamma_{\mu \rho}^{\lambda} \Gamma_{\lambda \sigma}^{\nu} \, u^{\rho} u^{\sigma} \hat{e}_{\nu} -  \xi^{\mu} \Gamma_{\sigma \rho}^{\lambda} \Gamma_{\lambda \mu}^{\nu}\, u^{\rho} u^{\sigma} \hat{e}_{\nu},
\end{align}
where we used $\Gamma^{\nu}_{\mu \rho} = \frac{d \hat{e}_{\mu}}{d x^{\rho}} \hat{e}^{\nu}$ and $u^{\sigma} = \frac{d x^{\sigma}}{d\tau}$.  By combining this equation with Eq.~\ref{geodev}, we obtain
\begin{align}
\frac{D^2 \boldsymbol{\xi}}{d\tau^2} =&\, \xi^{\mu} \left( \partial_{\sigma} \Gamma_{\mu \rho}^{\nu} - \partial_{\mu}\Gamma^{\nu}_{\sigma \rho} +  \Gamma_{\mu \rho}^{\lambda} \Gamma_{\lambda \sigma}^{\nu}-  \Gamma_{\sigma \rho}^{\lambda} \Gamma_{\lambda \mu}^{\nu}\right) u^{\rho} u^{\sigma}\, \hat{e}_{\nu} \nonumber \\[4pt]
=& \,  \xi^{\mu} R^{\nu}_{\mu \sigma \rho} \,u^{\rho} u^{\sigma}\, \hat{e}_{\nu}.
\end{align}
 If we assume a non-relativistic motion of the test particles, thus $\frac{dx^i}{d\tau} << \frac{dx^0}{d\tau}$, and we notice from Eq.~\ref{riemann} that $R^i_{0j0} = \frac{-1}{2c^2} \Ddot{h}_{j}^{i\;TT}$, the geodesic deviation equation can be reduced to a simpler form:
\be
\Ddot{\xi}^i = -c^2 R^i_{0j0} \,\xi^j = \frac{1}{2}\Ddot{h}_j^{i\;TT} \xi^j.
\label{displacement}
\ee
As an example \cite{maggiore}, let us consider the + polarization and study the motion of test particles in the $xy$ plane. In this case,
\be
h_{ab}^{TT} = h_+ \sin{\omega t}
\begin{pmatrix}
1 & 0\\
0 & -1
\end{pmatrix}, \quad a,b = \{x,y\}.
\ee
The distance between the particles can generally be written as
\be
\xi_a(t) = (X_0+\delta X(t), Y_0+\delta Y(t)),
\ee
where $(X_0, Y_0)$ are the unperturbed coordinates and $\delta X(t), \delta Y(t)$ are the displacements from the GWs. Eq.~\ref{displacement} results in
\ba
\delta \Ddot{X}&=\frac{-h_+}{2} (X_0+\delta X) \omega^2 \sin{\omega t},\\
\delta \Ddot{Y}&=\frac{h_+}{2} (Y_0+\delta Y) \omega^2 \sin{\omega t},
\end{align}
where the linear terms $\delta X$ and $\delta Y$ on the right-hand side can be neglected, since $\delta X << X_0$ and $\delta Y << Y_0$. Integrating the equations, we get
\be
\delta X = \frac{h_+}{2} X_0 \sin{\omega t}, \quad \delta Y = \frac{-h_+}{2}Y_0 \sin{\omega t}.
\ee
The result for a ring of test masses is shown in Fig.~\ref{fig:pluspol}.

\begin{figure}[t]
    \centering
\begin{tikzpicture}
\node at (0,2.5) {\emph{+ polarization:}};
\draw[->] (0,0)--(5,0) node[anchor=north]{x};
\draw[->] (2.5,-2.5)--(2.5,2.5) node[anchor=west]{y};
\draw[dotted] (2.5,0) circle (1.5);
\draw[->] (2.3, 1.6)--(2.3, 1.8);
\draw[->] (2.7, 1.6)--(2.7, 1.8);
\draw[->] (2.3,-1.6)--(2.3,-1.8);
\draw[->] (2.7,-1.6)--(2.7,-1.8);
\draw[->] (1.1, 0.2)--(1.3, 0.2);
\draw[->] (1.1, -0.2)--(1.3, -0.2);
\draw[->] (3.9, 0.2)--(3.7, 0.2);
\draw[->] (3.9, -0.2)--(3.7, -0.2);
\draw[dashed] (2.5,0) ellipse (1 and 2);
\draw[->] (4.5,-2)--(6.5,-2);
\node at (5.5,-1.8) {Time};
\draw[->] (6,0)--(11,0) node[anchor=north]{x};
\draw[->] (8.5,-2.5)--(8.5,2.5) node[anchor=west]{y};
\draw[dotted] (8.5,0) circle (1.5);
\draw[->] (8.3, -1.4)--(8.3, -1.2);
\draw[->] (8.7, -1.4)--(8.7, -1.2);
\draw[->] (8.3, 1.4)--(8.3, 1.2);
\draw[->] (8.7, 1.4)--(8.7, 1.2);
\draw[->] (10.1, 0.2)--(10.3, 0.2);
\draw[->] (10.1, -0.2)--(10.3, -0.2);
\draw[->] (6.9, 0.2)--(6.7, 0.2);
\draw[->] (6.9, -0.2)--(6.7, -0.2);
\draw[dashed] (8.5,0) ellipse (2 and 1);
\draw[white](5,-3)--(5,-4);
\end{tikzpicture}
\begin{tikzpicture}
\node at (0,2.5) {\emph{$\times$ polarization:}};
\draw[->] (0,0)--(5,0) node[anchor=north]{x};
\draw[->] (2.5,-2.5)--(2.5,2.5) node[anchor=west]{y};
\draw[dotted] (2.5,0) circle (1.5);
\draw[->,rotate around={45:(2.5,0)}] (2.3, 1.6)--(2.3, 1.8);
\draw[->,rotate around={45:(2.5,0)}] (2.7, 1.6)--(2.7, 1.8);
\draw[->,rotate around={45:(2.5,0)}] (2.3,-1.6)--(2.3,-1.8);
\draw[->,rotate around={45:(2.5,0)}] (2.7,-1.6)--(2.7,-1.8);
\draw[->,rotate around={45:(2.5,0)}] (1.1, 0.2)--(1.3, 0.2);
\draw[->,rotate around={45:(2.5,0)}] (1.1, -0.2)--(1.3, -0.2);
\draw[->,rotate around={45:(2.5,0)}] (3.9, 0.2)--(3.7, 0.2);
\draw[->,rotate around={45:(2.5,0)}] (3.9, -0.2)--(3.7, -0.2);
\draw[dashed,rotate around={45:(2.5,0)}] (2.5,0) ellipse (1 and 2);
\draw[->] (4.5,-2)--(6.5,-2);
\node at (5.5,-1.8) {Time};
\draw[->] (6,0)--(11,0) node[anchor=north]{x};
\draw[->] (8.5,-2.5)--(8.5,2.5) node[anchor=west]{y};
\draw[dotted] (8.5,0) circle (1.5);
\draw[->,rotate around={45:(8.5,0)}] (8.3, -1.4)--(8.3, -1.2);
\draw[->,rotate around={45:(8.5,0)}] (8.7, -1.4)--(8.7, -1.2);
\draw[->,rotate around={45:(8.5,0)}] (8.3, 1.4)--(8.3, 1.2);
\draw[->,rotate around={45:(8.5,0)}] (8.7, 1.4)--(8.7, 1.2);
\draw[->,rotate around={45:(8.5,0)}] (10.1, 0.2)--(10.3, 0.2);
\draw[->,rotate around={45:(8.5,0)}] (10.1, -0.2)--(10.3, -0.2);
\draw[->,rotate around={45:(8.5,0)}] (6.9, 0.2)--(6.7, 0.2);
\draw[->,rotate around={45:(8.5,0)}] (6.9, -0.2)--(6.7, -0.2);
\draw[dashed,rotate around={45:(8.5,0)}] (8.5,0) ellipse (2 and 1);
\end{tikzpicture}

\caption{\textit{A gravitational wave traveling in the $z$-direction with a $+$ polarization will curve spacetime such that a ring of test masses (gray dots) is alternating between a vertical and horizontal elliptical shape, creating a $+$ sign. A $\times$ polarized gravitational wave creates $\times$ sign as shown in the bottom figure.}}
\label{fig:pluspol}
\end{figure}

\section{Emission of gravitational waves}\label{sec:Emission of gravitational waves}
Having solved the linearized Einstein equation for the vacuum case in the last section, here we focus on the Einstein equation with a source term, i.e.,

\begin{equation}
\Box\Bar{h}_{\mn} = -\frac{16 \pi G}{c^4}T_{\mn}.
\label{withsource}
\end{equation}
After solving the equation above, we will discuss GWs in a curved background and derive the energy-momentum tensor for GWs. Later, we derive Einstein's quadrupole formula and the power spectrum of tensor perturbations. For a more detailed description, see the materials by Michele Maggiore \cite{maggiore}, Sean Carroll \cite{carroll}, and Charles W. Misner, Kip S Thorne, and John Archibald Wheeler \cite{MisnerThorneWheeler1973}.

\subsection{Gravitational waves emitted by a source}

Eq.~\ref{withsource} is solved by using the Green function for the d'Alembertian operator $\Box$,
\ba
\Box_x G(x^{\sigma}-y^{\sigma}) = \delta^{(4)}(x^{\sigma}-y^{\sigma}),
\end{align}
in exact analogy with the electromagnetic case.
The coordinates $x^{\sigma}$ and $y^{\sigma}$ are depicted in Fig.~\ref{fig:source}. The general solution is 
\ba
\Bar{h}_{\mn}(x^{\sigma})= - \frac{16 \pi G}{c^4} \int G(x^{\sigma}-y^{\sigma})\; T_{\mn}(y^0, \Vec{y}) \;d^4y,
\end{align}
with $G(x^{\sigma}-y^{\sigma}) = - \frac{1}{4 \pi |\Vec{x}-\Vec{y}|}\delta[|\Vec{x}-\Vec{y}| - (x^0-y^0)] \theta(x^0-y^0)$, and the theta function equals one when $x^0>y^0$ \cite{carroll}. After integrating over $y^0$, we obtain

\ba
\Bar{h}_{\mn}(t,\Vec{x})= \frac{4 G}{c^4} \int \frac{1}{|\Vec{x}-\Vec{y}|} T_{\mn}(t-|\Vec{x}-\Vec{y}|/c, \vec{y} ) d^3y,
\end{align} where $c t=x^0$.\\

In the following, we make the assumption that the source is far away (i.e., the far zone) and slowly moving. Hence, the distance to the source satisfies $r \approx \Vec{x} \gg \Vec{y}$ as is shown in Fig.~\ref{fig:source}. Expanding the spatial separation yields,
\begin{align}
    |\vec{x} - \vec{y}| = r - \hat{n} \cdot \vec{y} + \order{\frac{y^2}{r}},
\end{align}
with $\hat{n} \equiv \frac{\vec{x}}{r}$. Therefore, the retarded time is
\begin{align}
    t'= t - \frac{|\vec{x} - \vec{y}|}{c} = t- \frac{r}{c} + \frac{\hat{n} \cdot \vec{y}}{c} + \order{\frac{y^2}{cr}},
\end{align}
with the leading term $t_r \equiv t- \frac{r}{c}$, and
\begin{align}
    T_{\mu \nu} \left( t-\frac{|\vec{x} - \vec{y}|}{c}, \vec{y} \right) = T_{\mu \nu}\,(t_r, \vec{y}) + \frac{\hat{n} \cdot \vec{y}}{c} \,\partial_{t_r} T_{\mu \nu}\,(t_r, \vec{y}) + \dots .
\end{align}
We keep the leading time dependence such that the retarded time is reduced to $t'= t_r = t - \frac{r}{c}$, with $r=|\vec{x}|$.

\begin{figure}[t]
    \centering
\begin{tikzpicture}
\filldraw[black] (0,0) circle (1pt);
\draw[->](0,0)--(0.7,-0.65);
\draw (0.05,-0.4) node{\small $\Vec{y}$};
\draw plot [smooth cycle] coordinates {(-0.65,-0.6) (-1,-0.1) (-0.65,0.75) (-0.1,0.5) (0.3,0.8) (0.7,0.7) (1,0) (0.75,-0.75) (-0.3,-1) };
\draw[->] (5,0) -- (0.2,0);
\draw[->] (5,0) -- (0.9,-0.7);
\draw (2.5,0.3) node {$\Vec{x}$};
\draw (2.7,-0.8) node {$\Vec{r}=\Vec{x}-\Vec{y} \approx \Vec{x}$};
\draw[] (5.8,0)--(5.2,0.2);
\draw[] (5.8,0)--(5.2,-0.2);
\filldraw[black] (5.25,0) circle (1pt);
\draw (5.3,0.15) .. controls (5.2,0.1) and (5.2,-0.1) .. (5.3,-0.15);
\end{tikzpicture}
    \caption{\textit{An observer observes the center of a gravitational wave source at a distance $\vec{x}$. The outer edge of the source is at a distance $\vec{y}$ from the center and thus observed at a distance $\Vec{r}=\Vec{x}-\Vec{y}$. The source is a large distance from the observer, hence $\vec{r} \approx \vec{x}$.}}
    \label{fig:source}
\end{figure}
In turn, the gravitational wave takes the form
\be
\Bar{h}_{\mn}(t,\vec{x})=\frac{4G}{r c^4} \int d^3y\, T_{\mn}(t_r, y).
\ee
Hence, the wave observed at a time $t$ depends on its source at an earlier time $t_r$, considering that a wave travels at the speed of light over a distance $r$ before reaching the observer.
As seen before, in the vacuum case, the temporal components are set to zero, and thus we are only interested in the spatial components. This is given by
\be
\int d^3y \,T_{ij}(t_r, y) = \frac{1}{2} \partial^2_0 \int d^3y \,y_i\, y_j\, T_{00}(t_r, y).
\ee
To prove this relation, note that the energy/momentum conservation implies $\partial_{\mu} T^{\mn}=0$, and thus we can derive
\ba
\partial_{\mu} T^{0\mu} &= \partial_0 T^{00} + \partial_k T^{0k} =0 \nonumber\\
\partial_0^2T^{00} &= -\partial_k \partial_0 T^{0k} = \partial_k \partial_l T^{lk}\\
y_i y_j \partial_0^2T^{00} &= y_i y_j \partial_k \partial_l T^{lk} = 2 T_{ij} \nonumber 
\end{align}
In the second line, energy/momentum conservation was used again and the last step is obtained by partial integration and $\partial_k y_i=\delta_{ki}$. Thus 
\be
\Bar{h}_{ij}(t,\vec{x}) = \frac{2 G}{c^4} \frac{1}{r} \partial_0^2 \int d^3 y \; y_i y_j T_{00}(t_r, y),
\ee where it is conventional to define the integral as the tensor moment of the source $I_{ij}(t_r)$. The resulting formula,
\be
\Bar{h}_{ij}(t,\vec{x}) = \frac{2 G}{c^4} \frac{1}{r} \partial_0^2 I_{ij}\left(t_r \right),\ee
is known as the quadrupole formula.\\

The transverse traceless gauge for GWs outside the sources and propagating in $\hat{n}$ direction is found by projecting the solution onto the TT gauge given by Eq.~\ref{eq:tt_proj}. By projecting the metric perturbation, we obtain its traceless transverse version,
\be
h_{ij}^{TT} = \Lambda_{ij,kl}\; h_{kl}.
\ee
In the TT gauge, the metric perturbation is given by the quadrupole formula
\be
h_{ij}^{TT} (t_r, r) = \frac{2G}{c^4} \frac{1}{r} \Lambda_{ij,kl} \Idot{}^{\;kl}\left(t-\frac{r}{c}\right),
\ee
where the quadrupole moment is defined as
\be
\I{}_{kl} \equiv \int d^3 y (y_k y_l - \frac{1}{3} y^2 \delta_{kl}) T_{00} = I_{kl} - \frac{1}{3} I^m_{\;m} \delta_{kl}.
\ee 
This is the trace-free version of $I$. It is a bit redundant with the projector $\Lambda$, although it can be useful in practice to keep in this form. 

\subsection{Backreaction of gravitational waves on the background}
So far, we have considered linearized Einstein equations as an expansion around the flat spacetime metric $\eta_{\mn}$. Here, the definition of GWs is clear: The background, $\eta_{\mu\nu}$ is fixed and curvature-free. The fluctuations around the static flat background are associated with the GWs. This setting is useful in many occasions, although not suitable to study the backreaction of gravitational waves on the background geometry or act as a source of gravity. This is since the flat background is \emph{fixed} by assumption, and the equations describe waves propagating \emph{on} flat spacetime. The waves produce curvature at linear order, but they do not gravitate, and they do not carry energy–momentum as a source for the Einstein’s equations. In other words, GWs curve spacetime \emph{kinematically} (they create tidal forces), but they do not curve spacetime \emph{dynamically} (they do not change the background evolution).

Instead, we look at a curved, dynamical, background metric. In a general curved spacetime, however, with a metric 
\be
g_{\mn}(x) = \Bar{g}_{\mn}(x) + h_{\mn}(x),\ee
the separation of the metric into background and gravitational-wave perturbations is not unique. Both the background metric $\Bar{g}_{\mn}(x)$ and the perturbation $ h_{\mn}(x)$ can contribute to curvature. Since both slowly and rapidly varying metric components contribute to curvature, we can reshuffle $x-$dependent terms between $\Bar{g}_{\mn}$ and $h_{\mn}$. A scale separation is required to unambiguously identify gravitational waves \cite{maggiore}. 

A natural splitting arises by introducing a scale. Denoting a background length scale by $L_b$, and the gravitational-wave wavelength by $\lambda_{GW}$.\footnote{ Note that the typical length scale for the GW is $\lambda_{GW} =\frac{\lambda}{2 \pi}$ instead of $\lambda$ and is known as the reduced wavelength.} A suitable length scale $d$ is large enough to observe $\lambda_{GW}$ and small enough such that the background is approximately flat. This method of separation of the metric into a smooth background and perturbations is called short-wave (or high-frequency) expansion.

\begin{figure}[H]
\centering
\begin{tikzpicture}
\draw[] (-0.2,0) .. controls (1,1) and (2,-1) .. (3,0);
\draw[] (0,0) sin (0.05,0.4);
\draw[] (0.05,0.4) cos (0.1,0.1);
\draw[] (0.1,0.1) sin (0.15,-0.2);
\draw[] (0.15,-0.2) cos (0.2,0.2);

\draw[] (0.2,0.2) sin (0.25,0.5);
\draw[] (0.25,0.5) cos (0.3,0.2);
\draw[] (0.3,0.2) sin (0.35,-0.1);
\draw[] (0.35,-0.1) cos (0.4,0.3);
\draw[|-|] (-0.2,-0.5) --(3,-0.5) node at (1.5,-0.75) {$L_b$};
\draw[|-|] (-0.2,-1) --(0.8,-1) node at (0.3,-1.25) {$d$};
\draw[|-|] (0.2,0.75) --(0.45,0.75) node[anchor=south] {$\lambda_{GW}$};

\node at (7,0) {Order of scales: $\lambda_{GW} \ll d \ll L_b$.};
\end{tikzpicture}
\caption{\textit{A visualization of short-wave expansion. The spacetime is separated in a background with a length $L_b$, and a GW with wavelength $\lambda_{GW}$. This separation arises natural when considering a scale length $d$, such that $\lambda_{GW} \ll d \ll L_b$.}}
\end{figure}

\subsubsection{Splitting gravitational waves from the background}
How do gravitational-wave perturbations act as a source for the background evolution, and in turn affect the background geometry \cite{maggiore}? To address this question, we expand the Einstein equations around a background metric  $\Bar{g}_{\mn}$ and retain second-order terms in $h_{\mn}$. In this expansion, there are typically two small parameters, the amplitude $h$, and the ratio of length scales $\frac{\lambda_{GW}}{L_b}$ (equivalently, the frequency ratio $\frac{f_b}{f}$). 

Expanding the Einstein tensor $G_{\mn}$ in powers of $h_{\mn}$, we write
\be
G_{\mn} = G_{\mn}^{(B)} +G_{\mn}^{(1)} +G_{\mn}^{(2)}+ \dots,
\ee
where $G^{(B)}$ depends \emph{only} on the background metric $\Bar{g}_{\mn}$. The term $G_{\mn}^{(1)}$ is linear in $h_{\mn}$ and contains only high-frequency (short-wavelength) modes, while $G_{\mn}^{(2)}$ is quadratic and contains \emph{both} high and low frequencies. This mixing of scales arises naturally from quadratic terms. For example, consider a product $h_{\mn}h_{\rho \sigma}$, where $h_{\mn}$ and $h_{\rho \sigma}$ contain a mode with wave-vectors $\vec{k}_1$ and $\vec{k}_2$, respectively, with $|\vec{k}_1|, |\vec{k}_2|\gg\frac{1}{d}$. Their sum can generate a mode with $|\vec{k}_1 +\vec{k}_2| \ll\frac{1}{d}$, corresponding to a slowly varying (low-frequency) contribution. This allows the Einstein equations to be consistently separated into equations governing high-frequency gravitational waves and equations governing the slowly varying background.

We will focus on the small $\vec{k}$ part (low-frequency regime) of Einstein's equation
\begin{align}
    G_{\mn}^B &= -[G_{\mn}^{(2)}]^{small \,\Vec{k}} + \frac{8\pi G}{c^4}[T_{\mn}]^{small \,\Vec{k}}\nonumber\\
    &= - \expval{G_{\mn}^{(2)}}_d + \frac{8\pi G}{c^4}[T_{\mn}]_d,
    \label{g_mn}
\end{align}
where in the second line we have performed an average over a spatial region of characteristic size $d$. On the one hand, this does not affect the modes with a wavelength of order $L_b$, since these are more or less constant over a distance $d$. This averaging leaves modes with wavelength of order $L_B$ essentially unchanged, while rapidly oscillating contributions with wavelength $\lambda_{GW} \ll d$ average to zero. The resulting equations are known as the coarse-grained Einstein equations.

\subsubsection{Energy-momentum tensor of gravitational waves}
The averaged second-order $G_{\mn}$ is defined as the energy-momentum Pseudo-tensor of GWs
\begin{equation}
t_{\mn} = - \frac{c^4}{8 \pi G} \expval{G_{\mn}^{(2)}}_d =- \frac{c^4}{8 \pi G} \expval{R^{(2)}_{\mn} - \frac{1}{2}\Bar{g}_{\mn} R^{(2)}}.
\label{t}
\end{equation}
Thus, GWs carry energy that modifies the background, because of the way it enters in Eq.~\ref{g_mn}. An explicit computation of $G_{\mn}$ to second order in the TT gauge will give
\be
R_{\mn}^{(2)} = \dots = \frac{1}{4} \partial_{\mu} h_{\alpha \beta} \partial_{\nu} h^{\alpha \beta} + 12\, \text{terms},
\ee
\be
\expval{R_{\mn}^{(2)}}_d = - \frac{1}{4} \expval{\partial_{\mu} h_{\alpha \beta}^{TT} \partial_{\nu} h^{TT \alpha \beta}}, \quad \expval{R^{(2)}}=0.
\ee
Here, $\expval{R^{(2)}}$ vanishes after integration by parts and by using the equation of motion $\Box h_{\mn}=0$. The explicit expression for $t_{\mn}$ is then found by substituting $R_{\mn}^{(2)}$ into Eq.~\ref{t},
\be
t_{\mn} = \frac{c^4}{32 \pi G} \expval{\partial_{\mu} h_{\alpha \beta}^{TT} \partial_{\nu} h^{\alpha \beta TT}}.
\ee
Furthermore, the energy density of GWs is defined as the $00$ - component of the energy stress tensor and is given by
\be
\rho_{GW}=t_{00} = \frac{c^4}{32 \pi G} \expval{\Dot{h}_{ij}^{TT} \Dot{h}^{ij TT}}.
\label{eq:rho_00}
\ee

\subsection{Einstein's quadrupole formula}
Given the energy density of GWs, the energy of the gravitational radiation in a volume $V$ is given by
\be
E_{GW} = \int_V d^3x \, t^{00},
\ee
The effective stress–energy tensor of gravitational waves, defined at second order in the metric perturbations, is locally conserved, $\partial_{\mu} t^{\mn}$=0, as a consequence of the linearized Einstein equations and the background Bianchi identities.\footnote{This is possible because the Einstein equations are invariant under coordinate transformations (diffeomorphisms), which is manifested into a gauge symmetry at the field theory level. Once we have a symmetry, by following Noether's theorem, we can then obtain the associated conversed charges.} This conservation law allows one to convert the volume integral of $\partial_t t_{00}$ into a surface flux. First, we have

\be
\int_V d^3x (\partial_0 t^{00} + \partial_i t^{i0})=0, \ee
and then we can write
\be
\frac{d E_{GW}}{c \;dt}=- \int_V d^3x \,\partial_i t^{0i} = - \int_S dA\, n_i \,t^{0i},\ee
where $n_i$ is the outer normal to the surface and $dA$ the surface element of the volume $V$. Now, let S be a spherical surface at a large distance $r$ from the source. For a spherical volume, the surface element is $dA = r^2 d\Omega$, and its normal is $\hat{n}=\hat{r}$. Then
\be
\frac{d E_{GW}}{dt}= - \frac{r^2}{c} \int d \Omega \,t^{0r} = \frac{r^2}{c} \int d\Omega \, t^{00}.\ee
Hence, we have
\be
P_{GW}= \frac{d E_{GW}}{dt} = \frac{r^2 c^3}{32 \pi G} \int d\Omega \expval{\Dot{h}_{ij}^{TT} \Dot{h}^{ij\,TT}} = \frac{G}{8 \pi c^5} \int d\Omega \,\Lambda_{ij,kl}(\hat{n}) \expval{\dddot{I}^{ij} \dddot{I}^{kl}},\ee
which is known as Einstein's quadrupole formula:
\be
P_{GW} = \frac{G}{5 c^5} \expval{\dddot{I}_{ij} \dddot{I}_{ij}},\ee
describing the power emitted by a source with tensor moment $I_{ij}$. This allows us, for example, to compute the GW emitted by a black hole binary. Some physics intuition and analogy with the electromagnetic case can be gained from \cite{Dorsch:2022bhq}.

{\subsection{The power spectrum of tensor perturbations}

It is important to realize that the GW energy density depends on the two-point function of the time derivative of the tensor $h_{ij}$. In general, we can write the tensor $h_{ij}$ in Fourier space, at least in the following two ways, depending on whether we want to integrate the modes over a 3D volume or over a line of scalar modes:\footnote{Notice that the first Fourier transform maps us to the $h_a(\vec{k})$ modes, while the second transform maps us to $H_a(k)$ modes. The first modes can depend on the angular coordinates, while for the second modes, we assumed isotropy and they only depend on the radial coordinate $k$. This distinction is important because, as we see below, in the literature, the power spectral of density perturbations is defined in two different ways. Here we also emphasize that the Fourier modes $ h_a(\vec{k}) $ and $ H_a(k) $ have different units. In our notation, $[h_{ij}]=M^0$, therefore $ [h_a(\vec{k}) ] = M^{-3}$, while $ [H_a(k) ] = M^{-1}$. This observation is useful to understand the parameterizations of the power spectrum in the expression of the two-point correlation function in Eqs.~\ref{eq.:2ptTensor} and \ref{eq.:2ptSh}.}
\begin{align}
h_{ij}(t,\vec{x}) = & \sum_{a=+,\times} \int_V \frac{d^3\vec{k}}{(2\pi)^3} \, h_a(\vec{k})\; e^{i\,k_\mu x^\mu} \hat{e}_{ij}^a(\hat{n});
\label{eq:hij_Fourier} \\
= &  \sum_{a=+,\times} \int \frac{dk}{(2\pi)} \int_{S^2} d\Omega_{\hat{n}} \, H_a(k,\Omega)\; e^{i\,k_\mu x^\mu} \hat{e}_{ij}^a(\hat{n}),
\end{align} 
where $H_a(k, \Omega) \equiv \frac{k^2}{(2\pi)^2} h_a(\vec{k})$. We emphasize that the above equations are equivalent, and we are free to choose one based on convenience. In particular, with this definition, we have $\expval{H_a(k, \Omega) H_b(k',\Omega')} \propto \delta_{ab} \delta(k-k')\delta^2(\Omega-\Omega'),$ which is exactly what we want for an isotropic stochastic GW background. In these expressions, we already assumed that there are only two propagating degrees of freedom, the $+$ and $\times$ modes. The polarization tensors are given by $\hat{e}_{ij}^a$, which are the components of the polarization tensor that maps the Cartesian coordinates of the tensor $ h_{ij}^{TT}$ to the polarization modes $+,\times$. \\

The two-point function of $h_{ij}$ then depends on the two-point correlation function of the Fourier modes. In turn, we can relate the correlation functions to the \emph{power spectrum} of the theory \cite{Mukhanov:1990me}. Hence, we write
\begin{equation}
\langle  h_a(\vec{k}_1)  h_b(\vec{k}_2)  \rangle =  (2\pi)^3 \delta^3(\vec{k}_1+\vec{k}_2)  \delta_{ab} \dfrac{2\pi^2}{k_1^3} \mathcal{P}_t(k_1),\label{eq.:2ptTensor}
\end{equation}
where $\mathcal{P}_t(k)$ is the dimensionless \emph{power spectrum of tensor perturbations}.\footnote{Here, we can run into a circular argument if we are not careful with the definition. Indeed, we can also invert Eq.~\ref{eq.:2ptTensor} to define the power spectrum in terms of the correlation function of the Fourier modes. This is useful when we already know the details of the theory we work with. But we can also study model-independent scenarios in which we choose some benchmark scenarios, defined by some $\mathcal{P}_t(k)$, and then we obtain predictions for the GW spectrum, a strategy that has been quite used in the primordial GW phenomenology literature.}  Now, with  the Fourier expansion Eq.~\ref{eq:hij_Fourier}, and $\dot{h}_a \sim k h_a$, we can derive Eq.~\ref{eq:rho_00}. Assuming isotropy, the spatial average gives
\begin{align}
\rho_{\rm GW}
&= \frac{c^2}{32\pi G}
\sum_{a}\int \frac{d^3k}{(2\pi)^3}\,
k^2\,\langle |h_a(\vec k)|^2\rangle,\nonumber \\
&= \frac{c^2}{32\pi G}
 \int \frac{d^3k}{(2\pi)^3}\;
\frac{2\pi^2}{k}\mathcal P_t(k),
\end{align}
where one can perform the angular integral explicitly, $d^3k = 4 \pi k^2 dk.$ This becomes,
\be
\rho_{\rm GW}
= \frac{c^2}{32\pi G}
\int d\ln k \; k^2 \,\mathcal{P}_t(k).
\ee

It is helpful to parameterize the GW radiation emitted by a source through the \emph{GW spectrum}, defined by
\be 
\Omega_{\rm GW}\equiv\dfrac{1}{\rho_c}\dfrac{d \rho_{\rm GW}}{d \ln k}, \quad \text{with}\; \rho_c = \frac{3 H_0^2 c^2}{8 \pi G}.\label{eq:spectralshape}
\ee
Here, $\Omega_{GW}$ is normalized with the critical density $\rho_c$, which is the density required for a spatially flat universe. The symbol $H_0$ is the Hubble parameter, defined as $H(t) \equiv \frac{\dot{a}(t)}{a(t)}$, with $a(t)$ the scale factor, and the subscript $0$ denotes it is the Hubble parameter measured today. The Hubble parameter is derived from the Friedmann equation, $H^2 = \frac{8 \pi G}{3 c^2} \rho_{tot}$, it measures the expansion rate of the Universe, and shall be discussed further in section \ref{subsec:intro-cosmo}. 
Hence, we have
\begin{equation}
    \Omega_{\rm GW} =\frac{k^2}{12 H_0^2} \mathcal{P}_t(k).
\end{equation}
This dimensionless quantity, $\Omega_{\rm GW}$, tells how the GW energy density is distributed across the spectrum of frequency modes.\footnote{Here we define a frequency mode $f$ by using the following definition for the wavenumber length, $k=2\pi (f/a_0)$.} By knowing the power spectrum of tensor perturbations, we can fully characterize the GW spectrum. For instance, if the power spectrum is proportional to $k^{-2}$, the energy density is predicted to be logarithmically flat. Otherwise, the emitted spectrum is either blue--tilted (if it increases) or red--tilted (if it decreases) as a function of frequency. These concepts will be particularly useful for us in the next sections. We give some concrete examples in the last sections of these notes. \\

Alternatively, if our problem contains a background that is assumed to be Gaussian, stationary, unpolarized, spatially homogeneous, and isotropic, we can parameterize the background with the \emph{one-sided power spectral density} $S_h(k)$,
\begin{equation}
\langle  H_a(k_1,\Omega_1)  H_b(k_2,\Omega_2)  \rangle = \frac{1}{2} \delta_{ab} \frac{\delta(\Omega_1,\Omega_2)}{4\pi} (2\pi)\delta(k_1-k_2) S_h(k_1).\label{eq.:2ptSh}
\end{equation}
The one-sided power spectral density is a dimensional quantity. In the GW literature, $S_h$ is often multiplied by a frequency factor, giving rise to the \emph{characteristic strain}, a dimensionless quantity,
\begin{equation}
    h_c (f) = \sqrt{f S_h(f)}.
    \label{eq.:strain}
\end{equation}
}
The one-sided power spectral density and the characteristic strain are usually preferred in the GW and astrophysics literature, in comparison to the power spectrum of tensor perturbations, which appears more often in cosmology and particle physics papers.\footnote{If we scrutinize it more carefully, even though the GW background from inflation is Gaussian, it is actually non-stationary. However, for all practical purposes, it can be considered as stationary \cite{Allen:1999xw}.}

\section{The stochastic gravitational wave background}\label{sec:The stochastic gravitational wave background}

In the last sections, we generically described GWs, identifying the two propagating degrees of freedom from general relativity and the energy contained in GWs. Now, we specialize our studies on \emph{stochastic gravitational waves} by deriving the main properties and describing current searches. For more details, we encourage the reader to take a look at the references: \cite{Taylor:2021yjx} for the SGWB and PTAs, \cite{LIGOScientific:2016aoc} for ground-based interferometers,  \cite{LISA:2017pwj} for space-borne observatories, \cite{Regimbau:2011rp} for the astrophysical background,  and \cite{Caprini:2018mtu} for the cosmological background.

\subsection{The gravitational wave background}

Shortly, we can define the stochastic gravitational wave background (SGWB) as the composition of gravitational waves with different wavelengths, amplitudes, and phases, possibly emitted by astrophysical and cosmological sources.  If these waves are emitted with a certain frequency and with comparable amplitude, we are not able to tell where such waves are individually coming from, and the signals sum incoherently to produce a stochastic background. 

\begin{figure}[h]
\centering 
\includegraphics[width=0.75\textwidth]{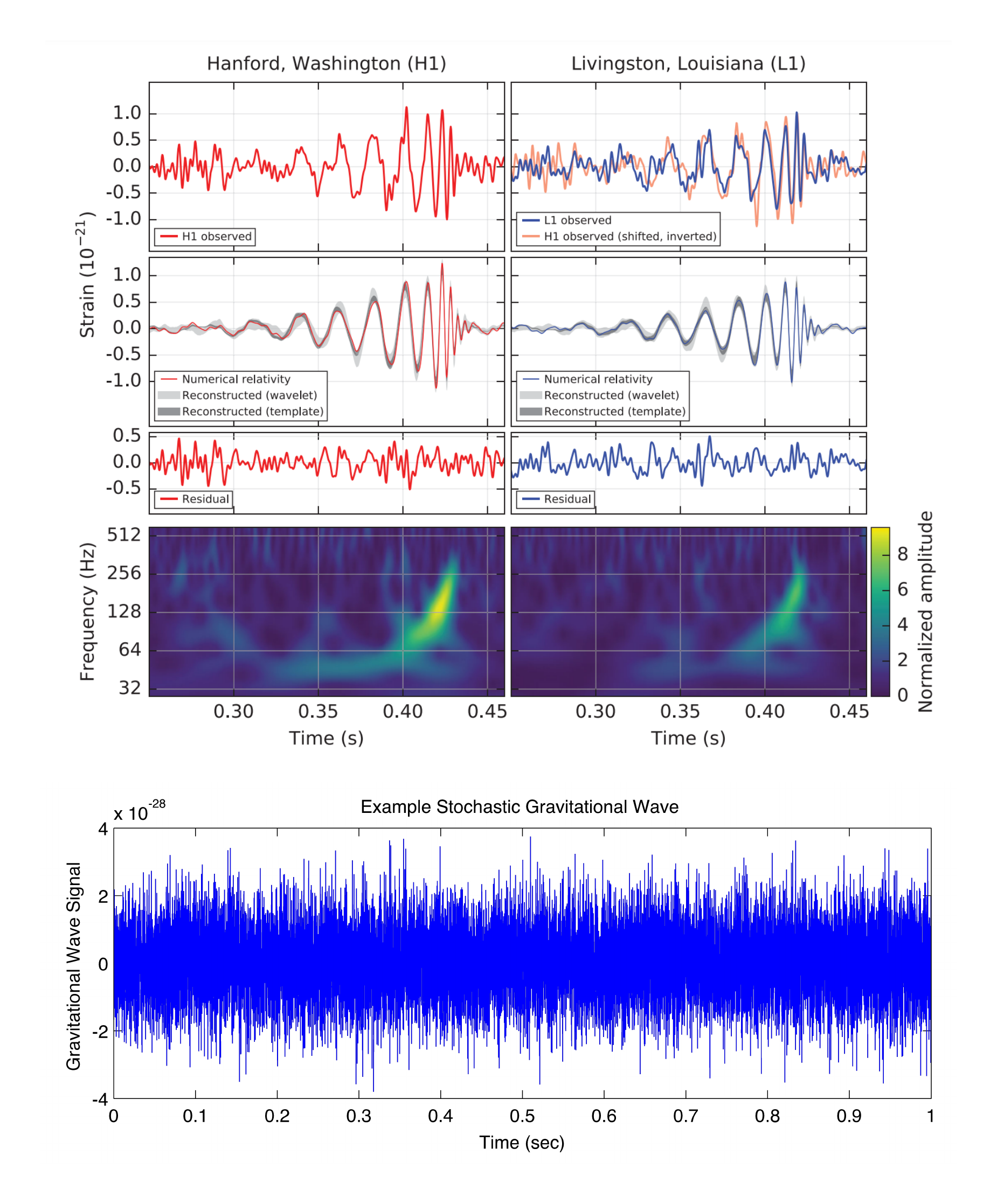} 
\caption{\textit{On top, we show the time-frequency data of the transient event GW150914 observed by the LIGO detectors in Hanford and Livingston. Notice that the event is well localized in time. In contrast, below, we sketch what we would expect from an SGWB signal in the time domain. We took the upper image from \cite{LIGOScientific:2016aoc} while the lower image is taken from the LIGO Science Collaboration website and is credited to A. Stuver/LIGO.}}
\label{fig:LIGO-waves}
\end{figure}

Notice that SGWB signals are different than the signals detected by LIGO-Virgo (transient signals) in their recent GW observations from binary mergers \cite{LIGOScientific:2016aoc,LIGOScientific:2017vwq}. These are transient signals that are very well characterized in time and are produced by the merger of black holes and/or neutron stars. By contrast, the signals of the SGWB are continuously reaching us, coming from all directions in the sky. See Fig.~\ref{fig:LIGO-waves}.

The term “stochastic” refers to the fact that this background is random noise and, therefore, can only be studied in terms of its statistical properties: the waves cannot be individually resolved by a detector, and, as a function of time, the background behaves like random noise. Interestingly, however, this noise is not useless or unwanted.\footnote{In some engineering processes, data processing, or even in daily life, the presence of noise is redundant or even unwanted: one needs to remove or filter the noise in order to access the desired piece of information. Just think about listening to your favorite song on a radio station on an old device with a lot of noise, or downloaded directly from the official album on the best earpods.} When we look at the spectral information (for instance, the dependence of the background on the frequency), we may start seeing non-trivial pieces of information, potentially telling something about the source(s) \cite{Taylor:2021yjx}. In this regard, in comparison to what is done with the cosmic microwave background (CMB) from the electromagnetic spectrum, the SGWB can allow us to study much higher frequencies and, therefore, stages in which the CMB cannot be used, since GWs could travel freely through the hot plasma of the early universe, while photons could not.

In these notes, we focus on the spectral properties and sources of the SGWB. We refer to the \emph{astrophysical background} as the background produced by astrophysical processes such as compact binary mergers, stellar activities, and bursts, in contrast to the \emph{cosmological background}, which might have been produced by several different types of cosmic sources, some of which are described in more detail in Secs.~\ref{sec:Primordial gravitational waves} and ~\ref{sec:Probing cosmology and BSM physics with the SGWB}.

\subsection{Sources}

In the case of an \emph{astrophysical background}, we do not expect the signal to be stationary, unpolarized, and isotropic \cite{Regimbau:2011rp}, assumptions often valid for a cosmological background. However, because there is a large number of sources in the sky and the time interval between events is small compared to the duration of a single event, it can result in a continuous background that obeys Gaussian statistics due to the central limit theorem \cite{Allen:1997ad}. As examples of these continuous waves, we mention the signal produced by a population of supermassive black hole binaries (SMBHBs) and the signal produced by individual, loud SMBHBs. Such a background could be detected by PTA searches \cite{NANOGrav:2023hfp,NANOGrav:2023pdq,EPTA:2023xxk,EPTA:2023gyr}. In addition, shot noise (small number of sources producing events with long time intervals between events) and popcorn noise (intermediate number of sources producing events with short duration followed by a second shot noise contribution) are astrophysical processes that might also be detected by ground-based interferometers \cite{Coward:2006df,Regimbau:2011rp,Cusin:2019jpv}.

We expect the astrophysical background to be present because we know these astrophysical sources are present in the sky -- for instance, the GW collaborations have already observed a diverse population of black hole binaries (no supermassive yet) \cite{LIGOScientific:2025slb}, while the Event Horizon Telescope (EHT) has directly imaged compact objects (most likely supermassive black holes) at the center of galaxies \cite{EventHorizonTelescope:2019dse,EventHorizonTelescope:2022wkp}, so we expect SMBHBs to be there. 

The detection of SGWB could also point out to the \emph{cosmological background}. Possible early-universe sources that emitted GWs in the past are related to the physics of cosmic inflation, scalar-induced gravitational waves,  phase transitions, and topological defects such as cosmic strings and domain walls. Signals from such SGWB are expected to be very small and will be challenging to detect because what arrives in the detector must be filtered out from all the other noise sources. Therefore, such waves serve as a cosmological history book, which is tricky to decipher.

These primordial sources are often related to beyond Standard Model theories. These theories usually rely on energy scales that are far beyond what Earth-based experiments can achieve. However, because the universe went through a high-temperature regime in the past, the signature of such BSM theories would be unavoidable in the SGWB, making their exploration very appealing to the BSM and cosmology communities.\footnote{For instance, current particle physics accelerators can reach center-of-mass energies of the order $\sim 10^{4}$ GeV, which is enough to probe the electroweak phase transition scale, $\sim 100$ GeV, and supersymmetric models that predict particles with mass of a few hundreds or thousands of GeV, but are far beyond the energy scale necessary for probing many BSM models, including some products of grand-unification theories, whose unification scale is of order $\sim 10^{16}$ GeV.}

Notice that the background does not need to respect all of the simplifying properties (stationary, unpolarized, Gaussian, and isotropic background). (Non-)detection of polarization, non-Gaussianity, and anisotropies would be ways to discern the origin of the signal, as each cosmological and astrophysical source has a different amount of these features. For example, cosmological sources are in many instances isotropic, while astrophysical sources contain not-so-small anisotropies; see, for example, \cite{LISACosmologyWorkingGroup:2022kbp,Pol:2022sjn,Sato-Polito:2023spo,NANOGrav:2023tcn,Gardiner:2023zzr} for a few works in this direction.

\subsection{Experiments}

\begin{figure}[H]
\centering 
\includegraphics[scale=0.5]{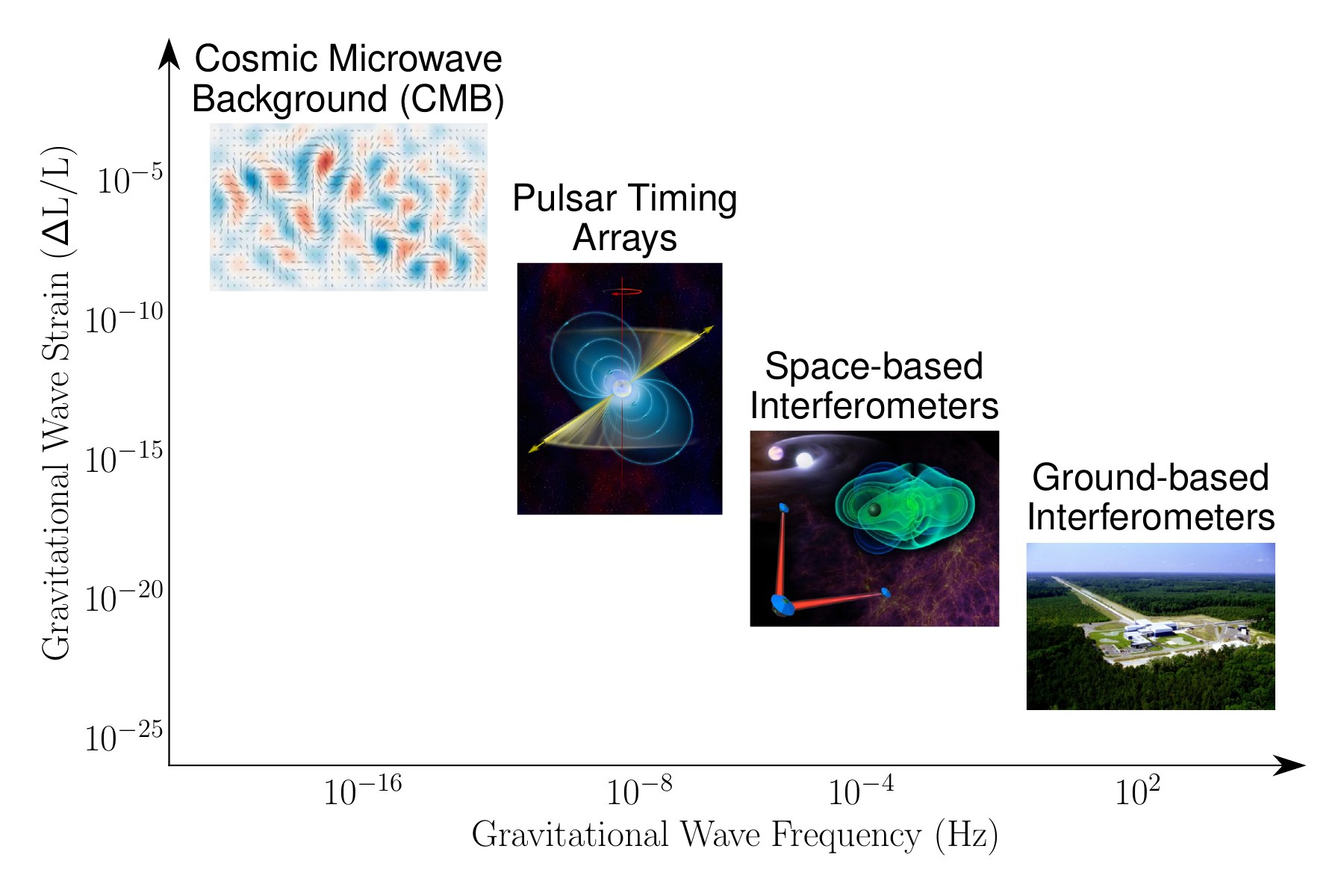} 
\caption{\textit{Scheme of different experiments probing the gravitational wave spectrum. Thanks to Michael Lam from the NANOGrav collaboration for sharing the template. From left to right, the images are credited to BICEP2, Bill Saxton at NRAO, eLISA, and LIGO, respectively.}}
\label{fig:experiments}
\end{figure}

Different experimental configurations can probe the SGWB: i) ground-based interferometers, ii) space-based observatories, and iii) pulsar timing arrays. Together, they cover different frequency ranges of the GW spectrum, see  Fig.~\ref{fig:experiments}. 
To understand how these observatories can detect GWs, we introduce the experiments in this subsection, so that we can discuss the idea behind interferometers and PTA searches in the coming section.

\subsubsection{Ground-based interferometers}\label{sec:interf}

LIGO, Virgo, and KAGRA \cite{LIGOScientific:2014pky, VIRGO:2014yos,KAGRA:2020cvd} (and their joint LIGO-Virgo-Kagra --- LVK ---collaboration) are collaborations based on ground-based interferometers. Their interferometers follow the idea behind the Michelson-Morley interferometers, see Sec.~\ref{sec:exp} and Fig.~\ref{fig:Michelson}.  Their detectors are sensitive to probe the background at frequencies in the range 10--1000 Hz. 

Up to the first part of the fourth LVK observing run, the LVK collaboration reported 218 events (128 new candidates) in their fourth gravitational-wave transient catalog (GTWC), which lists all events having at least $50\%$ of probability of origin from an astrophysical source rather than of terrestrial noise origin \cite{LIGOScientific:2020ibl,KAGRA:2021vkt,LIGOScientific:2025slb}. These astrophysical events are produced by the merging of binaries consisting of binary black hole coalescences, binary neutron star coalescences, and neutron star-black hole coalescences. Transient signals are well localized in time and space, as is shown in the famous Fig.~1 of Ref.~\cite{LIGOScientific:2016aoc} (upper image in our Fig.~\ref{fig:LIGO-waves}), in contrast to the SGWB. The LVK collaboration also put upper bounds on the background at LVK scales (around 100 Hz) \cite{LIGOScientific:2016wof}.

As the GW perturbs the spacetime, it delays photon beams traveling across the interferometer arms. In turn, this small time delay disrupts the perfect interference pattern that would exist without GWs. It is a common analogy to match the resulting GW strain to the ratio $\frac{\Delta L}{L} \sim \mid h_{\mu\nu} \mid \sim 10^{-21}$, where $L$ is the length of the arm and $(L \pm \Delta L) $ is the corresponding effective arm length, that would be necessary to cause such time delay. In the first GW detection in 2015 \cite{LIGOScientific:2016aoc}, the two LIGO interferometers observed a transient GW signal whose characteristic strain was of the order $| h_{\mu\nu} | \sim 10^{-21}$, with arm lengths of the order $L\sim 10^3$ m, resulting in a very small $\Delta L \sim 10^{-18}$ m. Compared with the proton dimension $10^{-15}$ m, we see that the GW interferometer must be extraordinarily sensitive! 

These experiments rely on very large arm lengths, and their detectors possess other enhancements to achieve higher sensitivities. Consider, for instance, a signal whose peak frequency is at $f\sim 100$ Hz, typical for binary mergers, such that $\lambda\sim L \sim \frac{1}{2\pi f} \sim 750$ km. For such wavelength, construction would be impossible. The trick is to use resonant Fabry-Perot resonant optical cavities, which reduce the required size to $L\sim 4$ km, by making the laser beam bounce between the two mirrors around 300 times. Moreover, the characterization of the fringe patterns in the photodetector is strongly dependent on the laser power. The detectors of LIGO operated close to 750 kW to detect those GWs in 2015. 

The combination of detectors improves the experimental power and sensitivity. Indeed, if one uses two detectors, one can analyze auto and cross-correlations: in the first GW detection by the Ligo-Virgo collaboration \cite{LIGOScientific:2016aoc}, there were two LIGO interferometers in the USA. For three detectors, one can analyze the 3D localization and polarization of isotropic SGWB: in the first multi-messenger GW detection from neutron-star binary merger \cite{LIGOScientific:2017vwq}, there were three interferometers (two LIGO detectors in the USA, one Virgo detector in Italy). Beyond LVK, the future generation of ground-based interferometers includes the Einstein Telescope \cite{Maggiore:2019uih}, and the Cosmic Explorer \cite{Reitze:2019iox}. These experiments probe the astrophysical and cosmological background in the $10-10^3$ Hz range, probing cosmology, new physics, and fundamental physics, see, for instance, \cite{Jenkins:2018nty,Belgacem:2018lbp,Nishizawa:2019rra,Bonilla:2019mbm,Cusin:2019jpv,Mastrogiovanni:2020gua,Nunes:2020rmr,LIGOScientific:2021aug,LIGOScientific:2021sio,Goncharov:2023woe}.

Because of the length of their arms ($\lambda_{\text{peak}}\sim L_{\text{arm}}$), the LVK collaboration can only probe frequencies in the $10-10^3$ Hz range. In turn, this range limits the mass of the black hole binaries they can probe. There are at least two ways to circumvent this problem: \emph{space-borne observatories}, by building a laser interferometer space antenna with huge arm lengths, such as LISA \cite{eLISA:2013xep}; and \emph{PTA searches}, by cross-correlating photon signals emitted by different pulsars -- PTAs are not interferometers, but if intuitively we imagined each pair pulsar-Earth as one arm, then in a PTA setting we would have an interferometer with galactic-sized correlated arms able to probe very large wavelengths.

\subsubsection{Space-borne observatories}

The idea behind a space-based GW observatory is to use laser interferometry in space, where three identical satellites composing a regular triangle follow Earth's orbit around the Sun. In these satellites, there are very stable test masses that free-fall in space while laser beams travel along the edges of the triangle (the arms of the interferometer). Therefore, likewise ground-based interferometers, when a GW passes by, there is a measurable time delay.

The Laser Interferometer Space Antenna (LISA) \cite{LISA:2017pwj} is the future space-based GW observatory to be launched in the 2030s. LISA will have three $2.5\times 10^{6}$ km arms that can probe frequencies in the mHz range (from below 0.1 mHz to above 0.1 Hz), with three detectors able to detect two independent signal channels. The LISA mission has already been adopted by the European Space Agency. Moreover, the Japanese Deci-hertz Interferometer Gravitational-Wave Observatory (DECIGO) \cite{Kawamura:2018esd,Kawamura:2020pcg} (probing GWs in the frequency band of 0.1 Hz to 10 Hz), the Chinese TianQin project \cite{TianQin:2015yph,TianQin:2020hid} and Taiji program \cite{Hu:2017mde,Luo:2021qji} (probing 0.1 mHz to 1 Hz) are other similar space-borne missions. 

These experiments would probe small frequencies that would require unrealistically large arms on Earth. For such frequencies, it will be possible to observe signals of ultra-compact binaries in our galaxy, supermassive black hole mergers, extreme mass ratio inspirals, and another plethora of cosmological possibilities associated with early universe physics \cite{LISACosmologyWorkingGroup:2019mwx,LISA:2022yao,LISACosmologyWorkingGroup:2022jok,LISA:2022kgy}, constituting the space-based GW observatories a promising probe of fundamental and new physics. 

\subsubsection{Pulsar timing arrays}

Pulsar timing array (PTA) searches are also based on time delay, but these searches use a different measurement method in comparison to the two interferometer strategies presented above: PTAs rely on the measurement of the time of arrival of radio pulses from millisecond pulsars. These pulses are disturbed by spatially correlated fluctuations induced by GWs.  As GWs perturb the metric along the Earth-pulsar lines, they modify the time of arrival of the radio pulses on Earth. A set of PTAs creates correlations across the baselines, while other noise sources are uncorrelated. Searches using PTAs then compare the measured spatial correlations with the expected values from the Hellings-Downs curve, the smoking gun for the isotropic, unpolarized background of quadrupole GW radiation \cite{1983ApJ...265L..39H,Jenet:2014bea,Taylor:2021yjx}. The so-called Hellings-Downs correlation is a prediction of GR that is crucial to characterize the SGWB and is studied in Sec.~\ref{sec:PTA_searches}.

PTAs can probe frequencies in the nHz range. These frequencies are lower than what ground- and space-based observatories can probe. In this range, supermassive black hole binaries are the primary source of GWs. Supermassive black holes have masses larger than $10^5$ solar masses, are present in the center of galaxies, and are much heavier than those producing the transient signals detected by the LVK collaboration. Since the frequency of GWs scales as the inverse of the binary chirp mass, neither LVK nor LISA can detect such supermassive black holes. On top of this astrophysical background, a cosmological background produced by early-universe and BSM physics can also be probed with PTAs \cite{NANOGrav:2023hvm}. 

NANOGrav \cite{Brazier:2019mmu}, PPTA \cite{Kerr:2020qdo}, EPTA \cite{Desvignes:2016yex}, as well as their joint international consortium IPTA \cite{Perera:2019sca} are among the most extensive PTA collaborations.\footnote{At the same frequency range, observations with the Square Kilometre Array (SKA)  \cite{Janssen:2014dka,Bacon:2018dui} promises to increase the sensitivity further in the following years.} Before June 2023, pulsar timing array (PTA) collaborations had claimed statistical evidence for excess noise that early-universe GWs could potentially explain, see, for instance, the reviews \cite{Caprini:2018mtu, Christensen:2018iqi,  Caldwell:2022qsj}. This excess noise was common red noise. In short, a common red noise process can be modeled with a frequency-dependent red-shifted spectrum across scales. Detection of GWs, however, depends on evidence for the Hellings-Downs correlation \cite{1983ApJ...265L..39H}. At the end of June 2023, enormous attention was given to the latest data release of the PTA collaborations. We could watch exciting results from NANOGrav, CPTA, EPTA, and PPTA. Even though we still do not have a $5-\sigma$ detection, for the first time, the collaborations found significant evidence for Hellings-Downs correlations\cite{NANOGrav:2023gor,EPTA:2023fyk,Reardon:2023gzh,Xu:2023wog}, strongly suggesting the presence of a GW signal in their datasets. We comment more on PTA searches in Sec.~\ref{sec:PTA_searches}.

For details on the past, present, and future of PTA collaborations, see, for instance, \cite{Taylor:2021yjx,Renzini:2022alw}.

\section{Searching for the background}
\label{sec:searching}

In this section, we discuss the searches for the background with terrestrial interferometers and PTA searches.  For more details, we encourage the reader to take a look at the references:\cite{Abbott:2016xvh} for searches with interferometers and \cite{Taylor:2014isa,Taylor:2021yjx} for PTA searches. A recent and useful reference for comparing searches for the SGWB in interferometers with those in PTAs is the ``FAQ'' by Romano and Allen \cite{Romano:2023zhb}.

\subsection{Searches with interferometers}

Here we describe how we can detect GWs using Michelson interferometers. This is the technique used by LIGO-Virgo in their first detections. The experiment consists of probing the interference pattern induced by GWs. 

\subsubsection{Experimental setting}\label{sec:exp}

\begin{figure}[H]
\centering 
\includegraphics[scale=0.5]{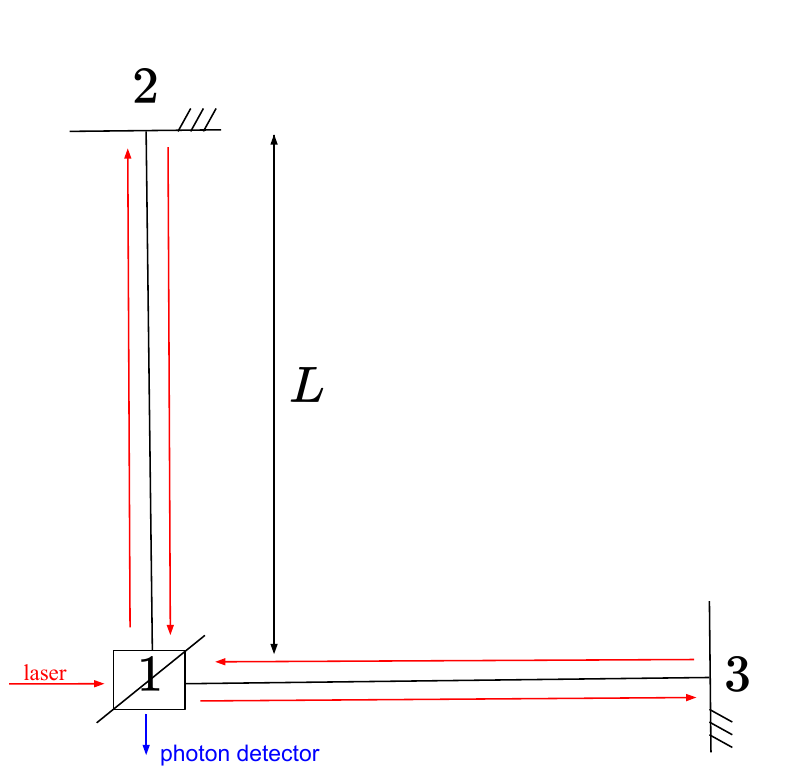} 
\caption{\textit{Michelson interferometer: the experiment consists in splitting laser light beams through a beam splitter at 1 in two different paths (here each arm has length $L$), reflecting them with two mirrors located at 2 and 3, and recombining them back in 1 so that interference patterns can be created (in particular, perfect destructive interference) in the photon detector (dark fringe).}}
\label{fig:Michelson}
\end{figure}

According to the setting in Fig.~\ref{fig:Michelson}, we need to compute the time delay associated with light departing and returning to 1. There are at least two possible frames: in the TT frame, GWs change the photon propagation, and the ``free-falling'' mirrors do not move; in the proper detector frame, GWs change the distance between the beam-splitter and mirrors. The frames are equivalent, and we are free to choose one based on convenience.

Let us work in the TT frame. If light travels with $c=1$, it takes $L$ to travel from 1 to the mirror and $2L$ back to 1. In the $\hat{l}=\vec{x}$ direction, the time delay for a light signal emitted at time $t$ is given by:\footnote{Here, it suffices to use flat spacetime. There is no relevant backreaction of the GWs in the spacetime where they are propagating, according to Eq.~\ref{sec:Emission of gravitational waves}. }
\be 
\Delta T (t) =  \dfrac{1}{2} \hat{l}^a  \hat{l}^b\int_0^L h_{ab}(t+s,\vec{x}+s\hat{l})ds, \qquad a,b=1,2,3;
\label{eq:DeltaT}
\ee
where 
\be 
h_{ab}(t,\vec{x})=\int d^3k \; e^{-2\pi i \vec{k}\cdot \vec{x}} \sum_\lambda \hat{e}_{ab,\lambda}(\hat{k})h_\lambda(t,\vec{k}).
\ee
The time delay, to be measured at time $t$, associated with the return trip is $$\Delta T_{12} (t-2L) + \Delta T_{21} (t-L). $$ Apart from GWs, other sources can also provoke interference, and therefore, we add some noise $n_1(t)$.\footnote{This noise can be seismic noise, thermal noise (fluctuation of individual atoms in the mirror), quantum noise (statistical uncertainty of photon counting), residual gas noise (interaction of gas particles with the mirror and light), among others. They must be quantified and define the limiting sensitivity of a detector. See, for instance, Fig. 5 in \cite{Abbott:2016xvh}.} Thus, we write 
\bea 
 s_{12}(t)&=&\Delta T_{12} (t-2L) + \Delta T_{21} (t-L)  + n_1(t) \\
 & = & L \int \dfrac{d^3k}{(2\pi)^3}\sum_\lambda I_\lambda^{12}(\vec{k})h_\lambda(t-L,\vec{k})+ n_1(t),
\eea
where we have defined the single-arm detector response as
\begin{align}
I_\lambda^{12}(\vec{k}) = & \dfrac{1}{2}  \hat{l}^a  \hat{l}^b\hat{e}_{ab,\lambda}(\hat{k}) \;e^{-2\pi i \vec{k}\cdot\vec{x}} \nn \\
&\times\left( e^{-\pi i k L ( 1 + \hat{k}\cdot\hat{l})}\dfrac{ \sin(\pi  k L ( 1 - \hat{k}\cdot\hat{l}))}{\pi k L ( 1 - \hat{k}\cdot\hat{l})} + e^{\pi i k L ( 1 - \hat{k}\cdot\hat{l})}\dfrac{ \sin(\pi  k L ( 1 + \hat{k}\cdot\hat{l}))}{\pi k L ( 1 + \hat{k}\cdot\hat{l})} \right).
\end{align}
Notice that this expression tells us the direction in which the detector is more sensitive! It also tells us something about frequency modes: the terms in the last bracket tend to be 2 for $k L << 1$ (small-frequency modes) and 0 for $k L >> 1$ (large-frequency modes). From the response function, there is suppression for both small and large frequency modes. For the large ones, the signal drops as $ (\sin x)/x  $ for $ k>>1/L $. For the small ones, the response function is constant for $ k<<1/L $. As the noise grows at low frequencies, sensitivity is lost.

\subsubsection{Overlap reduction function}
\label{sec:overlap_red}

\begin{figure}[t]
\centering 
\includegraphics[scale=0.5]{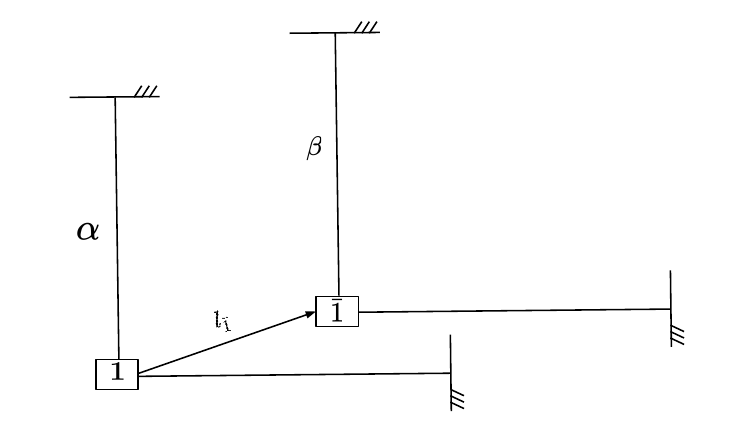} 
\caption{\textit{Two Michelson interferometers. We put the origin of the coordinate system at 1. To describe the location of the second detector, we need to include the vector from 1 to $\bar{1}$, where $\bar{1}$ is the central point of the second detector. Therefore, $ \expval{s_\alpha s_\beta} $ depends on the distance between the interferometers.}}
\label{fig:Michelson2}
\end{figure}

Now we are ready to correlate more than one interferometer. For this purpose, we distinguish interferometers with Greek labels. Assuming an isotropic SGWB, the measured time delay in the interferometer $\alpha$, $s_\alpha$, can be averaged 
\be 
\expval{s_\alpha^2}=L^2\int  \dfrac{d^3k}{(2\pi)^3}\sum_\lambda P_\lambda (k) \mid I_\lambda^{12} - I_\lambda^{13}\mid^2 + \expval{n^2},
\ee
where $ P_\lambda $ is the non-normalized isotropic power spectrum defined by
 \be 
\langle h_\lambda(\vec{k}_1) h_{\lambda^{\prime}}(\vec{k}_2)\rangle = (2\pi)^3 \delta_{\lambda\lambda^{\prime}}\delta^3(\vec{k}_1-\vec{k}_2)P_\lambda(\mid\vec{k}_1\mid), \label{eq:powerspectrumdef}
\ee
whose normalized version was introduced in Eq.~\ref{eq.:2ptTensor}. Above, $ I_\lambda^{12} - I_\lambda^{13} = I_\lambda^\alpha $, and $\expval{n^2} $ is the instrumental noise from different possible sources. While $ P_\lambda $ only depends on the GW signal, $ I_\lambda^\alpha $ depends on the detector response. In such cases, the main obstacle is noise. Consider that noise severely affects the data for very tiny time delays (which we expect for GWs). Now, assume that there are two interferometers $ \alpha $ and $ \beta $, as shown in Fig.~\ref{fig:Michelson2}. 
Then,
\be 
\expval{s_\alpha s_\beta} = L^2 \int  \dfrac{d^3k}{(2\pi)^3} \sum_\lambda P_\lambda (k) I_\lambda^\alpha{}^{*}(\vec{k},\vec{x}_1) I_\lambda^\beta(\vec{k},\vec{x}_2), \label{eq:sasb}
\ee
i.e., there is cross-correlation. Ideally, the two detectors are far away, so their instrumental noises are not correlated $\expval{n_\alpha n_\beta} = 0 $. The signal is reduced by \emph{overlap reduction function}, which is essential for detecting stochastic signals. Note that, although we got rid of the noise by considering two detectors, the measured time delay now depends on the distance between them, see Fig.~\ref{fig:Michelson2}. The overlap reduction function, therefore, depends on the response of the individual detectors and their relative geometry.\footnote{For pulsar timing arrays, the overlap reduction function is precisely what is known as the \emph{Hellings-Downs curve} \cite{Finn:2008vh}, which we study in the Sec.~\ref{sec:PTA_searches}.}

\subsubsection{Monopole response function}

\begin{figure}[t]
\centering 
\includegraphics[scale=0.4]{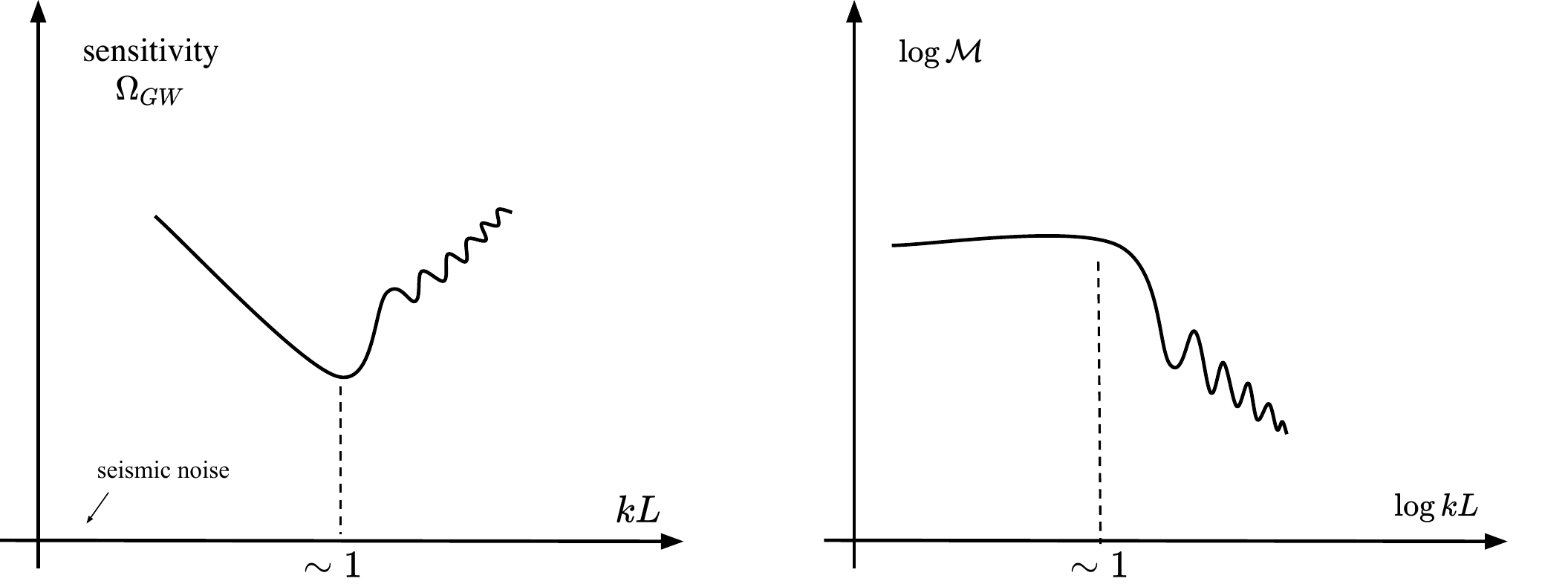} 
\caption{\textit{On left, sensitivity of gravitational waves versus $kL$ (length of detector $L$), given dependence of signal on the correlation $\expval{s_\alpha^2}$ through Eq.~\ref{eq:signaldetector}. For small wavelengths, seismic noise is an obstacle. For large wavelengths, the monopole response oscillates too much. On the right, log-log plot of the monopole detector response versus $kL$. The detector response drops for $kL>>1$ (averaging over many oscillations, see Eq.~\ref{eq:monopole}), and it is constant for $kL<<1$.}  }
\label{fig:sensitivity}
\end{figure}

Consider an isotropic, unpolarized SGWB, for which $P_+ = P_\times $. For this case, we can define the monopole response function as
\be  \mathcal{M}(k) \equiv \sum_\lambda \int d\Omega \mid I_\lambda^\alpha \mid^2. \label{eq:monopole}
\ee 
Then, the average time delay is
\be 
\dfrac{\expval{s_\alpha^2}}{L^2}=\dfrac{1}{8\pi}\int d\,k \sum_\lambda \left( k^2  P_\lambda (k) \right)  \left ( \int d\Omega \mid I_\lambda^\alpha \mid^2 \right ) = \dfrac{1}{4\pi}\int d(\ln k) \underbrace{\left( k^3  P_+ (k) \right)}_{\mathcal{P}_t(k)} \mathcal{M}(k). \label{eq:signaldetector}
%  = \dfrac{1}{8\pi}\int d(\ln k) \Delta_t^2 \left ( \sum_\lambda \int d\Omega \mid I_\lambda^\alpha \mid^2 \right ) = \dfrac{1}{8\pi}\int d(\ln k) \Delta_t^2 \mathcal{M}(k)
\ee
The dependence on the source is due to the power spectrum $ \mathcal{P}_t(k) $, while the dependence on the configuration of the detector is due to $ \mathcal{M}(k) $. For instance, for the LIGO-Virgo detectors, the two detector arms are oriented perpendicularly to each other. Their sensitivity and monopole response function depend on the wave number, as shown in Fig.~\ref{fig:sensitivity}. \\

The result in Eq.~\ref{eq:signaldetector} is quite representative of how we can probe sources with GWs, and we highlight it here: by measuring the time delay ($\expval{s_\alpha^2}$) and knowing the configuration of the detector ($ \mathcal{M}(k) $), \emph{it is possible to probe the source!} A similar expression will be derived for PTA searches below.\\

\begin{figure}[t]
\centering 
\includegraphics[scale=0.5]{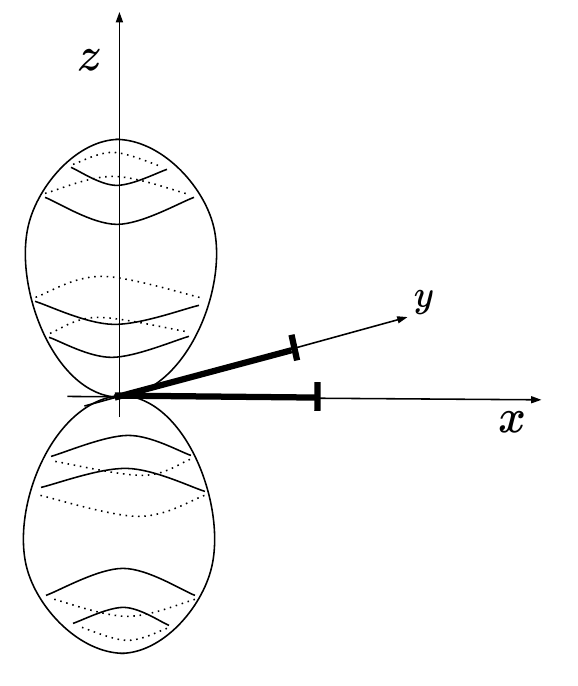} 
\caption{\textit{We sketch one example of an antenna pattern, in which we can see two main lobes oriented along the $z-$axis. The interferometer and detector are located in the $xy$ plane. The antenna pattern is a property of the detector that can be useful for detecting anisotropy, as the detector can be ``tuned'' to receive a signal emitted with its own antenna pattern.} }
\label{fig:nonisot}
\end{figure}

Finally, we note that for polarized sources, $P_\times (k) \neq P_+ (k)$, we should repeat the computation and keep $\expval{ h h }$ inside the sum $\sum_\lambda$. We can search for polarization through the multipoles that could be present in the signal. For non-isotropic sources, we should repeat the computation and keep $\expval{ h h }$ in the integral over $d\Omega$. We can search for anisotropies through antenna patterns, see Fig.~\ref{fig:nonisot}.\footnote{Antenna radiation patterns are graphical representations of the antenna directional characteristic, i.e., how the relative intensity of the energy radiation (amount of  electric and magnetic field strength) emitted by the antenna or source depends on the angular dependence. Due to the Rayleigh-Carson reciprocity theorem \cite{Carson:1924,Carson:1930,Potton:2004}, antennas have a receiving pattern that is identical to the far-field radiation pattern of the transmitting antenna, i.e., the radiation source. This is a property highly used in telecommunications. In our case, we would get information about the source once an anisotropic GW signal matches a given antenna pattern in the GW detector.}

\subsection{PTA searches}
\label{sec:PTA_searches}

\begin{figure}[t]
\centering 
\includegraphics[scale=0.5]{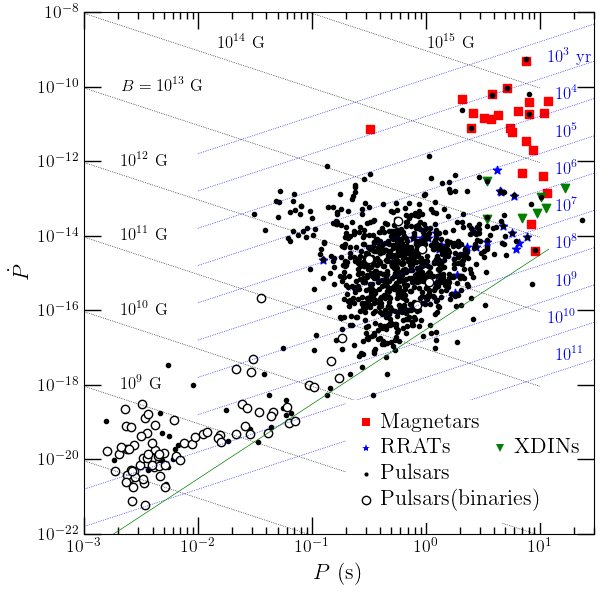} 
\caption{\textit{In this plot we can see the distribution of pulsars according to their slow-down rate (in the vertical axis, $\dot{P}$), their spin period (in the horizontal axis ($P$), their surface magnetic field (constant diagonal contours decreasing from left to right) and age (constant diagonal contours decreasing from right to left). Millisecond pulsars are located close to the bottom left of the plot. Image credited to A. W. Steiner in \url{https://neutronstars.utk.edu/code/nstar-plot/ppdot.html}}}
\label{fig:pulsars-msp}
\end{figure}

The idea behind PTA searches is to use millisecond pulsars in the nearby galactic environment and their precise time dependence. 
Pulsars are rotating neutron stars that are born in supernova explosions. Due to their rotation, pulsars appear as lighthouses, although the radiation they emit is generally not in the visible light part of the spectrum \cite{1968Natur.217..709H,1968Natur.219..145P,Gold:1968zf,1969ApJ...157..869G,Sturrock:1971zc}. Pulsars usually start rotating with periods 10--20 ms and then gradually slow down to periods of order 1 s, while \emph{millisecond pulsars} \cite{1982Natur.300..615B,1982Natur.300..728A} possess short rotations with period $P  < 100$ ms, and with a very low slow-down rate $\dot{P} < 10^{-17}$ \cite{Manchester:2017azn}. This is shown in Fig.~\ref{fig:pulsars-msp}, where we can also see that they are relatively older and have weaker magnetic fields at the surface. Due to their extreme stability of spin period (time between the flashes we observe on Earth), millisecond pulsars can be used as extremely precise clocks. Here, the idea is that GWs would interact with the photons emitted by millisecond pulsars and, consequently, redshift the pulses that arrive in a radio telescope on Earth.\footnote{From now on, we omit the name ``millisecond''. It must be understood that pulsars used in PTA searches are millisecond pulsars. } Here, we also highlight that the first indirect demonstration of the existence of GWs dates back to the discovery of binary pulsars in 1974 by Hulse and Taylor\cite{1975ApJ...195L..51H}, following up on the discovery of pulsars led by J. Bell just a few years before, in 1967 \cite{1968Natur.217..709H}.

\subsubsection{Setting}

The frequency of the GW that can be observed with PTA searches is constrained by how long observations are performed, and by how small the observing cadence is. The minimal accessible frequency is $f_{\text{min}}\sim 1/T_{\text{obs}} \sim 10^{-9}$ Hz, for $T_{\text{obs}}\sim10$ years of total observing time. The maximal accessible frequency depends on data sampling patterns but is usually assumed to be half the sample rate (Nyquist frequency) given by $f_{\text{max}}\sim 1/(2\,T_{\text{obs. cadence}}) \sim 10^{-6}$ Hz if the telescope can be used once every week.\footnote{The reconstructed signal is different than the original (continuous) one because the original frequency is too high to be able to be reconstructed with the given device. In order to record higher frequencies, the device needs higher resolution. In signal processing, this problem is known as aliasing. In the context of PTAs, it means one needs higher-cadence observations if one wants to increase the upper-limit frequency. In practice, one would not be able to indefinitely increase this frequency because of Poisson noise (shot or self noise). Optimistically assuming the same pulsar can be observed once per week, it gives an upper limit of $8\times10^{-7}$ Hz; if we consider LISA works around the $m\text{Hz}$ frequency band, it gives an unobservable gap around the $\mu\text{Hz}$ frequency band ($10^{-7}$---$10^{-4}$ Hz). See, for instance, \cite{Blas:2021mqw} on how to bridge this gap with binary resonant searches, or \cite{Wang:2020hfh} on how one could go beyond such upper limit without high-cadence observations.}

As described in Sec.~\ref{sec:overlap_red}, cross-correlation helps to distinguish a GW signal from other noise sources. In the PTA context, this means that signals of multiple pulsars must be observed. Huge radio telescopes on Earth have observed these pulsars for several years. Then, by cross-correlating the accumulated timing observations and designing precise noise models, we can investigate whether the overlap reduction function is consistent with the so-called Hellings-Downs correlation \cite{1983ApJ...265L..39H}, crucial to characterize the SGWB, studied in the next subsection.

Importantly, every pulsar has a \emph{timing model} that accounts for every known influence on the photon propagation between the pulsar and the radio telescope. The difference between the time of arrival and the predictions of the timing model for a pulsar gives the \emph{timing residual} of that pulsar. Everything that is not included in the timing model accounts for timing residuals. This includes the desired signal of GWs, but also other sources of noise that must be removed.

In general, the timing model includes timing corrections from pulsar dynamics (pulsar rotation period, rotation period variation), interstellar dispersion (delay by the column of electrons in the interstellar space known as dispersion measure (DM), DM variations, solar electron density), pulsar-Earth system dynamics  (Keplerian and relativistic orbital elements, kinematic perturbations) and Solar System dynamics (Solar System ephemeris, parallax, and relativistic corrections such as Roemer delay, Shapiro delay, and Einstein delay). For more details, we refer to \cite{Pirani:1956tn,1975GReGr...6..439E,Taylor:2021yjx,NANOGrav:2023hde}. 

Phenomenologically, to efficiently characterize the noise of the detector and to model noise from unknown sources, in real data analyses, noise sources are modeled as Gaussian processes (known as \emph{noise models}) \cite{vanHaasteren:2008yh,vanHaasteren:2014qva,Taylor:2021yjx,NANOGrav:2023ctt}. Noise sources are usually classified as \emph{white noise} or \emph{red noise} processes. The latter is modeled by a frequency-decreasing power spectrum (time-dependent processes), whereas the former is modeled by a constant power spectrum over the frequency domain. It is a common feature of pulsar timing data that there are noise sources whose origin is unknown, probably related to irregularities in the pulsar dynamics. That is the reason why millisecond pulsars are used in PTA searches (they have less intrinsic timing noise).\footnote{Like interferometer searches, notice that PTA searches rely primarily on Bayesian inference analysis with model comparison, but it can also be complemented by frequentist detection statistic techniques (for example, the \emph{optimal statistic} \cite{Anholm:2008wy,Demorest:2012bv,Chamberlin:2014ria}, in which a null hypothesis test is obtained by averaging over the Bayesian posteriors of the noise parameters \cite{Vigeland:2018ipb}), see more details, for instance, in \cite{Romano:2016dpx,NANOGrav:2023gor,NANOGrav:2023icp}). This topic would deserve its own set of lecture notes and we refrain from presenting further comments. A summary of Bayesian analysis methods can be found in \cite{Taylor:2014isa}.} 

In the past few years, we experienced considerable advances in PTA searches. A realistic search requires huge radio telescopes and enormous computational power. And it is also very challenging to reach the very high required precision to time and correlate the signals amid so much noise. The trailblazing idea behind finding evidence of the GWB within PTA searches and all the timing and noise modeling is, therefore, only possible thanks to joint technological and intellectual efforts.

\subsubsection{Timing residuals}
\label{sec:timing residual}

Here we prepare the stage to derive the correlation function of the timing residuals for a pair of pulsars, which will allow us to obtain the famous Hellings-Downs correlation. Further clarifications can be found on Refs.~\cite{Taylor:2021yjx,Romano:2023zhb}.\\

We start by defining the residual of the pulse arrival time for a single pulsar as the projection of the metric tensor coordinates onto the photon direction (pulsar-Earth axis) integrated over the trajectory (from the emission of the pulsar until detection on Earth), here neglecting all other noise sources. Therefore, for a pulsar distant $L$ from the Earth in a direction $\hat{p}$ (a unit vector pointing from Earth to the pulsar), the residual (time delay) at time $t$ (time of arrival on Earth) is
\begin{equation}
R(t) = \dfrac{1}{2} p^i p^j \int_0^L ds \, h_{ij}(x(s)),
\label{eq:residual}
\end{equation}
where $x(s)=(t-(L-s),(L-s)\hat{p})$. The coordinates $h_{ij}(x)$ are the solutions of the Einstein equations, after fixing the gauge, see the discussion in Sec.~\ref{sec:Linearized Einstein equations}. In the TT gauge, for example, we saw that there were only two independent components ($h_+$ and $h_\times$). Here, we choose the synchronous gauge,\footnote{This name follows from the fact that the time components of $h_{ij}(x)$ vanish in this gauge. The reader can check that such a solution obeys the equation of motion and the transverse and traceless conditions.} 
\begin{equation}
 h_{ij}(t,\vec{x}) \equiv \sum_A \, h^A(u(t,\vec{x}))\,\hat{e}^A_{ij}(\hat{\Omega}),
\label{eq:h-synchronous}
\end{equation}
where $u(t,\vec{x})=t-\hat{\Omega}\cdot\vec{x}$, the sum over index $A$ denotes the sum over the two independent modes for a GW traveling along a direction $\hat{\Omega}$, and $\hat{e}_{ij}^a (\hat{\Omega})$ are the coordinates of the GW polarization tensor defined in Eq.~\ref{eq:hij_Fourier}. This expression holds as long as we deal with a single GW source at a fixed $\hat{\Omega}$. That is not the envisaged scenario in PTA searches, in which the typical background sources are too weak to be seen isolated in the data, so we look for a common GW signal coming from an ensemble of GW sources and correlate different Earth-pulsar baselines. 

When we look at such an ensemble, it is useful to define the GW in the frequency domain. Therefore, we take the Fourier transform of Eq.~\ref{eq:h-synchronous}.\footnote{Below we omit the integration limits in the frequency domain when integrating from $-\infty$ to $\infty$.}
\begin{equation}
 h_{ij}(t,\vec{x})  \equiv \sum_A \, \int_{-\infty}^{\infty}\, \frac{df}{2\pi} e^{i2\pi f\,(t-\hat{\Omega}\cdot\vec{x})}\; \tilde{h}^A(f,\hat{\Omega})\,\hat{e}^A_{ij}(\hat{\Omega}).
\label{eq:h-f}
\end{equation}
Next, to compute the residual we notice that is also related to the redshift by 
\begin{equation}
R(t) = \int_0^t d\tilde{t} \left(\frac{\nu_0-\nu(\tilde{t})}{\nu_0}\right) =  \int_0^t d\tilde{t} \, z(\tilde{t}).
\label{eq:timingresidual-redshift}
\end{equation}
Here, the photon frequency in the presence of some GW passing by is $\nu(t)$, the frequency without GW is $\nu_0$, and the corresponding redshift at a given time is $z(t)$. Now, we can compute the redshift by taking the time derivative of the residual in Eq.~\ref{eq:residual} and using Eq.~\ref{eq:h-f}. This is just a mathematical trick to simplify the computation. Indeed,
\begin{equation}
z(t) = \dfrac{1}{2} p^i p^j \int_0^L ds\,\frac{\pt \, h_{ij}(x(s))}{\pt t}
\end{equation}
can be simplified if we observe that as function of $s$, $u(s) = t -(L-s)(1+\hat{\Omega}\cdot \hat{p})$, so that 
\begin{equation}
\frac{d \, h_{ij}(x(s))}{d t} = \frac{\pt \, h_{ij}(x(s))}{\pt s}\left( \frac{\pt \, u(s)}{\pt s} \right)^{-1} = \frac{1}{(1+\hat{\Omega}\cdot \hat{p})} \frac{\pt \, h_{ij}(x(s))}{\pt s},
\end{equation}
and the integration above becomes trivial. We obtain
\begin{equation}
z(t) = \frac{1}{2}\frac{\hat{p}_i \hat{p}_j}{(1+ \hat{\Omega}\cdot \hat{p})} \Delta h_{ij},
\end{equation}
where
\begin{align}
     \Delta h_{ij} &= \left[ h_{ij}(t,\textbf{x}_{\rm Earth}) - h_{ij}(t-L,\textbf{x}_{\rm pulsar})\right] = \left[ h_{ij}(u(L)) - h_{ij}(u(0))\right] \nonumber \\ &= \sum_A \int \frac{df}{2\pi}\, \tilde{h}^A(f,\hat{\Omega})e^{i2\pi f(u(L)-u(0))}\,\hat{e}^A_{ij} (\hat{\Omega})
     \nonumber  \\
     & = \sum_A \int \frac{df}{2\pi}\, \tilde{h}^A(f,\hat{\Omega}) e^{i 2\pi f t}\left(1-e^{-i 2\pi f L (1+\hat{\Omega}\cdot \hat{p})}\right)\,\hat{e}^A_{ij}(\hat{\Omega}).
\end{align}
Next, we define the $A^{\rm th}$-GW \emph{antenna response pattern} mode for the pair Earth-pulsar,
\begin{equation}
F^A (\hat{\Omega}) = \frac{1}{2}\frac{\hat{p}_i \hat{p}_j}{(1+ \hat{\Omega}\cdot \hat{p})} \hat{e}_{ij}^A(\hat{\Omega}),
 \label{eq:antennapatern}
\end{equation}
and the \emph{response function}
\begin{equation}
R^A (f,\hat{\Omega}) = \left(1-e^{-i 2\pi f L (1+\hat{\Omega}\cdot \hat{p})}\right) F^A (\hat{\Omega}),
 \label{eq:responsefunction}
\end{equation}
which, quoting \cite{Romano:2023zhb}, elegantly captures the difference between the Earth term (unity) and pulsar term (complex phase). With these expressions, we can also write an elegant expression for the redshift, 
\begin{equation}
z(t) = \sum_A \int \frac{df}{2\pi}\, R^A (f,\hat{\Omega}) \tilde{h}^A(f,\hat{\Omega}) e^{i 2\pi f t},
\label{eq:redshift-response}
\end{equation}
and according to Eq.~\ref{eq:timingresidual-redshift}, the timing residual is, therefore,
\begin{align}
    R(t) &= \int_0^t d\tilde{t}\sum_A \int \frac{df}{2\pi}\, R^A (f,\hat{\Omega}) \tilde{h}^A(f,\hat{\Omega}) e^{i 2\pi f \tilde{t}}, \nonumber \\
    &= \sum_A \int \frac{df}{2\pi}\, R^A (f,\hat{\Omega}) \frac{\tilde{h}^A(f,\hat{\Omega})}{2\pi i\,f} (e^{i 2\pi f t}-1).
\end{align}
Therefore, the timing residual is the Fourier transform (frequency domain) of  $\tilde{h}^A(f,\hat{\Omega})/(2\pi i f)$ weighted by the response function plus a constant factor over $t$.  We highlight the appearance of this extra factor $1/f$, coming from the integration of the time variable $t$. Also, in the literature, the constant factor is removed because it is an unobservable constant offset in $R(t)$. Constants like this can be absorbed, for instance, in an arbitrary pulsar phase when defining the timing model. The pulsar phase is an unknown quantity and therefore is marginalized over, so the pulsar phase can be typically omitted.\footnote{The derivation of the timing residual is not given so explicitly in the literature, and this point about the constant offset is not straightforward. So we rather be careful here. Thanks to Kai Schmitz for pointing this out in private communication.} As we see below, the important quantity linked to observations is the correlator between timing residuals at time $t$ and $t+\tau$. For this purpose, we  use the expression \cite{Taylor:2014isa}
\begin{equation}
    R(t) = -\frac{i}{2\pi} \sum_A \int \frac{df}{2\pi}\, R^A (f,\hat{\Omega}) \frac{\tilde{h}^A(f,\hat{\Omega})}{f} e^{i 2\pi f t},
    \label{eq:residual-response}
\end{equation}
which is the Fourier transform of the function
\begin{equation}
    \tilde{R}(f) = -\frac{i}{2\pi} \sum_A R^A (f,\hat{\Omega}) \frac{\tilde{h}^A(f,\hat{\Omega})}{f}.
    \label{eq:residual-fourier}
\end{equation}

\subsubsection{A small detour on correlation functions}

With Eq.~\ref{eq:residual-response}, we can compute the residuals of a pulsar signal perturbed by GWs. Large GW amplitudes could be directly detected, but, in practice, the smallness of the measured timing residual tells us that the GW amplitudes cannot be too large. Because the sources of typical backgrounds are too weak, they are not directly visible in the data as they cannot be separated from noise. In Sec.~\ref{sec:overlap_red} we have already seen that if we used two different interferometers we could significantly reduce the impact of noise in the measurement of timing delays. Here, we follow a similar strategy for PTAs: we look for the GW background by cross-correlating the timing residuals from two or more Earth-pulsar baselines.

The correlation between two noise-free timing residuals from two pulsars $a$ and $b$ is given by the time-averaged product over the observation time.\footnote{ We can also make a similar derivation for redshift measurements. In this case, the correlation between two noise-free redshift measurements, for two pulsars $a$ and $b$ is given by the time-averaged product over the observation time,
\begin{equation}
    \langle z_a(t) z_b(t) \rangle \equiv \frac{1}{T}
\int_{-T/2}^{T/2} {} dt\, z_a(t) z_b(t).
\end{equation}} Since the frequency domain is useful for studying the properties of the background, we go to Fourier space. First, we compute the following expectation value
\begin{align}
\langle \tilde{R}^*_a(f_a)\tilde{R}_b(f_b) \rangle  = \int \frac{d\hat{\Omega}_a}{4\pi}\frac{d\hat{\Omega}_b}{4\pi}  \sum_A \sum_B R_a^{*A} (f_a,\hat{\Omega}_a) R_b^B (f_b,\hat{\Omega}_b)\frac{\langle \tilde{H}^{*A}(f_a,\hat{\Omega}_a),\tilde{H}^B(f_b,\hat{\Omega}_b)\rangle}{(2\pi)^2\,f_af_b}.
\label{eq:our_exp_value}
\end{align}
Now, we assume the background is stationary, Gaussian, and unpolarized so that we can use Eq.~\ref{eq.:2ptSh} and straightforwardly reduce the expectation value of the Fourier transform of timing residuals to 
\begin{equation}
\langle \tilde{R}^*_a(f_a)\tilde{R}_b(f_b) \rangle   = \frac{1}{2} \delta(f_a-f_b) S_t(f_b)_{ab},
\label{eq:correlator_redshift}
\end{equation}
where $S_t(f)_{ab}$ is the \emph{cross-power spectral density} of the noise-free timing residual defined by
\begin{align}
        S_t(f)_{ab} =& \frac{S_h(f)}{4\pi^2f^2}\int \frac{d\hat{\Omega}}{8\pi} \sum_{A=+,\times}  R_a^{*A} (f,\hat{\Omega}) R_b^A (f,\hat{\Omega}) \nonumber \\
            =& \frac{S_h(f)}{4\pi^2f^2}\int\frac{d\hat{\Omega}_a}{8\pi} \left(1-e^{i 2\pi f L_a (1+\hat{\Omega}\cdot \hat{p}_a)}\right)\left(1-e^{-i 2\pi f L_b (1+\hat{\Omega}\cdot \hat{p}_b)}\right) \sum_{A=+,\times}
            F^A_a(\hat{\Omega})  F^A_b(\hat{\Omega}),
\label{eq:St}
\end{align}
where we used the expression of the response function in Eq.~\ref{eq:responsefunction}.

Finally, in our prescription for the expectation value \cite{Jenet:2014bea},\footnote{It can be confusing to the reader to relate different definitions of correlators and expectation values in the literature. We want to compute the correlator $\langle R^*_a(t), R_b(t+\tau) \rangle$. For this, we use Eqs.~\ref{eq:residual-response} and \ref{eq:residual-fourier} in $R^*_a(t) R_b(t+\tau)$, and just integrate the corresponding expression over solid angles. There is no integration over $t$. This is different than the definition of the cross-correlator of two complex functions $f$ and $g$ as the convolution of $f$ and $g$, which leads to the cross-correlation theorem
\be
\langle R^*_a(t), R_b(t) \rangle \equiv 
\int_{-\infty}^{\infty} {} dt\, R^*_a(t) R_b(t+\tau)
 = \int_{-\infty}^{\infty} {} \frac{df}{2\pi} \tilde{R}^*_a(f) \tilde{R}_b(f) e^{i2\pi f\tau},
\label{eq:two-pt-correlator}
\ee
which, in the case of autocorrelation $a=b$, is the Wiener-Khinchin theorem. We are not using Eq.~\ref{eq:two-pt-correlator}.} we want to integrate the product of the response functions over all GW solutions $h^A(f,\hat{\Omega})$ in the region between the pulsar and the Earth. In Fourier space, we integrate it over $f$ and $\Omega$ (sky-averaged) and sum over the polarizations (polarization-averaged). Then, assuming that the background is isotropic, we have
 \begin{align}
    \langle R^*_a(t), R_b(t+\tau) \rangle &\equiv  \int \frac{d\hat{\Omega}_a}{4\pi} \frac{d\hat{\Omega}_b}{4\pi}  
    \int_{-\infty}^{\infty} {} \frac{df_a}{2\pi}  \int_{-\infty}^{\infty} {} \frac{df_b}{2\pi}\tilde{R}^*_a(f_a)e^{-2\pi if_a t} \tilde{R}_b(f_b)e^{2\pi if_b(t+\tau) }\nonumber \\
    & =\int\frac{df_a}{2\pi} \frac{df_b}{2\pi}  \langle \tilde{R}^*_a(f_a),\tilde{R}_b(f_b) \rangle e^{2\pi i(f_b-f_a)t} e^{2\pi if_b\tau}, \nonumber \\
    & = \int\frac{df}{2\pi} \frac{S_t(f)_{ab}}{4\pi} e^{2\pi if\tau},
      \label{eq:two-pt-residual}
\end{align}
where we used Eqs.~\ref{eq:residual-response},~\ref{eq:residual-fourier},~\ref{eq:correlator_redshift}, and ~\ref{eq:St}. The cross-correlation, $C_{ij}(\tau)$, induced by the GWB can then be obtained by extracting the real part of this last expression \cite{Taylor:2014isa}.   

\subsubsection{Hellings-Downs correlations}
Now we are very close to finding the Hellings-Downs correlations. From the expression for $S_t(f)_{ab}$, we conveniently define 
\begin{align}
        \tilde{\Gamma}_{ab}(f,\Omega,\hat{p}_a,\hat{p}_b) &=  \frac{1}{2} \left(1-e^{i 2\pi f L_a (1+\hat{\Omega}\cdot \hat{p}_a)}\right)\left(1-e^{-i 2\pi f L_b (1+\hat{\Omega}\cdot \hat{p}_b)}\right) \sum_{A=+,\times}
            F^A_a(\hat{\Omega})  F^A_b(\hat{\Omega}) \label{eq:gamma_til}, \\
        \Gamma_{ab}(f,\xi) &= \int\frac{d\hat{\Omega}_a}{4\pi} C_1 \tilde{\Gamma}_{ab}(f,\Omega,\hat{p}_a,\hat{p}_b), 
            \label{eq:ORF}
\end{align}
where $\xi = \xi_{ab} = \cos^{-1} (\hat{p}_a \cdot \hat{p}_b)$ is the angular separation between two pulsars, and we introduced a constant, normalization factor $C_1$. Then, we can rewrite Eq.~\ref{eq:two-pt-residual} as 
 \begin{align}
    \langle R^*_a(t), R_b(t+\tau) \rangle &= \frac{1}{8 \pi C_1}\int\frac{df}{2\pi} \frac{S_h(f)}{4\pi^2f^2} \int  \frac{d\hat{\Omega}_a}{4\pi} C_1 \tilde{\Gamma}_{ab}(f,\Omega) \,e^{2\pi if\tau} \nonumber \\
    &= \frac{1}{8\pi C_1}\int\frac{df}{2\pi} \left(\frac{S_h(f)}{4\pi^2f^2}\right) \Gamma_{ab}(f,\xi) \,e^{2\pi if\tau}.
    \label{eq:exp_value_RR}
\end{align}
What this equation shows is that we can separate the spectral information about the background from the geometric dependence. The dependence on the angular separation between two pulsars $a$ and $b$, $\xi_{ab}$, is encoded in $\Gamma_{ab}(f,\xi)$, while $ S_h(f) $ encodes information about the source of the background and it is related to the power spectrum of the tensor perturbations, see Eq.~\ref{eq.:2ptSh}. The quantity $\Gamma_{ab}(f,\xi)$ is the overlap reduction function for PTAs, in analogy with the overlap reduction function described in Sec.~\ref{sec:overlap_red}; compare, for instance, with Eq.~\ref{eq:signaldetector}. It contains information about the geometry of the system.  This quantity is the famous fingerprint of an SGWB in PTA searches, known as \emph{Hellings-Downs correlation} \cite{1983ApJ...265L..39H}. In particular, it tells how the GW radiation power is distributed along the sky according to the angular distribution of pulsars. 

To determine the normalization factor $C_1$ and the shape of $\Gamma_{ab}(f,\xi)$, we follow \cite{Taylor:2021yjx}. All pulsars probed by the PTA collaborations are millisecond pulsars that belong to our cosmic galactic neighborhood. They are at least 100 parsecs (about $3\times 10^{18}$ m) distant from us, while the typical frequency values are of order $10^{-9\cdots-7}$ Hz. Therefore, $fL \gg10 $, and the dependence on the frequency in the brackets of Eq.~\ref{eq:gamma_til} can be neglected \cite{Mingarelli:2014xfa}. In this approximation, we need to treat separately the cases $a=b$ (identical pulsars, i.e. same $L$, $\hat{\Omega}$,\,$\hat{p}$) and $a\neq b$ (distinct pulsars) because the products of the brackets in Eq.~\ref{eq:gamma_til} are 1 for distinct pulsars and 2 for identical pulsars. Using a Kronecker delta to take into account both cases, the overlap reduction function therefore reduces to 
\begin{equation}
 \Gamma_{ab} (\xi) = C_1 \int \frac{d^2\Omega}{4\pi} \sum_A F_a^A (\hat{\Omega}) F_b^A (\hat{\Omega})(1+\delta_{ab}).
\label{eq:hd_delta}
 \end{equation}

Then, we use the antenna pattern in Eq.~\ref{eq:antennapatern} to explicitly write $\Gamma (\xi) $ as a function of the angular separation.\footnote{See an explicit parametrization of the angular dependence of the antenna pattern on $\hat{p}_i$ and the computation of the corresponding antenna pattern and response functions, for instance, in the appendices A of \cite{Jenet:2014bea,Allen:2022dzg}. To simplify, we can define $\hat{p}_1=\hat{z}$ and  $\hat{p}_2 = \sin \gamma \hat{x} +  \cos \gamma \hat{z}$ so that $F_1^{\times}=0$. Therefore, the overlap reduction function only depends on the average of $F_1^+ F_2^+$ over the angular separation.} Then we can show that, for a pair of different pulsars, say $1$ and $2$, whose angular separation is $\xi_{12}\equiv\xi$, the  \emph{Hellings-Downs} curve is
\begin{equation}
\Gamma_{\text{HD},12}(\xi) = \frac{C_1}{12} \left(4 +  (\cos\xi - 1) + 6(1-\cos\xi)\ln\frac{1-\cos\xi}{2}\right).
\end{equation}
According to Eq.~\ref{eq:hd_delta}, if we compute the Hellings-Downs curve for identical pulsars, we must obtain twice the result as for distinct pulsars whose angular separation is zero ($\xi=0$), with the same line of sight. In this case, all terms proportional to $(1-\cos\xi)$ vanish and  $\Gamma_{\text{HD},12}(0)=C_1/3$. So, taking into account the numerical factor of two, we conclude that for an identical pair of pulsars, $\Gamma_{\text{HD},11}=2C_1/3$. Therefore, for any pair of pulsars $i$ and $j$, we can write
\begin{equation}
\Gamma_{\text{HD},ij}(\xi_{ij}) = \frac{C_1}{12} \left(4 (1 +\delta_{ij}) +  (\cos\xi_{ij} - 1) + 6(1-\cos\xi_{ij})\ln\frac{1-\cos\xi_{ij}}{2}\right).
\end{equation}
Next, we define $x_{ij}=(1-\cos\xi_{ij})/2$. A usual convention is to set $\Gamma(0)=1/2$ for distinct pulsars at the same line of sight and $\Gamma(0)=1$ for identical pulsars. This fixes the normalization factor to $C_1=3/2$. Finally, in this case, the Hellings-Downs correlation is 
\begin{equation}
\Gamma_{\text{HD}, ij} = \frac{1}{2} (1+\delta_{ij}) -\frac{1}{4} x_{ij} +  \frac{3}{2}x_{ij} \ln x_{ij}.
\label{eq:HD}
\end{equation}

\begin{figure}[t]
\centering 
\hspace*{-.5cm}\includegraphics[height=7.0cm]{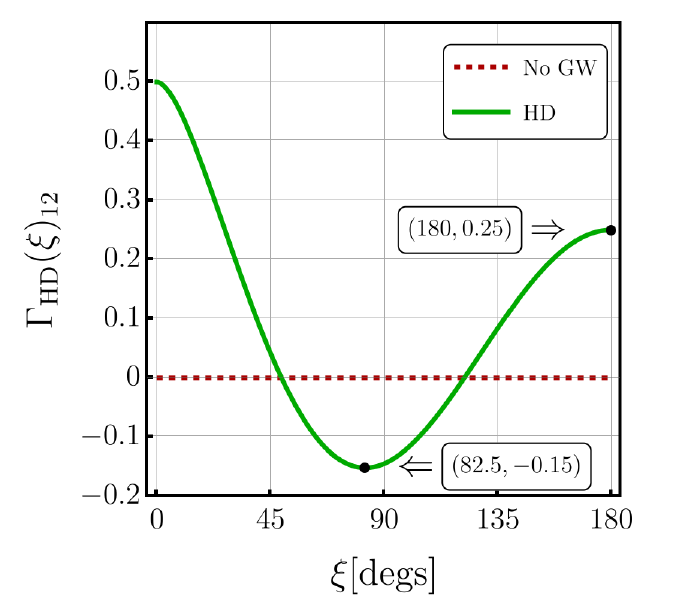} 
\caption{We show the normalized Hellings-Downs curve -- overlap reduction function for an isotropic SGWB in PTA searches -- for two different pulsars as a function of the pulsar separation angle $\xi$. We also show the minimum and the value at 180 degrees. }
\label{fig:hd_curve}
\end{figure}

We plot the corresponding curve for $i\neq j$ as a function of the pulsar separation angle in Fig.~\ref{fig:hd_curve}. It is the famous fingerprint that should be hidden in pulsar timing data if the signal comprises an isotropic GWB. We emphasize that the presence of the Hellings-Downs correlation in the data is independent of the source of the GWB, as the source contributes to $S_h(f)$, as long as the background is isotropic and unpolarized, and the GWs are luminal. 

Since the function depends on the cosine of the angle, it is symmetric only under $\xi\rightarrow 2\pi - \xi$. Moreover, to understand how the Hellings-Downs correlation depends on multipolar modes, we can expand it in a Legendre polynomial basis
\begin{equation}
\Gamma_{\text{HD}, ij}  = \sum_{l=0}^{\infty} a_l P_l(\cos\xi_{ij}) = \sum_{l=2}^{\infty} \frac{3(l-2)!}{2(l+2)!} (2l + 1) P_l(\cos\xi_{ij}).
\end{equation}
The main contribution is \emph{quadrupolar} ($l=2$,  $a_2/a_3 \sim 25/7$), and there are no monopole ($a_0=0$) or dipole ($a_1=0$) contributions, as expected for GWs in GR. By contrast, if we repeated the calculation for short-arm L-shaped interferometers, as is the case of LVK detectors, we would have obtained $\Gamma_{\text{LVK}}(\gamma)\propto P_2(\cos \gamma) = 1/2 (3 \cos^2 \gamma)$, symmetric only under $\gamma\rightarrow \pi - \gamma$ \cite{Romano:2023zhb}. Differently from PTAs, in this case, the quadrupolar deformation of spacetime produced by a passing GW affects in the same way two test masses $180^\circ$ apart.
\\

By comparing Eqs.~\ref{eq:two-pt-residual} and \ref{eq:exp_value_RR}, we can write the cross-power spectral density of the noise-free timing residuals as 
\begin{equation}
    S_t(f)_{ij} = \frac{1}{3}\frac{S_h(f)}{4\pi^2 f^2}\Gamma_{\text{HD},ij}
    = \frac{S_h(f)}{12\pi^2 f^2} \Gamma_{\text{HD},ij} = \frac{h^2_c(f)}{12\pi^2 f^3}\Gamma_{\text{HD},ij},
\end{equation}
where in the last equality we used the definition of the characteristic strain from Eq.~\ref{eq.:strain}. Since what is measured are timing residuals, we can write the cross-correlation induced by the GWB in the time domain $C_{ij}(\tau)$. According to Eq.~\ref{eq:two-pt-residual}, we have
\begin{equation}
    C_{ij}(\tau) =\int_0^\infty df\, S_t(f)_{ij} \cos(2\pi f\tau)  =  \Gamma_{\text{HD},ij}\left( \int_0^\infty df\, \frac{h^2_c(f)}{12\pi^2 f^3} \cos(2\pi f\tau) \right).
\end{equation}
If we now assume that we have Hellings-Downs correlations in our signal, we can use the previous expression to inquire which sources could have produced such a signal, since different sources have different strains $h^2_c(f)$. Whether they fit available data better or worse than each other, one can perform a Bayesian inference analysis to find out. 

Finally, we comment that throughout this section we have assumed that GR holds. Then, a detection would also confirm one more prediction of GR, this time encoded in the Hellings-Downs correlation. Instead, we could also derive an expression for the overlap reduction function in models beyond GR, see, for instance, \cite{Cornish:2017oic,Qin:2020hfy,Liang:2023ary,Cordes:2024oem}. In these models, we may not only have the two GR degrees of freedom because modified theories of gravity can contain more propagating degrees of freedom. We also assumed isotropy, but more realistic astrophysical scenarios and even some cosmological models may include anisotropies. These anisotropies would induce small perturbations to the signal and modify the overlap reduction function. It is expected that if anisotropies should exist for cosmological sources, they would be very small, as in the CMB, but an astrophysical signal should contain some degree of anisotropy, potentially observable in future datasets \cite{LISACosmologyWorkingGroup:2022kbp,Pol:2022sjn,Sato-Polito:2023spo, NANOGrav:2023tcn, Gardiner:2023zzr}. In this regard, the observation of anisotropy would support an astrophysical interpretation over cosmological ones.

\subsubsection{The latest PTA data releases: HD on the radar!}

It is impossible to finish the discussion on PTA searches without commenting on the latest results concerning the detection of Hellings-Downs correlations. On June 29th, 2023, the main PTA collaborations released their latest datasets and a set of new papers exploring the data. A complete set of papers released that day by these collaborations is: by NANOGrav \cite{NANOGrav:2023gor, NANOGrav:2023pdq,NANOGrav:2023ctt,NANOGrav:2023hfp,NANOGrav:2023hvm,NANOGrav:2023hde,NANOGrav:2023tcn,NANOGrav:2023icp}, by EPTA (with InPTA data) \cite{EPTA:2023fyk,EPTA:2023sfo,EPTA:2023akd,EPTA:2023gyr,EPTA:2023xxk,EuropeanPulsarTimingArray:2023egv},  by PPTA \cite{Reardon:2023gzh,Zic:2023gta,Reardon:2023zen}, and by CPTA \cite{Xu:2023wog}. These papers created considerable excitement because evidence for HD spatial correlations has been found for the first time  \cite{NANOGrav:2023gor,EPTA:2023fyk,Reardon:2023gzh,Xu:2023wog}, suggesting a GW origin of the signal.

Even though the collaborations obtained different confidence levels for HD and used different methods to analyze their data, they all agree that the latest datasets favor HD correlations over a model without HD. The NANOGrav collaboration, for instance, claimed evidence at the 3--4 $\sigma$ level. The current results are highly encouraging to PTA searches and GW multimessenger astronomy, and it is expected that evidence will increase further in the next IPTA data release DR3. In the meantime, the reader can see \cite{InternationalPulsarTimingArray:2023mzf} for preliminary results in a joint analysis.\footnote{Notice that combining data is not a simple task. The collaborations use different detection techniques, from timing to statistical methods. That is why the cited work is not yet conclusive and some time is needed to finish a more robust analysis. }

Besides searches for the HD correlation in the pulsar data, the collaborations also searched for signals of SMBHBs \cite{NANOGrav:2023hfp,NANOGrav:2023pdq, EPTA:2023xxk}, new physics \cite{NANOGrav:2023hvm,EPTA:2023xxk,EuropeanPulsarTimingArray:2023egv}, and anisotropies \cite{NANOGrav:2023tcn}.\footnote{Many follow-up papers exploring the data also appeared. We do not cite them here because they are many, and we would easily forget one or the other.} At this moment, it is not possible to tell the source of the observed GWB. The interpretation of the origin of the signal is not conclusive because we still need to understand better how to model the astrophysical source (SMBHBs). The simplest scenario is modeled with a power-law power spectral density $S_h(f) \sim f^{-\gamma}$,  where $\gamma=13/3$ is a constant. The resulting GW spectrum, $\Omega_{\rm GW}\sim f^{5-\gamma}$, is too much red-tilted to fit the pulsar timing data, explaining why many new physics models could explain the data better, for instance, in \cite{NANOGrav:2023hvm,Figueroa:2023zhu}. These papers exemplify how powerful PTA searches can be when it comes to probing fundamental physics and cosmology. We dedicate the following sections of our lecture notes to examine these searches.

\section{Cosmological backgrounds}\label{sec:Primordial gravitational waves}

In Secs.~\ref{sec:Emission of gravitational waves}, \ref{sec:The stochastic gravitational wave background}, and \ref{sec:searching}, we saw how to compute the GW energy density $\rho_{\rm GW}$ and its spectrum $\Omega_{\rm GW}$, and we introduced interferometers and PTA searches. In the next two sections, we will learn how GW detectors can probe fundamental physics and different stages of the universe. First, in this section, we will review some concepts in cosmology, mostly related to cosmic inflation and the propagation of GWs in an expanding and flat manifold. After establishing some notions regarding the characteristic frequency of primordial GWs and bounds coming from other cosmological probes, like CMB and BBN, then, in the next section, we will explore some cosmological sources of the SGWB.
We encourage the reader to take a look at these references for more details: \cite{Baumann:2009ds} for cosmic inflation and \cite{Domcke:2013pma,Saikawa:2018rcs} for primordial GWs.

\subsection{A very short review on cosmology}\label{subsec:intro-cosmo}

At very large length scales, our universe is isotropic and homogeneous \cite{Planck:2018vyg}. Comoving observers can describe this universe through the Friedmann-Robertson-Lemaître-Walker (FRLW) metric,
\begin{equation}
ds^2 = - dt^2 + a^2(t) \left(\frac{dr^2}{1-\kappa r^2} + r^2 d\Omega^2\right),
\end{equation}
in which $\Omega$ is the 3D solid angle, and $a(t)$ is the scale factor. The quantity $\kappa$ is a constant describing the curvature of the universe, assuming the values $+1$ if the universe is closed and has positive curvature, $0$ if the universe is open and flat, and $-1$ if the universe is open and has negative curvature. Here, we can confidently set $\kappa=0$ and work with a flat universe.

Next, we define the Hubble parameter as (dot denoting time derivative)
\begin{equation}
H (t) \equiv \frac{\dot{a}}{a}.
\end{equation}
When we substitute the FRLW metric into the Einstein equations, we derive the Friedmann equations 
\begin{align}
& H^2 = \left(\frac{\dot{a}}{a}\right)^2 = \frac{8\pi G\, \rho}{3},
\label{eq:Friedmann_equations_1}\\
& \dot{H} + H^2 = \left(\frac{\ddot{a}}{a}\right) = - \frac{1}{6} (\rho + 3 p),
\label{eq:Friedmann_equations_2}
\end{align}
where $\rho$ is the total energy density and $p$ is the momentum (given by the diagonal entries of the energy-momentum tensor), for a perfect fluid given by $T^\mu{}_\nu = \text{diag} (\rho,-p,-p,-p)$. 

\subsubsection{Properties of the FRLW solution}

We can rewrite the fluid equation, derived from the conservation of the energy-momentum tensor, $\nabla_{\mu} T^{\mn} =0$, as
\begin{equation}
\frac{d\rho}{dt} + 3H(\rho + p) = 0.
\end{equation}
With an equation of state $p = w \rho$, its solution is $\rho \propto a^{-3(1+w)}$. By substituting this result in one of the Friedmann equations above, we find a solution for the scale factor
\begin{equation}
\qquad a(t)\propto 
    \begin{cases}
			t^{\frac{2}{3(1+w)}}, & \text{if $w \neq -1$}\\
            e^{Ht}, & \text{otherwise} \label{eq:budget_inflation}
    \end{cases}
\end{equation}
 for a constant $H$. Therefore, matter ($w=0$), radiation ($w=1/3$), and any other type of fluid for which $w>-1/3$ do not produce accelerated expansion. Finally, we also write the energy budget as 
\begin{equation}
\rho = \sum_i \rho_i, \quad \Rightarrow \quad \Omega_i = \frac{\rho_i^0}{\rho_c},
\end{equation}
where $\rho_c = 3H_0^2/(8\pi G)$ is the critical energy density, $H_0$ is the Hubble parameter measured today, and $\Omega_i$ describes the fraction of each component making up the budget of the observable universe.
 
 We make two important observations at this point:
 \begin{itemize}
     \item Since the supernovae observation in the '90s, we know that the expanding universe is accelerating \cite{SupernovaSearchTeam:1998fmf,SupernovaCosmologyProject:1998vns}, so we need to include some form of \emph{dark energy} ($w \sim -1$) that is part of the energy budget. The most robust solution to the FRLW equation in this context is the $\Lambda$CDM model, in which we use a non-evolving cosmological constant ($w = -1$) to describe the late-time acceleration.

     According to the latest PLANCK results \cite{Planck:2018vyg}, the FRLW solution describes the observable universe with preferred values $w_\Lambda \simeq -1$, $\Omega_\Lambda = 0.68$ (dark energy, cosmological constant), $\Omega_{\rm dm} = 0.27$ (dark matter), $\Omega_b = 0.04$ (baryonic matter), and $H_0=67.4$ km s${}^{-1}$ Mpc${}^{-1}$.\footnote{A very interesting direction in cosmology in this time is to understand the $H_0$ and $S_8$ tensions between values obtained with early-universe probes, such as the CMB \cite{Planck:2018vyg}, and late-time measurements, such as supernovae \cite{Riess:2021jrx}. See, for instance, more details in \cite{Schoneberg:2021qvd,DiValentino:2022fjm}.}
     
     \item Similarly, in inflationary cosmology, the universe should have undergone a period of very fast expansion so that we can explain why the universe looks so homogeneous today, without problems of causality and fine-tuning of initial conditions in the past. In this case, we use directly Eq.~\ref{eq:budget_inflation} with $H>0$.
 \end{itemize} 
  
In these notes, we use $H$ to define scales. The size of the observable universe is set by the comoving particle horizon,
\begin{equation}
\tau \equiv \int_0^t \frac{dt^{'} }{a(t^{'})} = \int_0^{a(t)} \frac{da}{H a^2} = \int_0^{a(t)} d\ln(a) \left(\frac{1}{aH}\right),
\end{equation}
which is the light-like distance traveled by light from $0$ to $t$  (the maximum distance an observer can travel). The Hubble horizon, $(aH)^{-1}$, gives a rough measure of the region that can be causally influenced from the center within a Hubble time. If we use the FRLW solution above, since $(aH)^{-1} = (1/H_0) a^{(1+3w)/2}$, the Hubble horizon increases over time for matter and radiation but decreases for standard slow-roll cosmic inflation, in which $H$ is approximately constant and $a$ increases. This notion of the Hubble horizon is fundamental to understanding the production of primordial GWs in Sec~\ref{sec:GWexpanding}. We comment more about this below.

\subsubsection{Cosmic inflation}

For simplicity, we set here $M_p=1, c=1$. According to the FRLW solution in Eq.~\ref{eq:budget_inflation}, we have decelerating expansion, $\ddot{a}<0$, if the equation of state parameter is $w=p/\rho>-1/3$.  There are at least two problems associated with this solution -- the horizon and flatness problems. The horizon problem is due to the  Hubble horizon growing faster than any other physical scale so that the observed CMB spectrum implies uniformity for regions that were not causally connected in the early universe. The flatness problem is about the fact that the curvature of the universe is very small today, requiring even smaller curvatures in the past and very fine-tuned initial conditions. For more details on these problems, see, for instance, these lecture notes on inflation \cite{Baumann:2009ds}. If we want to avoid these problems, the condition $w<-1/3 $ must be satisfied. This can be done, for instance, in a cosmic inflation model, in which the era of inflation ($w<-1/3$) occurred before BBN and the radiation-domination era. 

Next, we consider the simplest model of cosmic inflation, the \emph{single-field slow-roll} model,
\be
S=\int d^4x \sqrt{-g} \left( \dfrac{R}{2}-\dfrac{1}{2}g^{\mu\nu}\del_\mu\phi\del_\nu\phi - V(\phi) \right).
\ee
\begin{figure}[t]
\centering 
\includegraphics[scale=0.5]{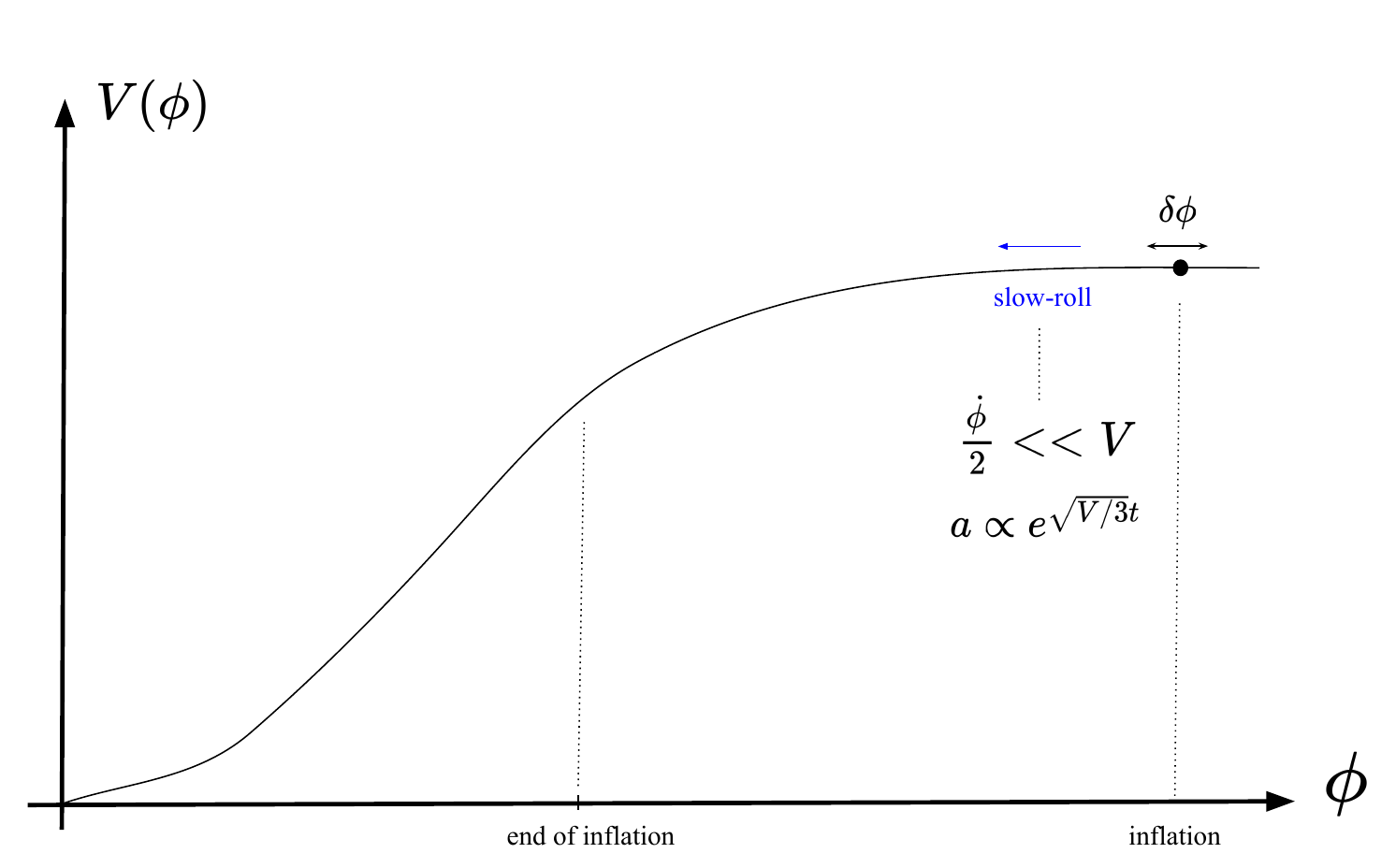} 
\caption{\textit{Slow-roll scalar field potential $V(\phi)$ as a function of $\phi$. A Slow-roll condition implies that the potential does not vary much with the evolution of the field, allowing for accelerating solutions $w<-1/3$. In the evolution of the scalar field, time runs from right to left. At the beginning of the evolution,  $V(\phi)>>\dot{\phi}^2$, implies a constant $H=\sqrt{V/3}$ and a de-Sitter solution for the scale factor. In addition, quantum fluctuations $\delta\phi$ source density scalar perturbations. Inflation ends when the slow-roll condition is not satisfied.}}
\label{fig:inflation}
\end{figure}
The energy-momentum tensor is given by\footnote{Our convention is $ g_{\mu\nu}=\bar{g}_{\mu\nu} + h_{\mu\nu}$, where $ \bar{g}_{\mu\nu}=\text{diag}(-1,a,a,a)$. } 
\be 
T_{\mu\nu}^{(\phi)}=-\dfrac{2}{\sqrt{-g}}\dfrac{\delta S_\phi}{\delta g^{\mu\nu}} = \del_\mu \phi \del_\nu \phi - g_{\mu\nu}\left(\dfrac{1}{2}\del_\alpha\phi\del^\alpha\phi+V(\phi)\right).
\ee
By assuming that the field is homogeneous, i.e. $\phi(x,t)=\phi(t), \del_i \phi = 0$,\footnote{At the classical level, this assumption is required to satisfy the symmetries of the FRLW spacetime (isotropy and homogeneity). Quantum perturbations can then introduce anisotropies and inhomogeneities. We do not cover these cases here.} we have
\be 
\rho_\phi = \dfrac{\dot{\phi}^2}{2} + V(\phi), \qquad p_\phi=  \dfrac{\dot{\phi}^2}{2} - V(\phi),
\ee
and then 
\be 
w_\phi = \dfrac{p_\phi}{\rho_\phi} = \dfrac{ \dot{\phi}^2 / 2 - V(\phi)}{ \dot{\phi}^2/2+ V(\phi)} \qquad \Rightarrow  \qquad w_\phi \rightarrow -1 \qquad  \text{if}\qquad  V(\phi) >>  \dot{\phi}^2 / 2.
\ee
The last condition is known as slow roll, see Fig.~\ref{fig:inflation}. That is why this model can describe a period of accelerated expansion, since $w_\phi<-1/3$. In particular, from the Friedmann equations, its solution is quasi de-Sitter,\footnote{The de Sitter solution is the maximally symmetric solution of vacuum Einstein equations with a constant positive cosmological constant $\Lambda$. Without any matter content, we have $\Omega_\Lambda = 1$. In this case, we can show that $H\propto\sqrt{\Lambda}$ and the Ricci curvature scalar is $R\propto\Lambda$, which is the only free degree of freedom of the theory (no longer two, as GR in vacuum, without cosmological constant). That means, to have GWs, a de Sitter period must finish.} i.e. it is an exponential expansion $a(t) \propto e^{Ht}$, for a constant positive Hubble parameter $H$, that ends by the end of inflation. Before inflation ends and the radiation-domination era starts, some form of reheating mechanism is necessary, but we skip these details here. For the scalar field $\phi$, the equation of motion is
\be 
\dfrac{1}{\sqrt{-g}}\del_\mu (\sqrt{-g}\del^\mu \phi) + V_{,\phi}=0\qquad \Rightarrow \qquad \ddot{\phi}+3 H \dot{\phi} +  V_{,\phi} = 0, 
\ee 
for homogeneous $\phi$ with $H^2=\dfrac{1}{3}(\dot{\phi}^2 / 2 + V(\phi) ) \approx \dfrac{V(\phi)}{3}$.

\subsubsection{Horizon modes}
\label{sec:horizon_modes}

So far, we have discussed inflation classically. In addition, quantum fluctuations of the scalar field $\delta \phi$ and of the metric $\delta g^{\mu\nu}$ source density perturbations and GWs. We explore this in Secs.~\ref{sec:GWexpanding} and \ref{sec:GW_inflation}. These quantum fluctuations can follow from a similar SVT decomposition \cite{Lifshitz:1945du,Mukhanov:1990me} as in Sec.~\ref{sec:svt}, in which the metric tensor and any other fields can be decomposed into tensor, vector, and scalar parts. These fields, in Fourier representation, are associated with modes of wavenumber $k$. The dynamics of the modes will determine the behavior of the field in a given era.

Generically, at earlier times during inflation, the Hubble horizon $H$ is constant, and the scale factor $a(t)$ grows exponentially.  As a result, the Hubble horizon $(aH)^{-1}$ decreases in time. A mode with wavenumber $k$ is said to be 

\begin{itemize}

    \item \emph{sub-horizon}  when $k^{-1}<(aH)^{-1}$, i.e., they are inside the horizon ($\lambda_k < (aH)^{-1}$);

    \item \emph{super-horizon} when $k^{-1}>(aH)^{-1}$, i.e., they are outside the horizon ($\lambda_k > (aH)^{-1}$);

    \item at \emph{horizon crossing} when $k^{-1}= (aH)^{-1}$ and the modes either exit or re-enter the horizon.

\end{itemize}
Since $(aH)^{-1}(t)$ is decreasing in time, modes with smaller $k$ -- larger $\lambda_k$ -- will exit (re-enter) the horizon before (after) modes with larger $k$ -- smaller $\lambda$ -- see Fig.~\ref{fig:horizons}. As we see below, at the moment the modes exit the horizon, they are frozen. When the mode re-enters the horizon, they start to evolve again. In particular, the tensor modes will then propagate as GWs. Therefore, if we manage to detect such primordial GWs in the form of a stochastic background, we will be observing an imprint of how things were at the time of horizon re-entrance. The higher the frequency of such GW background, the earlier the mode re-entered the horizon.  
This picture is very similar to what happens in the CMB case, where we can ``see'' the photons that propagated since decoupling.

\begin{figure}[t]
\centering 
\includegraphics[scale=0.5]{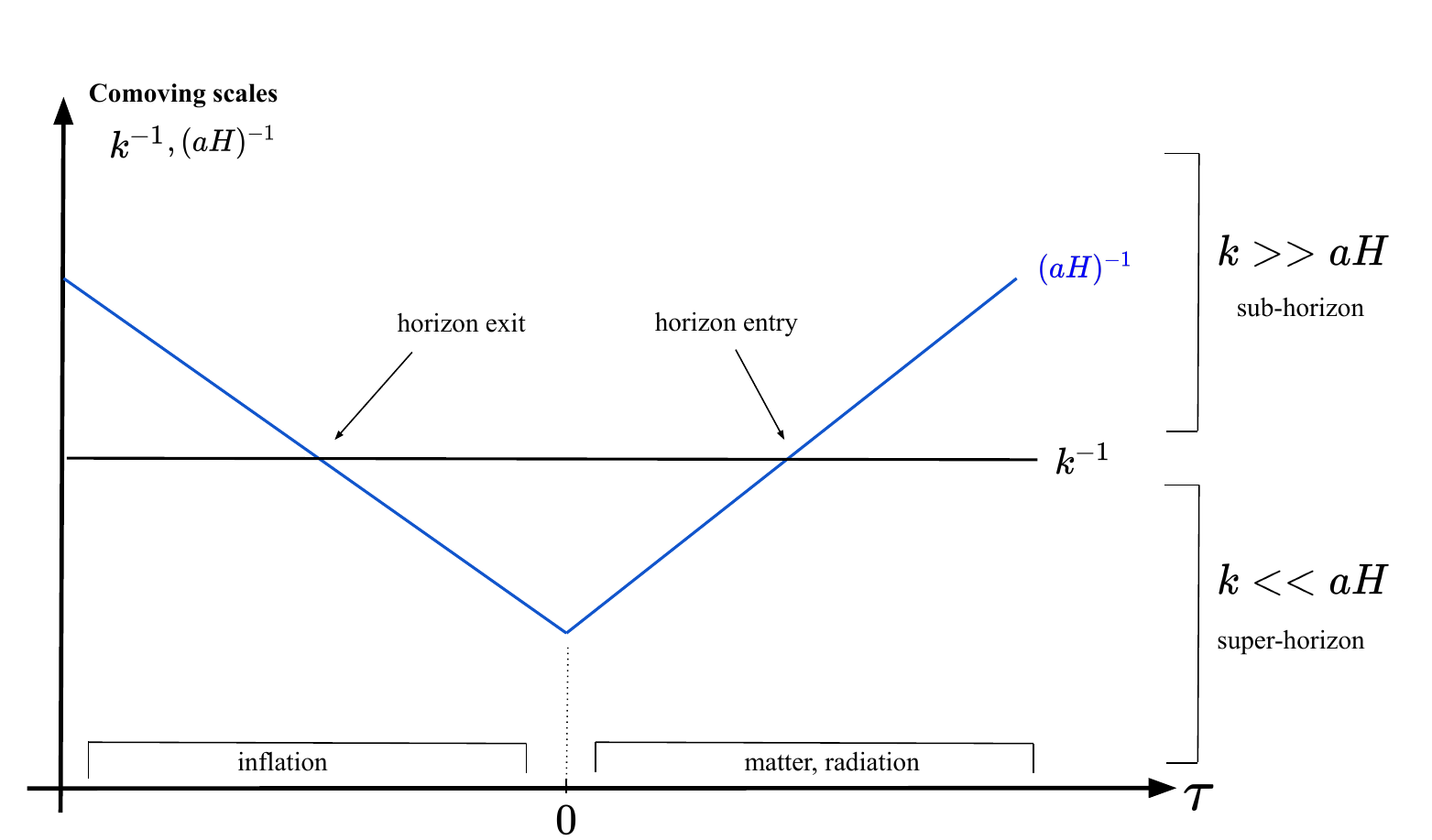} 
\caption{\textit{Diagram of the comoving scales $k^{-1}$ and $(aH)^{-1}$ as a function of conformal time, normalized so that inflation ends at $\tau=0$. For a mode with a given wavenumber $k$, at earlier times during inflation, the Hubble horizon $H$ is constant, and the scale factor $a(t)$ grows exponentially. As a result, $(aH)^{-1}$ decreases, so that the mode leaves the horizon when $k^{-1}= (aH)^{-1}$. After inflation, for both matter and radiation eras, $(aH)\sim 1/t$, and $(aH)^{-1}$ increases with time so that the mode re-enters the horizon when $k^{-1}= (aH)^{-1}$. Sub-horizon scales refer to modes in the horizon $k^{-1}<(aH)^{-1}$. Super-horizon scales refer to modes out of the horizon $k^{-1}> (aH)^{-1}$. Fluctuations only propagate after horizon re-entry since they are frozen in super-horizon scales. }}
\label{fig:horizons}
\end{figure}

\subsection{Gravitational waves in an expanding universe}\label{sec:GWexpanding}

Here we explore some properties of GWs in an expanding universe. The metric of an expanding FRLW universe in Cartesian coordinates with time $(t,x,y,z)$ and conformal time $(\tau,x,y,z)$ is given by
\be
ds^2=-dt^2+a(t)^2 dx^i dx_i = a(\tau)^2(- d\tau^2+g_{ij}dx^i dx^j),
\ee
where $\tau$ is the conformal time, defined as $dt = a d\tau $, and $a$ the scale factor. Expanding around a flat homogeneous cosmological background $ g_{ij}=\delta_{ij}+h_{ij} $, the linearized field equations are
\be 
\Box_s\, \bar{h}_{ij}(\vec{x},\tau)-\dfrac{2a^{\prime}}{a}\bar{h}^{\prime}_{ij}(\vec{x},\tau)=-16\pi G T_{ij}, \label{lin-field-eq-cosmo}
\ee
where derivatives with respect to conformal time are denoted by primes. We also defined the scalar d'Alembertian operator as
\[
\Box_s \phi \equiv \frac{1}{\sqrt{-g}} \partial_\mu (\sqrt{-g}\, \partial^\mu \phi).
\]
Rescaling the tensor perturbation as \(\bar{h}_{ij} \equiv a(\tau) h_{ij}\), the linearized Einstein equations then reduce to Eq. (\ref{lin-field-eq-cosmo}),
where \(\Box_s\) acts as a scalar wave operator.

Notice that the second term on the left-hand side vanishes for a static universe, and we recover the usual linearized Einstein GW equation. By Fourier transforming and defining $\tilde{h}_\lambda\equiv a h_\lambda$, where $\lambda=+,\times$ are the two polarization modes of the GW, we can rewrite the field equations as
\be 
\tilde{h}_\lambda^{\prime\prime}(\vec{k},\tau)+\left(k^2-\dfrac{a^{\prime\prime}}{a}\right)\tilde{h}_\lambda(\vec{k},\tau)=16\pi G a T_\lambda(\vec{k},\tau). \label{eq:waveeq}
\ee
We can approximate in the following two cases: $k^2>>(aH)^2 $  (sub-horizon modes) and $ k^2<<(aH)^2$ (super-horizon modes),  see Fig.~\ref{fig:horizons}. In vacuum, for the \emph{sub-horizon} case, \ref{eq:waveeq} reduces to 
\be 
\tilde{h}_\lambda^{\prime\prime}(\vec{k},\tau)+k^2\tilde{h}_\lambda(\vec{k},\tau)\approx 0,
\ee
which is a simple wave equation whose solution for $\tilde{h}$ is oscillatory. Consequently,
\be 
h_\lambda(\vec{k},\tau)\approx \dfrac{A_\lambda}{a}\cos (k\tau+\varphi), \label{standardgw}
\ee
where $A_\lambda=A_\lambda(\vec{k})$ is a constant in time. Notice that the wave amplitude decays as the universe (scale factor) expands. For the \emph{super-horizon} case, we have instead 
\be 
2 a^{\prime}h^{\prime}_\lambda + a h^{\prime\prime}_\lambda \approx 0,
\ee
whose solution is 
\be 
h_\lambda = A_\lambda + B_\lambda\int_{0}^{\tau}\dfrac{d\gamma}{a(\gamma)^2} \approx \text{constant},
\ee
where $A_\lambda=A_\lambda(\vec{k})$ and $B_\lambda=B_\lambda(\vec{k})$ are constant in time. In the approximation above, we used the fact that the integral decays as $a(t)$ increases. Thus, GWs are ``frozen'' outside the Hubble horizon, meaning their amplitude remains approximately constant while their wavelength exceeds the horizon scale. After re-entering the horizon, tensor perturbations become sub-horizon modes again, evolving according to the usual wave equation, as described in \ref{standardgw}. 

The re-entry time is the moment when a given mode’s wavelength becomes smaller than the Hubble radius. For modes generated during inflation, this corresponds to modes that were originally stretched to super-Hubble scales, where they were frozen with nearly constant amplitude. As the Universe expands and the Hubble radius grows, these modes eventually re-enter the horizon and start oscillating again. Since then, they have propagated through spacetime and can be detected today. Let us focus, therefore, on these sub-horizon modes.

A useful parametrization for the sub-horizon modes is \cite{Domcke:2013pma}
\be 
 h_{ij}^{TT}(\vec{x},\tau)=\sum_{\lambda=+, \times}\int \dfrac{d^3k}{(2\pi)^3}  \; h_\lambda(\vec{k})\mathcal{T}_k(\tau)\hat{e}_{ij}^{\lambda}(\hat{k})e^{-i(k\tau-\vec{k}\cdot\vec{x})} .
\ee
For this parametrization, we define an initial time $\tau_{\ast}$ as the time of formation or horizon entry, or, for sub-horizon sources, as the time of GW emission, i.e., when the decaying behavior ($1/a$) starts; $h_\lambda $ is the Fourier coefficient at time $\tau=\tau_{\ast}$; $\mathcal{T}{}_k(\tau) $ is the transfer function given here by the ratio $a(\tau_{\ast})/a(\tau)$ (notice we have factored out $1/a$); and $ \hat{e}_{ij}^{\lambda} $ are the components of the polarization tensor which map the Cartesian coordinates of the tensor $ h_{ij}^{TT}$ to its $+,\times$ degrees of freedom, as defined in the previous sections.

We use the expression for the energy density associated with GWs, Eq.~\ref{eq:rho_00},
\be 
\rho_{\rm GW}(\tau)=\dfrac{1}{32\pi G}\langle \dot{h}_{ij}^{TT}(\vec{x},t)\dot{h}^{TT*}{}^{ij}(\vec{x},t) \rangle.
\ee
Since we are using conformal time, $ \dot{h_{ij}}=(1/a) h_{ij}^{\prime},  \;\mathcal{T}_k^{\prime} = - \mathcal{T}_k (a^{\prime}/a)$, and $ \mathcal{H} = a^{\prime}/a $, for $ a = a(\tau)$. Therefore,
\be 
\dot{h}_{ij}^{TT}(\vec{x},t)=-\dfrac{1}{a}\sum_{\lambda}\int \dfrac{d^3k}{(2\pi)^3}  h_\lambda(\vec{k})\mathcal{T}_k(\tau)e_{ij}^{\lambda}(\hat{k})\left(  i k + \mathcal{H}\right)e^{-i(k\tau-\vec{k}\cdot\vec{x})},
\ee
and
\begin{eqnarray}
\rho_{\rm GW}(\tau) = \dfrac{1}{32\pi G} \dfrac{1}{a^2} \sum_{\lambda_1\lambda_2}\int \dfrac{d^3k_1}{(2\pi)^3} \int \dfrac{d^3k_2}{(2\pi)^3} \langle h_\lambda(\vec{k}_1) h_{\lambda^{\prime}}(\vec{k}_2)\rangle\hat{e}_{ij}^{\lambda}(\hat{k}_1)e_{ij}^{\lambda}(\hat{k}_2) \mathcal{T}_{k_1} \mathcal{T}_{k_2} \times \nonumber \\ 
\times \left(  i k_1 + \mathcal{H}\right)\left(  - i k_2+ \mathcal{H}\right)e^{-i(k_1-k_2)\tau}e^{-i(\vec{k}_2-\vec{k}_1)\cdot\vec{x}}.
\end{eqnarray}
The last term simplifies after assuming homogeneity and isotropy as in Eq.~\ref{eq:powerspectrumdef},
\be 
\langle h_\lambda(\vec{k}_1) h_{\lambda^{\prime}}(\vec{k}_2)\rangle = (2\pi)^3 \delta_{\lambda\lambda^{\prime}}\delta^3(\vec{k}_1-\vec{k}_2)P_\lambda(\mid\vec{k}_1\mid), 
\ee
where $P(k)$ is the non-normalized tensor power spectrum, which has mass dimensions $[M]^{-3}$. We can explicitly factor out the powers of momentum and write the following expression with the dimensionless power spectrum of tensor perturbations introduced in Eq.~\ref{eq.:2ptTensor}, through the replacement $ P_\lambda  = (2\pi^2/k^3)\mathcal{P}_t(k)$.

We can write the energy density associated with the SGWB as 
\be
\rho_{\rm GW}(\tau) =  \dfrac{1}{32\pi G} \dfrac{1}{a^2}\sum_{\lambda}\int \dfrac{d^3k}{(2\pi)^3}\mathcal{T}_{k}^2 \vert  \left(  i k + \mathcal{H}\right) \vert^2\hat{e}_{ij}^{\lambda}e^{\lambda}_{ij}(k)P_\lambda(k).
\ee
For each polarization mode, $\hat{e}_{ij}^{\lambda}\hat{e}^{\lambda}_{ij}$ = 1, since  $\hat{e}_{ij}^{\lambda_1}\hat{e}^{\lambda_2}_{ij}=\delta_{\lambda_1,\lambda_2}$. Since we are working with \emph{sub-horizon modes} and $\vert  \left(  i k + \mathcal{H}\right) \vert^2 = k^2 +  \mathcal{H}^2 = k^2  + (a H)^2 $, we can approximate $aH<<k$ so that 
\be 
\rho_{\rm GW}(\tau) =  \dfrac{1}{32\pi G} \dfrac{1}{a^2}\sum_{\lambda}\int \dfrac{d^3k}{(2\pi)^3}\mathcal{T}_{k}(\tau)^2 k^2 P_\lambda(k).
\ee
Hence, at a time $\tau_0$, 
\be 
\rho_{\rm GW}(\tau_0)=\dfrac{1}{32\pi G}\dfrac{1}{ a^2(\tau_0)}\dfrac{4\pi}{(2\pi)^3}\int (k^2 d k)\times k^2\sum_\lambda  P_\lambda(k) \dfrac{a^2(\tau_{\ast})}{a^2(\tau_0)}. \label{eq:energydensity}
\ee
The term $\sum_\lambda P_\lambda (k)$ can be identified with the primordial power spectrum, and $  \dfrac{a^2(\tau_{\ast})}{a^2(\tau_0)} $ tells the cosmological history. We can also relate the energy density $\rho_{\rm GW}$ to the GW density spectrum $\Omega_{\rm GW}$ through \ref{eq:spectralshape},
\be
\rho_{\rm GW}=\rho_c \int d(\log k) \dfrac{1}{\rho_c}\dfrac{\del \rho_{\rm GW}}{\del \log k} = \rho_c \int d(\log k)\Omega_{\rm GW}(k,\tau_0). 
\ee
Why is $\Omega_{\rm GW}(k,\tau_0)$ relevant? We use $\Omega_{\rm GW}(k,\tau_0)$ as a convenient measure of GW amplitude, which is directly related to the strain power spectrum $\expval{h_{ij} h^{ij}}$ measured by detectors today. We can obtain the spectrum $\Omega_{\rm GW}(k,\tau_0)$ by comparing the last expression with \ref{eq:energydensity}. The critical energy density is given by $  \rho_c = (3H_0^2)/(8\pi G)$. Therefore,
\be 
\Omega_{\rm GW}(k,\tau_0)=\dfrac{k^2}{12 H_0^2} \left( \frac{k^3}{2\pi^2} \sum_\lambda  P_\lambda(k) \right) \dfrac{a^2(\tau_{\ast})}{a^4(\tau_0)}.
\ee
By defining the power spectra of tensor fluctuations as 
\be 
\Delta_t^2 \equiv \dfrac{k^3}{2\pi^2} \sum_\lambda  P_\lambda (k) = \sum_t \mathcal{P}_t (k), \label{eq:tensorspec}
\ee
we can rewrite
\be 
\Omega_{\rm GW}^0(k)=\dfrac{\Delta_t^2}{12}\dfrac{k^2}{H_0^2}\dfrac{a^2(\tau_*)}{a^4(\tau_0)}=\dfrac{\Delta_t^2}{12}\left(\dfrac{k}{a_{\ast}H_{\ast}}\right)^2 \dfrac{a_{\ast}^4 H_{\ast}^2}{a_0^4 H_0^2},
\ee
where $a_*=a(\tau_*), H_*=H(\tau_*)$ and the index $\ast$ denotes time of horizon crossing (GW production) and $0$ denote today's time. At the time of horizon entry, $a_{\ast}H_{\ast} = k_*$, so we write the SGWB spectrum as
\be 
\Omega_{\rm GW}^0(k_*)=\dfrac{\Delta_t^2(k_*)}{12}\dfrac{a_{\ast}^4 H_{\ast}^2}{a_0^4 H_0^2}.
\label{eq:GW_spectrum_exp}
\ee

Take, for instance, the case of \emph{single-field slow-roll inflation}. In inflation, GWs are created from quantum fluctuations in quasi-de-Sitter spaces. Here we assume that the polarization modes are the same, i.e., $ P_+=P_- $, which is a valid assumption for single-field slow-roll inflation and shows no preference for either polarization, and then the SGWB is unpolarized. We assume $\tau_*$ occurs during the radiation domination era. The power spectrum for each polarization mode is
\be 
P_\lambda(k) = \left(\dfrac{2}{M_p}\right)^2\left(\dfrac{H_{\text{inflation}}^2}{2k^3}\right), 
\ee
where $M_p^2=1/(8\pi G)$ and $H_{\text{inflation}}$ is the Hubble parameter at the end of inflation. We will deduce this expression in the next section, where we discuss sources, see Eq.~\ref{eq:power_inflation}. It follows that $\Delta_t^2$ is constant, given by\footnote{In the inflationary period, $a(t) = e^{\Lambda t}$, for constant $\Lambda=H_{\text{inflation}}$. Since the $k^3$ from the power spectrum cancels the corresponding term in the definition of the tensor fluctuation, $\Delta_t^2$ is constant for any value of $k$.} 
\be 
\Delta_t^2=128 G^2H_{\text{inflation}}^2.
\ee 
So, according to ~\ref{eq:GW_spectrum_exp}, $ \Omega_{\rm GW}^0$ does not depend on $k$, and the GW spectrum and the energy density decrease with $a^{4} $, as expected for radiation. 

\begin{figure}[t]
\centering 
\hspace{-1cm}\includegraphics[]{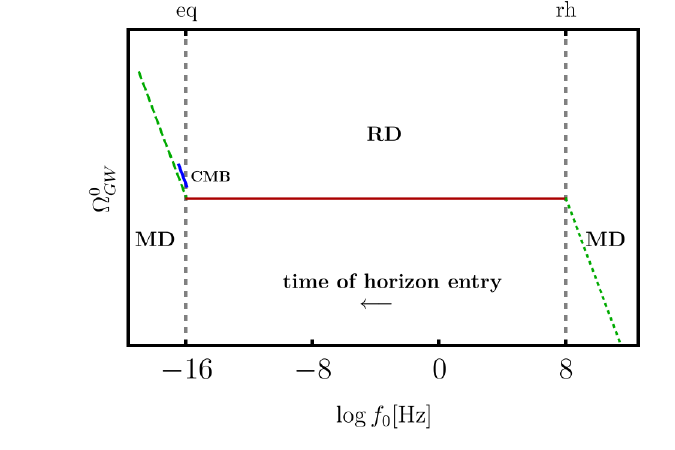} 
\caption{\textit{Plot in logarithm scale of the GW power spectrum from single-field slow-roll inflation versus frequency (observed today), in Hz. Notice that the time of horizon entry flows in the opposite direction of the frequency axis since $f_*\sim t_*^{-1}$. The spectrum does not depend on the frequency during the radiation-domination era (RD, red, solid line, $w_*=1/3$). During the matter-domination era (MD, green, dashed line, $w_*=0$), $\Omega_{\rm GW}^0 \sim f_0^{-2}$. The acronym ``eq'' stands for matter-radiation equality, and ``rh'' for reheating (the last stage of inflation, which takes into account all processes from the decay of the inflaton field to establish the hot thermal bath of the Big Bang). We assumed that reheating (MD, green, dotted line) occurred at $f_0 \sim 10^{8} \text{Hz}$. During reheating $w_*=\expval{w}=0$, as in matter-domination. On the left, we also sketch the power spectrum of the CMB (blue, thick line, $r=\Delta_t^2/\Delta_s^2<0.1$), related to tensor anisotropies on the last scattering surface, during the matter-domination era.
}}
\label{fig:gravitational wavespecInf}
\end{figure}

If a GW is emitted with some frequency $f_*$ at time $\tau_*$, then the observed frequency at time $\tau_0$ is red-shifted according to
\be 
f_0 = f_* \frac{a(\tau_*)}{a(\tau_0)}\label{eq:redshift-f} .
\ee
Using $k=2\pi(f/a)$, at horizon crossing $f_*=k_*a_*/(2\pi)=H_*/2\pi$, and $f_0\sim H_* a_*$.
Therefore $\Omega_{\rm GW}^0 \sim f_0^2 a_*^2$.  During radiation domination, $\rho \sim a^{-4}$, then  $H_*\sim a^{-2} $, $f_0\sim a_*^{-1} $ because $H_*^2\sim \rho  $, and then $\Omega_{\rm GW}^0 \sim (f_0)^0$, i.e., the GW spectrum does not depend on the observed frequency. Instead, if the GW was emitted during a matter-dominated era, then $H_*\sim a^{-3/2} $, $f_0\sim a_*^{-1/2} $, and finally   $\Omega_{\rm GW}^0 \sim f_0^{-2}$.  We can also plot the expected shape of such a GW spectrum as a function of the observed frequency, see Fig.~\ref{fig:gravitational wavespecInf}. This is the behavior expected for inflationary GWs. Different primordial sources will depend on $f_0$ in different ways.

\subsection{Characteristic frequencies of relic gravitational waves} \label{subsec:freq}

Next, we relate the GW spectrum produced at a time $t$ long ago to the observable spectrum that is probed nowadays. If a GW is emitted with some frequency $f_*$ at time $\tau_*$, then the observed frequency at time $\tau_0$ is redshifted due to the expansion of the universe,
\be 
f_0 = f_* \frac{a(\tau_*)}{a(\tau_0)} =  \frac{k_* a^2(\tau_*)}{2\pi a(\tau_0)}= \frac{H_* a_*}{2\pi a_0}.
\ee
The waves are produced during \emph{horizon re-entry} when the wavelength becomes comparable to the comoving Hubble horizon. The comoving Hubble horizon $(aH)^{-1}$ is controlled by the Hubble parameter, $H\sim f = 1/t$. The emitted frequency  evolves as $f_* = (\epsilon_* H_*^{-1})^{-1} $, where $\epsilon_* $ is a small value satisfying $\epsilon_* \leq 1$ and its exact value depends on the source. The subscript $*$ denotes horizon crossing. This parametrization sets the inverse of the Hubble factor $(H_*)^{-1}$ as the cosmological horizon. GWs from a source in the early universe cannot be correlated on time scales larger than $(H_*)^{-1}$; otherwise, it would break causality. The exception is cosmic inflation, which is motivated by such causality issues in the so-called homogeneity problem. We comment more about it in Sec.~\ref{sec:GWexpanding}.

Assume that a GW signal is produced during the radiation era, then we can use
\be
H_*^2= \frac{\rho_{r}}{3M_p^2} = \frac{\pi^2 g_\star T_*^4}{90 M_p^2},
\label{eq:rad_en_den}
\ee
where $T_*$ marks horizon re-entry at radiation domination era, in which $ a \sim 1/T $ and $ t \sim 1/T^2$. For relativistic degrees of freedom about $g_\star\sim 100$, we have
\begin{align}
& f_0\simeq 10^{-8} \epsilon_*^{-1} \left( \dfrac{T_*}{\text{GeV}}\right) \text{Hz}, \label{eq:peak_freq_char}\\
& t_*\simeq 10^{-22}  \epsilon_*^{-1} \left( \dfrac{1 \text{Hz}}{f_0}\right)^2 \text{s}.
\end{align}
Thus, it is possible to associate the observed frequency of GWs in the detectors with the epochs of the universe when such GWs had been produced. By operating in different frequency ranges, GW detectors can probe separated energy scales and cosmological epochs.\footnote{Take as a grain of salt that the stochastic GW signal is weak. As we have seen previously, there is a $\left(a_*/a_0\right)^2$ suppression for the GW density spectrum. This means that largely redshifted sources -- as is the case of cosmological sources -- are weak and challenging to detect.} As shown in Table~\ref{tablefreq}, in principle, we can access very high energy scales. These scales cannot be probed by other cosmological probes, for instance, those related to the CMB, to Big Bang nucleosynthesis (BBN), and large-scale structures (LSS), which can probe only much lower temperatures, $T_* \leq 1 $ MeV. 

\begin{table}[hbt!]
\center
\begin{tabular}{|c|c|c|}
\hline 
$\epsilon_*=0.1$ & $f_0$ (Hz) & $T_*$ (GeV) \\ 
\hline 
PTA & $10^{-8}$ & $0.1$ \\ 
\hline 
LISA & $10^{-2}$ & $10^5$ \\ 
\hline 
LVK & $10^2$ & $10^9$ \\ 
\hline 
\end{tabular} 
\caption{\textit{Typical peak frequencies and their associated temperature of emission, expected for PTAs, and ground-based and space-based GW experiments (LVK and LISA, respectively) for $\epsilon_*=0.1$.}}
\label{tablefreq}
\end{table}

In particular, when $T_*$ is associated with the reheating temperature at the end of inflation, we can see which kind of inflationary scenarios the different experiments can probe. For instance, PTAs could probe only inflationary GWs produced after a late period of reheating, i.e., low reheating temperature, when compared to LVK.

\subsection{The gravitational wave spectrum and the cosmological redshift} \label{subsec:gw_spectrum_today}

\begin{figure}[t]
\centering 
\includegraphics[scale=0.8]{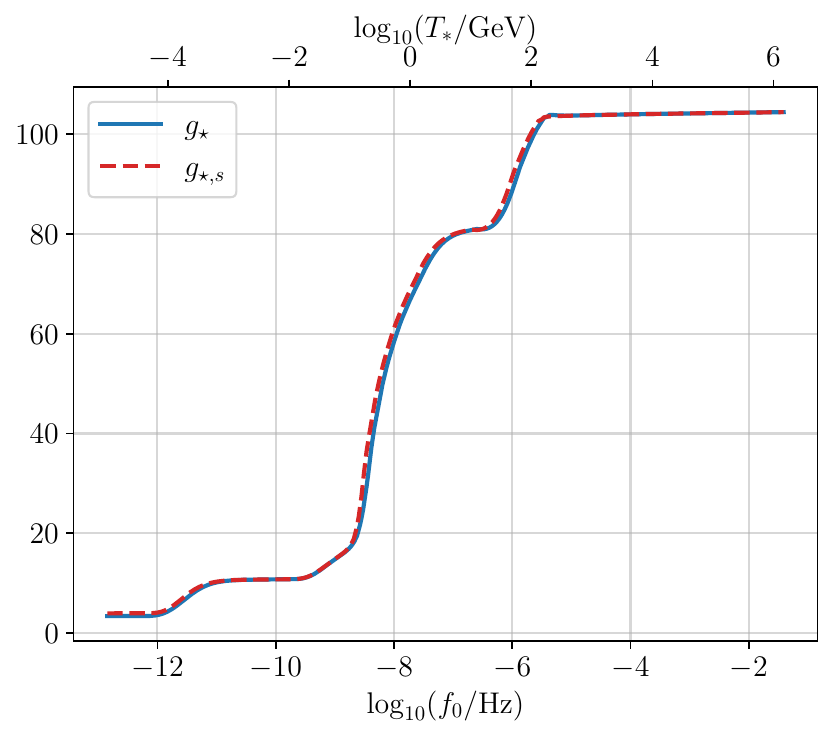} 
\caption{\textit{We plot the SM effective number of relativistic ($g_\star$, blue, solid line) and entropy degrees of freedom ($g_{\star,s}$, red, dashed line) as a function of observed frequency ($f_0$, bottom) and temperature ($T_*$, top) at horizon crossing. We used data provided by Ref.~\cite{Saikawa:2020swg}.}}
\label{fig:gstar}
\end{figure}

The correct correspondence between frequency and temperature does depend on the effective number of relativistic and entropic degrees of freedom, $g_\star$  and $g_{\star,s}$ respectively, as well as on the equation of state of the universe. For the SM, the correspondence can be found in \cite{Saikawa:2020swg} and is plotted in Fig.~\ref{fig:gstar}. We can see mainly three major events (jumps) changing these effective numbers. They are associated with the EW phase transition at $T\sim100$ GeV, the QCD phase transition at $T\sim0.1$ GeV, and the $e^+e^-$ annihilation at $T\sim1$ MeV. The starting point in the UV for the SM is at $g_\star=g_{\star,s}=106.75$, while the endpoint is approximately $(g_\star,g_{\star,s})=(3.38,3.93)$ after neutrino decoupling \cite{Saikawa:2018rcs}. We can also see that the QCD phase transition event coincides with the range that PTAs can probe. 

Since the universe expands and the degrees of freedom $g_\star$ and $g_{\star,s}$ change across different epochs, there is a cosmological redshift that must multiply the evolution of the GW spectrum. By knowing the relation between degrees of freedom, time, and frequency, we can write a final expression as
\begin{equation}
\Omega_{\rm GW}^{\rm \text{observed}}\left(k\right) = \Omega_{r}^0 \left(\frac{g_\star\left(k\right)}{g_\star^0}\right)\bigg(\frac{g_{\star,s}^0}{g_{\star,s}\left(k\right)}\bigg)^{4/3} \hspace{-0.25em}\Omega_{\rm GW}^{\rm \text{emitted}}(k)\,.
\label{eq:gw_today}
\end{equation}
To derive this equation, we change variables from $t_0$ to $t$ and assume that the GW energy density behaves like radiation, decaying with the fourth power of the scale factor $a(t)$. Then, we consider the case for which horizon re-entry occurred during the radiation domination era, in which $\Omega_r(t)=1$, and use the expression for the energy density given by Eq.~\ref{eq:rad_en_den}. Finally, we use conservation of the entropy $S=g_{\star,s} a^3 T^3$. Explicitly, we have
\begin{align}
\Omega_{\rm GW}\left(t_0\right) = \frac{1}{\rho_c^0}\frac{d\, \rho_{\rm GW} (t_0)}{d\, \ln k}  
= & \, \left(\frac{\rho_c(t)}{\rho_c^0}\right) \frac{1}{\rho_c(t)}\left(\frac{\del\, \rho_{\rm GW} (t_0)}{\del\, \rho_{\rm GW} (t)}\right)\frac{d\, \rho_{\rm GW} (t)}{d\, \ln k} \nonumber  \\
=& \,\frac{\Omega_r^0}{\rho_r^0}\frac{\rho_r(t)}{\Omega_r(t)} \left(\frac{a(t_0)}{a(t)}\right)^{-4} \frac{1}{\rho_c(t)} \frac{d\, \rho_{\rm GW} (t)}{d\, \ln k} \nonumber  \\
= & \, \frac{\Omega_r^0}{\Omega_r(t)} \left(\frac{\rho_r}{\rho_r^0}\right)\left(\frac{a(t_0)}{a(t)}\right)^{-4} \Omega_{\rm GW}(t)  \nonumber \\
= & \, \Omega_r^0 \left(\frac{g_\star(t) T^4}{g_\star(t_0) T_0^4}\right) \left(\frac{a(t_0)}{a(t)}\right)^{-4} \Omega_{\rm GW}(t) \nonumber \\
= & \,  \Omega_r^0 \left(\frac{g_\star(t)}{g_\star(t_0)}\right) \left(\frac{g_{\star,s}(t_0)}{g_{\star,s}(t)}\right)^{4/3} \Omega_{\rm GW}(t).
\end{align}
This proves Eq.~\ref{eq:gw_today}.

\subsection{Constraints from BBN and CMB}

Big Bang nucleosynthesis (BBN) and the cosmic microwave background (CMB) are also cosmological probes. They help us to answer the question: ``What is the maximum fraction $\Omega_{\rm GW}/\Omega_{r}$ we can observe today?'' The energy density due to GWs cannot be larger than the radiation density observed. And it also cannot be larger than the excess, $\Delta\rho_{r}=\rho_{r}^{\rm obs}-\rho_{r}^{\rm SM}$, the extra observed radiation due to neutrino species. 

After electron decoupling \cite{Saikawa:2018rcs}, 
\be \rho_{r}= \rho_\gamma + \rho_\nu = \dfrac{\pi^2}{30}\left(2+\dfrac{7}{4}N_{\rm eff}\left(\dfrac{4}{11}\right)^{4/3}\right)T^4.
\ee
The factor of $2$ corresponds to the two degrees of freedom of photon radiation, $7/4=(7/8)2$ to neutrinos and anti-neutrinos (fermions with one helicity state each), and $4/11$ to the heating of the photon bath relative to the neutrino bath due to $e^+e^-$ decay after neutrino decoupling. $ N_{\rm eff} $ is the neutrino effective number given by $ N^{\rm SM}_{\rm eff} + \Delta N_{\rm eff}$. Consequently, 

\begin{equation}
\left(\dfrac{\rho_{\rm GW}}{\rho_\gamma}\right)_{T=\text{MeV}} <  \dfrac{\rho_{r}^{\rm obs}-\rho_{r}^{\rm SM}}{\rho_\gamma} \leq  \dfrac{7}{8}\left(\dfrac{4}{11}\right)^{4/3}\Delta N_{\rm eff}.
\label{eq:Neff_bound}
\end{equation}

In the standard model, $N_{\rm eff}^{\rm SM} \simeq 3.0440$ \cite{Froustey:2020mcq,Bennett:2020zkv}, which is consistent with the CMB data, $N_{\rm eff}= 2.99^{+0.34}_{-0.33}$ at the $95\,\% $ confidence level \cite{Planck:2018vyg}. BBN \cite{Cyburt:2004yc} constrains $\Delta N_{\rm eff} \lesssim 0.5$ \cite{Pisanti:2020efz,Yeh:2020mgl}. Therefore, for $T<(T_{\text{BBN}},T_{\text{CMB}})$, the observed ratio is bounded by Eq.~\ref{eq:Neff_bound},
\be 
\left(\dfrac{\rho_{\rm GW}}{\rho_\gamma}\right)_T \lesssim 0.1\,.
\ee 
 Today, $\Omega_\gamma^0 = \frac{\rho_\gamma^0}{\rho_{c}^0} \sim 10^{-5}$, and $\rho_{\rm GW}\lesssim \Omega_{\gamma}\rho_{c} \Delta N_{\rm eff} $ constrains the GW spectrum, $\rho_{\rm GW} \lesssim 10^{-6}\rho_{c}$. Since $\rho_{\rm GW}=\rho_c \int d(\log f)\Omega_{\rm GW}(f)$, the BBN constraint implies, for a broad spectrum, that the observed GW spectrum today is bounded by
\be \Omega_{\rm GW}\lesssim 10^{-6}.
\ee This constraint holds for GWs inside the horizon at $T_{\text{BBN}}$ and $T_{\text{CMB}}$, i.e., the GWs were produced during radiation domination, before CMB decoupling. It was already used to constrain some early universe models.

\section{Probing cosmology and BSM physics with the SGWB}\label{sec:Probing cosmology and BSM physics with the SGWB}

In this section, we discuss some primordial sources of GWs, also described, for instance, in \cite{Caprini:2018mtu, LISACosmologyWorkingGroup:2022jok, Caldwell:2022qsj}: the cosmic gravitational microwave background, inflation, axion-inflation, scalar-induced gravitational waves, phase transitions, and cosmic strings. We organize the sources in a possible chronological order, based on the re-entry horizon time. 
We do not cover these sources in detail. Each one of them would deserve a dedicated set of lecture notes. To keep the notes concise, we aim for an executive summary by providing the main aspects and summarizing the status of the field, with more or fewer details depending on the model. To help the reader, we provide in advance an abridged list of recommended references for each considered source.  

\subsubsection*{Reference list sorted by sources:}
\begin{enumerate}
    \item[] Primordial Gravitational Wave Background: \cite{Ringwald:2020ist,Ghiglieri:2022rfp} 
\item[] Inflation: \cite{Domcke:2013pma,Guzzetti:2016mkm,Bartolo:2016ami}
\item[] Axion Inflation: \cite{Barnaby:2010vf,Barnaby:2011qe,Sorbo:2011rz,Garcia-Bellido:2023ser}
\item[] Scalar-Induced Gravitational Waves \cite{Ananda:2006af,Baumann:2007zm,Domenech:2023fuz}
\item[] First-order Phase Transitions: \cite{Caprini:2015zlo,Hindmarsh:2020hop}
\item[] Topological defects: \cite{Hindmarsh:1994re,Domcke:2013pma,Buchmuller:2023aus}
\end{enumerate}

\subsection{Primordial gravitational wave background}

In the primordial plasma, photons decoupled (recombination) at $T\sim$ eV (3000 K), leading to the CMB photon emission. This temperature of 3000 K is related to the temperature at which atoms of neutral hydrogen form, around $13.6$ eV. This means that the charged particles inside were no longer free, and therefore, the electrons could no longer be scattered by the photons. Today's CMB temperature is about 2.7 K due to redshift.

Primordial gravitational waves (GWs) are expected to be produced very early in the Universe, potentially up to the Planck scale. Unlike photons, which decoupled from the plasma at recombination ($T \sim 0.3$ eV) giving rise to the cosmic microwave background (CMB), GWs interact extremely weakly and essentially decouple immediately after production. 

The resulting stochastic background of high-frequency GWs can be characterized by its energy density fraction today relative to the total radiation density, $\Omega_{Rad}$. A simple estimate \cite{Ringwald:2020ist,Ghiglieri:2022rfp} gives
\begin{equation}
\Omega_{GW} \sim \frac{T_{max}}{M_P}\,\Omega_{Rad},
\end{equation}
where $T_{max}$ is the maximal temperature reached during radiation domination (sub-Planckian, typically $T_{max}\leq 10^{16}$ GeV due to BBN constraints) and $M_P\sim 10^{18}$ GeV is the Planck mass. This background is expected to peak at very high frequencies, around $\sim 100$ GHz, far beyond the sensitivity of current or planned GW detectors. Recent work on high-frequency GW detection methods (MHz to THz) can be found in \cite{Domcke:2023qle}.

\subsection{Gravitational waves from cosmic inflation}
\label{sec:GW_inflation}

Now we turn to inflation. We will focus on the tensor modes of the metric perturbation. The field equations in vacuum have been described in Sec.~\ref{sec:GWexpanding},
\be 
\tilde{h}_\lambda^{\prime\prime}(\vec{k},\tau)+\left(k^2+\dfrac{a^{\prime\prime}}{a}\right)\tilde{h}_\lambda(\vec{k},\tau)= 0.
\ee 
Here, $\tilde{h}_\lambda=a h_\lambda$, for the two polarization modes. In de Sitter spacetimes, at $\tau \rightarrow -\infty $, we have the Bunch-Davies vacuum solution \cite{Bunch:1978yq}
\be 
\lim_{\tau\rightarrow - \infty} \tilde{h}_\lambda = \dfrac{2 e^{-ik\tau}}{\sqrt{2k}}.
\ee
We use
\be 
\tilde{h}_\lambda = \dfrac{2 e^{-ik\tau}}{\sqrt{2k}}\left( 1 - \dfrac{i}{k\tau}\right)
\ee
and the correlation function
\be 
 \langle h_\lambda(\vec{k}) h_{\lambda^{\prime}}(\vec{k})\rangle = (2\pi)^3 \delta_{\lambda\lambda^{\prime}}\delta^3(\vec{k}-\vec{k}^{\prime})\left(  \dfrac{1}{M_p} \right) \dfrac{\mid \tilde{h}_\lambda \mid^ 2}{a^2} .
\ee
Then,
\be 
\dfrac{\mid \tilde{h}_\lambda \mid^ 2}{a^2} = \dfrac{4}{a^2 (2k)}\left(1+\dfrac{1}{k^2 \tau^2}\right) = \dfrac{4 H^2}{2k^3}(1+k^2\tau^2),
\ee
for a de-Sitter expanding universe solution, in which $ a H = -1 / \tau$.  The last term above can be neglected on super-horizon scales, since $ (a H )^{-1} << k^{-1} $. As we have seen in Sec.~\ref{sec:GWexpanding}, a super-horizon mode is frozen until it re-entries the horizon. We have
\be 
 \langle h_\lambda(\vec{k}) h_{\lambda^{\prime}}(\vec{k})\rangle = (2\pi)^3 \delta_{\lambda\lambda^{\prime}}\delta^3(\vec{k}-\vec{k}^{\prime})\left(  \dfrac{2}{M_p} \right)^2 \dfrac{H_*^2}{2k^3}, \label{eq:2ptInflation}
\ee
where $H_*=H(k\tau = -1)$ is the Hubble parameter when the mode $k$ left the horizon during inflation. Since the modes are frozen on super-horizon scales and inflation ends before horizon re-entry, the last imprint from inflation in the GW signal comes from the time when the mode had left the horizon, i.e., at $k\tau = - 1)$. From Eq.~\ref{eq:2ptInflation}, we get the power spectrum (compare with Eqs.~\ref{eq.:2ptTensor} and \ref{eq:powerspectrumdef})
\be P_\lambda(k) = \left(\dfrac{2}{M_P}\right)^2 \dfrac{H_*^2}{2k^3}. \label{eq:power_inflation} \ee
The slow-roll condition implies $H \approx \text{constant}$. The respective GW spectrum is given by Eq.~\ref{eq:GW_spectrum_exp}, with a constant scale-invariant spectrum
$ \Delta_t^2 \propto k^3 P_\lambda(k) $.

{\bf{How large are these tensor perturbations?}} CMB observations point to a tensor-to-scalar ratio $r=\Delta_t/\Delta_s<0.1$ \cite{Planck:2018vyg}. With $\Delta_s$ inferred, the upper bound on $r$ is an upper bound on $\Delta_t$ and hence on the GW signal. This would imply $\Omega_{\rm GW} \leq 10^{-15}$. The smallness of the parameters also makes it very difficult to detect these tensor perturbations in the CMB or any other indirect detection. In the simplest scenario of slow-roll inflation, scalar and tensor perturbations are modeled with a simple power-law spectrum that is red-tilted. So, if $\Omega_{\rm GW}$ due to these tensor perturbations is small at the CMB scale, it will be even smaller at the scales of PTAs, LISA, and LVK, too small to detect a signal. Beyond the simple power-law, one possible detectable scenario is to go beyond the single-power law parametrization and demand that the power spectrum is blue-tilted at CMB and PTA scales; then, there is an inflection point at which the spectrum becomes red-tilted at intermediate scales (before LVK scales), so that LVK and BBN bounds are respected. This scenario has been recently explored in \cite{Benetti:2021uea,NANOGrav:2023hvm}. For more details on GWs from inflation, see, for instance, the review \cite{Guzzetti:2016mkm}.

\subsection{Axion inflation}

Although the previous slow-roll scalar field model is the dominant paradigm in inflationary cosmology, the requirement of having a flat scalar potential is a challenge in particle physics. We can solve this problem by introducing a global symmetry, which protects the flatness of the potential from large radiative corrections, and it is spontaneously broken at large scales. One simple implementation is to assume that the inflaton $\phi$ is a pseudo-Nambu-Goldstone boson, whose model is invariant under \emph{shift symmetry}: $\phi \rightarrow \phi + c$, where $c$ is a constant. The resulting pseudo-scalar field model is an axion-like particle model:
\be
S=\int d^4 x\left\{ \sqrt{-g} \left[ \dfrac{R}{2} - \dfrac{1}{2}\del_\mu\phi \del^\mu \phi - V(\phi) - \dfrac{1}{4} F_{\mu\nu}F^{\mu\nu} \right] - \dfrac{\phi}{4\pi \bar{f}} F_{\mu\nu}\tilde{F}^{\mu\nu} \right\}, \label{eq:axionmodel}
\ee
where $ F_{\alpha\beta} = \del_\alpha A_\beta - \del_\beta A_\alpha $, $ \tilde{F}^{\mu\nu} = 1/2 \epsilon^{\mu\nu\alpha\beta}F_{\alpha\beta}  $ is the dual strength tensor, and the inflaton $\phi$ is coupled to the (dark) photon through a dimension five operator and the coupling $1/\bar{f}$. Shift symmetry\footnote{
The last term of the action can be rewritten as $  \phi F_{\mu\nu}\tilde{F}^{\mu\nu} =  2 \phi  \del_{\mu}(\epsilon^{\mu\nu\alpha\beta} A_\nu \del_\rho A_\sigma)=-2\del_\mu \phi (\epsilon^{\mu\nu\alpha\beta} A_\nu \del_\rho A_\sigma)$, which makes it explicitly shift-symmetric.} protects the flatness of $V(\phi)$, in the sense that symmetry breaking means departure of the flatness of the potential.

Next, we derive the equations of motion for $A_\mu$,  ($d\tau = a dt $)
\be 
\square \vec{A} = - \vec{A}^{\,''} + \nabla^2 \vec{A} = - \dfrac{\phi^{'}}{\pi \bar{f}}\nabla \times \vec{A},
\ee 
for the gauge-fixing $\nabla\cdot \vec{A} = 0$ and $ A_0 = 0$. The right-hand side contains the axion-like particle term and is interpreted as a source term. With the Fourier decomposition  
\be 
\vec{A}(\tau,\vec{x})= \sum_{\lambda=\pm}\int\dfrac{d^3 k}{(2\pi)^{3/2}}\left( A_\lambda(\tau,\vec{k})\vec{\varepsilon_\lambda}(\vec{k})\hat{a}_\lambda(\vec{k})e^{i\vec{k}\cdot\vec{x}} + \text{h.c.} \right),
\ee
polarization vectors
\be 
\vec{\varepsilon_\lambda}(\vec{k})\cdot \vec{\varepsilon_{\lambda^{'}}}(\vec{k})=\delta_{\lambda\lambda^{'}}, \qquad \vec{\varepsilon_\lambda}(\vec{k})\cdot\vec{k}=0, \qquad i\vec{k}\times\varepsilon_\lambda(\vec{k})=\lambda k \varepsilon_\lambda(\vec{k}),
\ee
and creation and annihilation operators which satisfy $[\hat{a}_\lambda(\vec{k}),\hat{a}^{\dagger}_{\lambda^{'}}(\vec{k})] = \delta_{\lambda\lambda^{'}}\delta^{(3)}(\vec{k}-\vec{k}^{'}) $.

The equations of motion in de-Sitter space are 
\be 
\left[ \dfrac{\del^2}{\del \tau^2} + k^2\left( 1 \mp \dfrac{2\xi}{k\tau}\right)  \right] A_{\pm} (\tau,k)=0,  \qquad \xi = \dfrac{\dot{\phi}}{2\pi \bar{f} H}. \label{eq:eqdiff}
\ee
There is a tachyonic instability\footnote{In other contexts, tachyons are associated with space-like propagations (propagation speed larger than the speed of light). We highlight that in these notes and the axion literature, this is \emph{not} what we mean by tachyonic instability. Instead, we mean a real exponential solution for the differential equation \ref{eq:eqdiff}, which leads to the amplification of the mode $A_+$. The existence of the amplified solution does not depend on the usage of a de-Sitter space.} for $A_+(\dot{\phi}>0)$ if $-k\tau=k(aH)^{-1} < 2 \xi $. In this case, there is an exponential gauge field production for $A_+(\dot{\phi}>0)$ when $ (aH)^{-1}\sim k^{-1} $, i.e., around horizon exit (assuming $\xi \sim \mathcal{O}(1)$). Therefore, with this model, we produce gauge fields with a preferred helicity (helical gauge fields). 

Next, we want to understand the phenomenology behind this model.
\begin{figure}[t]
\centering 
\includegraphics[scale=1]{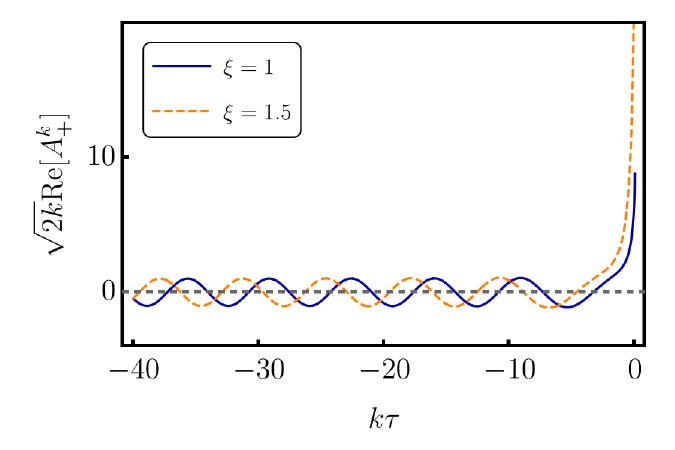} 
\caption{\textit{Exponential gauge field production for the $A_+(\xi>0)$ solution at around horizon exit. The larger $\xi$, the larger the gauge field production.}}
\label{fig:axioninflation}
\end{figure}
Assuming Bunch-Davies vacuum at $\tau\rightarrow -\infty$, slow-roll inflation (constant $\xi$), the solutions are
		\be 
		A_\lambda^k (\tau) = \dfrac{e^{\lambda\pi\xi/2}}{\sqrt{2k}} W_{-i\lambda\xi,1/2}(2ik\tau),  \label{eq:Aaxion}
		\ee
where we have used a Whittaker function W. These solutions are only valid under the assumption of a small backreaction of the gauge field to the dynamics of the axion field. They are also known in the literature as the Amber-Sorbo solutions \cite{Anber:2006xt,Anber:2009ua,Sorbo:2011rz}. The gauge field backreacts to the equation of motion of the axion field via the friction term proportional to $\langle F \tilde{F} \rangle$ in
\begin{equation}
\ddot{\phi}+3 H \dot{\phi} +  V_{,\phi} = \frac{1}{\pi\bar{f}} \langle EB \rangle,     
\end{equation}
and $\xi$ itself. The largest contribution is at the end of inflation since $\xi\sim\dot{\phi}$ grows, as $\phi$ rolls down its potential, see Fig.~\ref{fig:axioninflation}. Therefore, the system can leave the regime of small backreaction, in which our analytic treatment is no longer valid and more robust techniques are needed. In general, the exponential production could overcome the exponential de-Sitter expansion of the background, which dilutes matter/radiation. This way, this mechanism has phenomenological consequences, giving axion inflation signatures that can be explored by CMB observations, primordial black hole searches, and GW experiments \cite{Domcke:2020zez}.\footnote{See \cite{Garcia-Bellido:2023ser} for a GW template in axion-inflation, which is the current state of the art, by the time of writing.}

Next, we comment on the GW signals. The field equations from the action \ref{eq:axionmodel} for the rank-2 tensor are \cite{Cook:2011hg}
\be 
h^{\prime\prime}_{ij}(\vec{x},\tau)+\dfrac{2a^{\prime}}{a}h^{\prime}_{ij}(\vec{x},\tau) - \nabla^2  h_{ij}(\vec{x},\tau)=2\Pi_{ij}{}^{ab} T_{ab},
\ee
whose solution is 
\be 
h_{ij}(\vec{k},\tau) = 2 \int d\tau^\prime G_k(\tau,\tau^{'}) \Pi_{ij}{}^{ab} T_{ab},
\ee
where the Green functions satisfy $ [\del_\tau^2 + 2 a^{\prime}/a \del_\tau + k^2]G_k = 0 $, $\Pi_{ij}{}^{ab}$ is the traceless and transverse projection operator, and  $T_{ab}$ is the energy-momentum tensor from the axion-like particle model. By using the solution $A_+$ from Eq.~\ref{eq:Aaxion} in  $T_{ab}$, we can compute the contribution for the GW spectrum. 

 \begin{figure}[t]
\centering 
\includegraphics[scale=0.5]{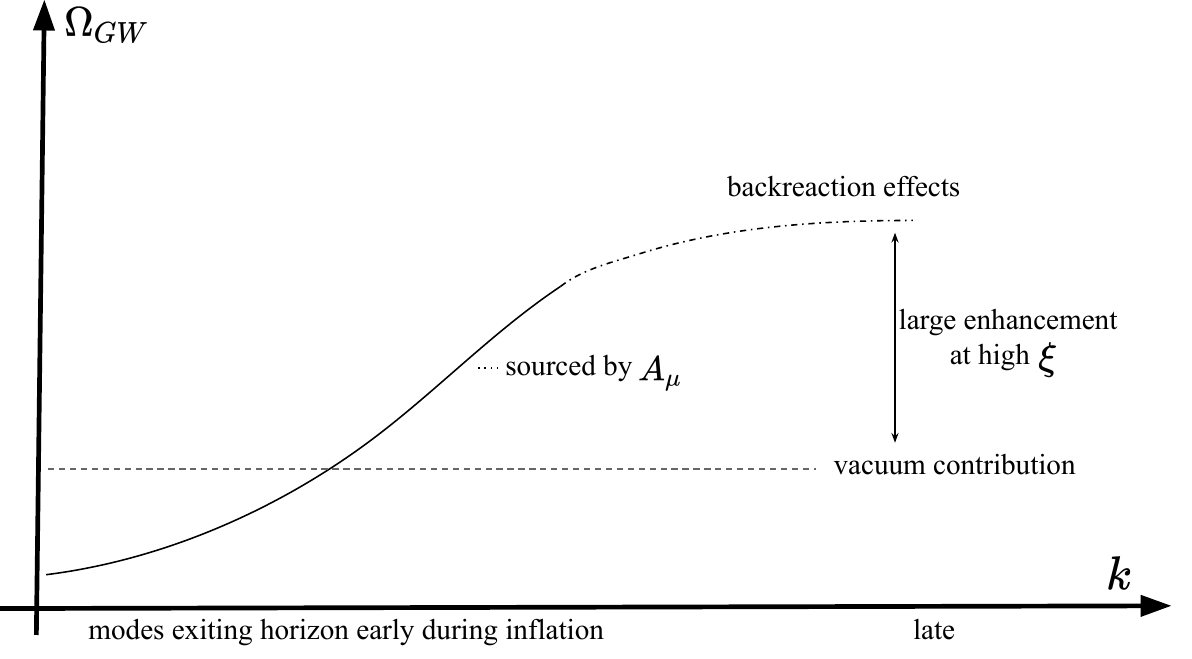} 
\caption{\textit{The GW spectrum for the axion-inflaton model. The vacuum contribution accounts for inflationary GWs without a source. Larger wavelengths exit the horizon earlier and contribute less to the spectrum since they are frozen. But there is also a large enhancement of gauge boson production, due to the velocity of the scalar field, proportional to $\xi$, producing helical GWs. The precise balance requires taking into account the backreaction of the gauge boson onto the axion field.}}
\label{fig:axionGW}
\end{figure}

Since $\xi\propto\dot{\phi}$, and $\phi$ is a pseudo-scalar field, we have parity violation. The chiral contribution is a consequence of the enhancement of only the $A_+$ mode above. Indeed, we have two helicities ($h_+$ and $h_\times$) whose contributions to the GW spectrum are different. The larger contribution to the spectrum is
\be 
\Omega_{\rm GW}= \left(\dfrac{H}{\pi M_p}\right)^2\left(1+10^{-7}\dfrac{H^2}{M_p^2}\dfrac{e^{4\pi\xi}}{\xi^6}\right).
\ee
The first term corresponds to the vacuum contribution (solution of the homogeneous equation \ref{eq:eqdiff}, without source, computed in the previous sections) and the second term is sourced by $A_\mu$. This last contribution is chiral, and its signature can be probed by LISA, the third-generation Einstein telescope, and the Cosmic Explorer detectors by searching for anisotropies. For results from the LVK collaboration, see \cite{Martinovic:2021hzy}, and using PTA data, see \cite{Figueroa:2023zhu,Guo:2023hyp,Niu:2023bsr,Unal:2023srk,Inomata:2023drn}.

The GW spectrum is sketched in Fig.~\ref{fig:axionGW}. As we previously commented, there is a large enhancement for large values of $\xi$ due to the exponential production, where backreaction effects from the gauge bosons onto the axion field become important \cite{Domcke:2020zez}. In particular, the spectrum peaks towards higher frequencies, corresponding to modes exiting the horizon just before the end of inflation. 

Since the equations of motion depart from the small backreaction regime and are highly non-linear, there are computational challenges in the field. In this regard, over the past years, we have seen continuous development in axion-inflation predictions. For instance, \cite{Domcke:2020zez,Gorbar:2021rlt,Peloso:2022ovc} showed the strongly non-linear regime experiences resonant enhancement of the gauge field production. Most recently, novel lattice techniques with a non-homogeneous axion field in \cite{Figueroa:2023oxc} confirmed the previous results to some extent, despite obtaining a different behavior after the backreaction effects become important. Additionally, the gradient expansion formalism \cite{Gorbar:2021rlt} is a recent and efficient alternative computation technique. Recently, a development \cite{vonEckardstein:2023gwk} showed that the analytical regime (in which the backreaction is small) is unstable, indicating that the system may even start within the strong backreaction regime. Beyond scalar and gauge fields, non-Abelian gauge fields \cite{Klose:2022knn} and fermion fields could also impact axion-inflation predictions, partly suppressing the exponential production, although exact solutions are not yet known. More details on axion-inflation can be found, for instance, in \cite{Barnaby:2011qe, LISACosmologyWorkingGroup:2022jok}.

\subsection{Scalar-induced gravitational waves}

Next, we describe another primordial source that is closely related to inflation: \emph{scalar-induced gravitational waves} (SIGWs), which are solutions of the Einstein equations in which the tensor modes are sourced by second-order scalar modes. 

Previously, we discussed sources that automatically fed the tensor modes at linear order. As a result, the GW spectrum is fully characterized by the power spectrum of tensor perturbations, as can be seen in Eq.~\ref{eq.:2ptTensor}, i.e.,
\begin{equation}
\expval{ h_a(\vec{k}_1)  h_b(\vec{k}_2)  }= \delta_{ab} (2\pi)^3 \delta^3(\vec{k}_1+\vec{k}_2)\dfrac{2\pi^2}{k_1^3} \mathcal{P}_t(k_1).
\end{equation}
Likewise, scalar modes are associated with the power spectrum of scalar perturbations,
\begin{equation}
\expval{\Phi(\vec{k}_1)\Phi(\vec{k}_2)} = (2\pi)^3 \delta^3(\vec{k}_1+\vec{k}_2)\dfrac{2\pi^2}{k^3_1}\mathcal{P}_\Phi(k_1),
\end{equation}
which are typically important in inflation.  Remember that inflationary GWs are produced by first-order tensor modes that re-enter the horizon after the end of inflation. They directly feed the tensor mode of the GW energy-momentum tensor. The scalar modes, however, cannot produce inflationary GWs because scalar and tensor modes are decoupled at linear order in GR.\footnote{Rather, all modes that re-entered the horizon before photon decoupling left imprints on the last scattering surface, making the CMB background a powerful probe of inflation and early-universe mechanisms that depend on these modes.} However, at higher order, scalar and tensor modes are coupled to each other. Then, scalar modes can source tensor modes and induce higher-order GWs \cite{Ananda:2006af,Baumann:2007zm}. From these higher-order contributions, if the scalar perturbations are large enough, the leading contribution to $\expval{h_a(\vec{k}_1)  h_b(\vec{k}_2)}  \rangle $ comes from terms quadratic in first-order scalar perturbations. This gives rise to the SIGWs.

\begin{figure}[t]
\centering 
\hspace{-1cm}\includegraphics[width=0.75\linewidth]{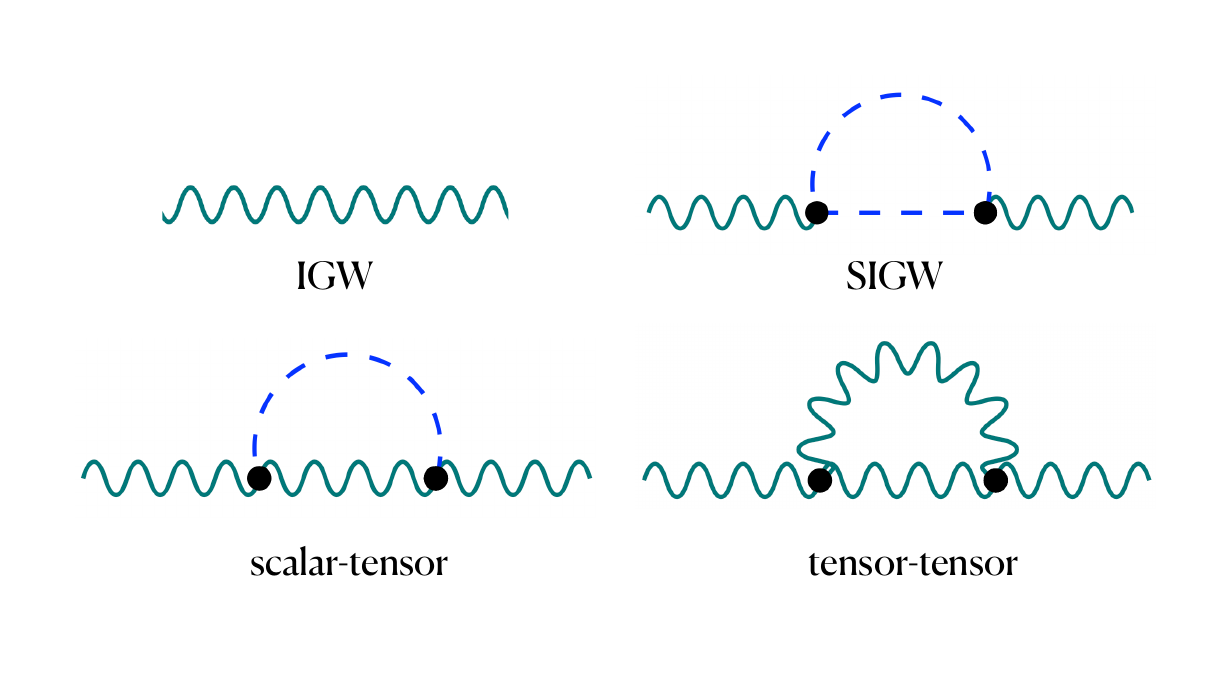} 
\caption{\textit{We show Feynman diagrams for the graviton propagator at tree level and its one-loop correction in the presence of first-order scalar (blue, dashed lines) and tensor (green, wavy lines) modes. The different combinations of internal lines correspond to the different GW scenarios mentioned in the text. IGW refers to the first-order GW whose contribution directly comes from tensor modes. SIGW refers to the second-order GW contribution coming from first-order scalar modes. Scalar-tensor and tensor-tensor are also second-order contributions induced by first-order tensor modes with and without another first-order scalar mode, respectively. The diagrams were produced with FeynGame 2.1 \cite{Harlander:2020cyh,Harlander:2024qbn}.}}
\label{fig:feynman_diagrams}
\end{figure}

Here, we illustrate the second-order contributions with Feynman-like diagrams, see Fig.~\ref{fig:feynman_diagrams}. 
We stress that these diagrams are used as a \emph{computational tool}: they organize the convolution integrals of first-order scalar and tensor perturbations in Fourier space. 
The underlying dynamics are \emph{completely classical}, corresponding to density waves propagating in the early Universe sourcing tensor modes. 
The first-order GW corresponds to the tree-level diagram (linear tensor perturbations), in which the propagator is the inverse of the second derivative of the action with respect to the tensor modes. The second-order GWs, such as scalar-induced GWs (SIGWs), correspond to \emph{sunset} diagrams with two internal scalar lines (loop), and two external graviton lines. There can also be second-order tensor-tensor (loop with two graviton propagators) \cite{Picard:2023sbz} and scalar-tensor (loop with one graviton and one scalar propagator) \cite{Yu:2023lmo,Bari:2023rcw} contributions. If the scalar perturbations are sufficiently enhanced, the scalar-scalar contribution prevails. The tensor-tensor contribution is subdominant, but the scalar-tensor contribution can become relevant on smaller scales \cite{Picard:2023sbz}. There is no tadpole diagram with one scalar loop because the 4-point function of two gravitons and two massless scalars vanishes. The same would apply for a tadpole with a graviton loop.\footnote{ This result holds for any tadpole of massless modes in dimensional regularization. These diagrams are not necessarily zero if we consider a non-vanishing cosmological constant, which induces a mass term for the graviton, or if we consider a method of regularization that introduces new mass scales to the problem.}\\

Back to cosmology, the resulting GW spectrum is (see, for instance, derivation in \cite{Yuan:2021qgz,Domenech:2021ztg})
\begin{align}
\Omega_{\text{GW}}(k,\eta)& = \dfrac{3}{4} \int_0^\infty dv \int_{\mid 1-v\mid}^{1+v} du\,\left( \dfrac{4v^2 - (1 + v^2 - u^2)^2}{4 u v} \right)^2  \dfrac{J(u,v)}{u^2 v^2} \mathcal{P}_\xi(vk)\mathcal{P}_\xi(uk), \label{eq.:Omegak}\\
J(u,v) &= \left( \dfrac{u^2 + v^2 - 3}{2 uv} \right)^4\left[ \left( \ln 	\left|  \dfrac{3-(u+v)^2}{3 - (u-v)^2} \right| - \dfrac{4uv}{u^2+v^2-3} \right)^2 + \pi^2 \Theta(u+v-\sqrt{3}) \right].
\end{align}
Above, the quantity $\mathcal{P}_\xi(k)$ is the power spectrum of the comoving curvature perturbation $\xi$. The curvature perturbation is a scalar and gauge-invariant quantity, that is related to the scalar perturbation $\Phi(k)$ in Eq.~\ref{eq:decomposition_svt} via $ \Phi (k)  = 2/3 \xi (k)$, and is the only surviving free parameter after gauge fixing (Newton gauge) and assuming negligible anisotropic stress. 

Because it is a second-order effect, the curvature perturbations must be immensely enhanced to produce a detectable GW spectrum. Scenarios of ultra-slow-roll inflation could produce such enhancements, creating GWs when the perturbation modes re-enter the horizon, usually assumed to occur during the radiation-domination era. Interestingly, these scalar modes can also produce \emph{primordial black holes} (PBHs) \cite{Carr:1974nx,Carr:1975qj} at horizon re-entry \cite{Sasaki:2018dmp,Carr:2020gox}. In turn, the primordial black hole population contributes to the budget of dark matter \cite{Carr:2016drx,Carr:2021bzv,Domenech:2023fuz}, and can be seeds of the supermassive black holes at the center of galaxies \cite{Carr:2018rid}. 

Recently, we have seen in the literature an increased interest in fitting pulsar timing data with signals from SIGWs.\footnote{See a more elaborated discussion in the thesis \cite{LinoSantos:2023ylg}. See also \cite{Romero-Rodriguez:2021aws}, for a search with LVK data.} From the latest datasets, among other primordial sources, one model for SIGWs had the best performance in terms of Bayes factor \cite{NANOGrav:2023hvm,Figueroa:2023zhu}, even though the viability and interpretation of these models depend on a precise formulation of the PBH DM abundance and the absence/presence of non-Gaussianities, see, for instance, \cite{Dandoy:2023jot,Franciolini:2023pbf,Inomata:2023zup,Wang:2023ost,Liu:2023ymk,DeLuca:2023tun}. It looks like non-Gaussianities cannot be ignored. Moreover, the employment of a perturbative approach can also be questioned. In this regard, an interesting question is the impact of non-perturbative effects on the abundance of PBHs \cite{Celoria:2021vjw} and the recent debate raised by \cite{Kristiano:2022maq} and \cite{Riotto:2023hoz}, in which the perturbativity of loop corrections in single-field ultra-slow-roll inflation is discussed \cite{Firouzjahi:2023ahg,Franciolini:2023agm,Davies:2023hhn}. The ultra-slow regime would be necessary to produce the necessary enhancement of curvature perturbations to make SIGWs detectable. If the loop corrections are not perturbative, this could jeopardize the SIGW and PBH interpretation.

We refer to the review \cite{Yuan:2021qgz,Domenech:2021ztg} and the lecture notes \cite{Domenech:2023fuz} for more details about SIGWs.

\subsection{First-order phase transitions} 

\emph{Phase transitions} are observed in many physical systems. One famous example is the different physical states of matter -- solid, liquid, or gas. According to changes in pressure and temperature, different thermodynamic quantities of the system -- such as free energy, energy, entropy, and volume -- change. Depending on how the change of state occurs, we have \emph{first-order} or \emph{second-order phase transitions}. The change is continuous for second-order phase transitions, even though their first derivatives are discontinuous, while the change is discontinuous for first-order phase transitions, associated with latent heat. In Fig.~\ref{fig:water_diagram}, we show the phase diagram for water. The first-order phase transition occurs where the separatrix is a continuous line. Upon this line, the free energy ($E-TS$) of the phases coincide, although some extra heat is necessary to actually change from one state to another. An example is the liquid-gas transition of water at constant pressure. Here, the temperature stays the same until all the water is boiled. In the same diagram, the second-order phase transition is depicted by a critical point, which marks the endpoint of the coexisting separatrix, beyond which there is no difference between the states, just a single phase.

\begin{figure}[t]
    \centering
    \begin{tikzpicture}
    \draw[thick,->] (0,0.5) -- (8,0.5) node[anchor=north west] {T};
    \draw[thick,->] (0,0.5) -- (0,8) node[anchor=south east] {P};
    \draw (1,1) .. controls (1.5,1.3) and (1.8,1.6)  .. (2.5,2.5) node[circle,fill,inner sep=1.2pt]{};
    \node[anchor = north west] at (2.5,2.5) { Triple point};
    \draw (2.5,2.5) .. controls (3.5,2.5) and (5.5,4.5)  .. (6.5,6) node[circle,fill,inner sep=1.2pt]{};
    \node[anchor = south] at (6.5,6) {Critical point};
    \node[anchor = north west, align = left] at (6.5,6) {Second-order\\ phase transition};
     \node[anchor = north west, align=left] at (2,8.5) {First-order\\phase transition};
     \draw (2.5,2.5) .. controls (1.8,3.5) and (1.8,4.5)  .. (1.8,9);
     \draw[dashed] (4,3)--(0,3) node[anchor =east]{ $1$ atm};
     \draw[dashed] (6.5,6)--(0,6) node[anchor =east]{ $218$ atm};
     \node at (0.9,4.4){\small{Solid}};
     \node at (3.2,4.8){\small{Liquid}};
     \node at (5.4,3.7){\small{Gas}};
\end{tikzpicture}
    \caption{\textit{Representation of the phase diagram of water as a function of pressure $P$ and temperature $T$. We highlight that a second-order phase transition is associated with a critical point, beyond which there is no longer a separation between the two phases, whereas a first-order phase transition is associated with a continuous line, on which the two phases are in equilibrium but the transition is not immediate, as some amount of latent heat is necessary to convert one state to another, while temperature remains constant.}}
    \label{fig:water_diagram}
\end{figure}

For a large class of phase transitions, the first and second-order transitions are related to symmetry-breaking patterns. An example is the solid-liquid water transition. There exist exemptions, such as the liquid-gas water transition, which does not break any symmetry. On the one hand, it does not matter how much pressure or temperature changes along the separatrix, the solid phase will always have certain symmetry properties that the liquid phase does not have. Then the change from one state to another cannot be continuous, as the symmetry must be broken or formed. On the other hand, since liquid and gas phases do not have these symmetry properties, it is possible to find a region where both phases can coexist.

\begin{figure}[t]
\centering 
\hspace{-1cm}\includegraphics[]{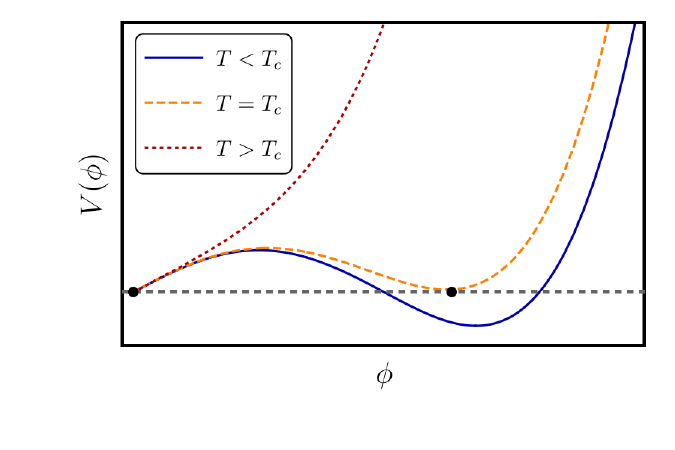} 
\caption{\textit{Representation of a first-order phase transition, for a potential $V(\phi)$ depending on a field $\phi$ (for instance, the Higgs field) and on the temperature $T$. $\phi=0$ is the true ground state only for $T>T_c$ (red dotted line). When the temperature decreases, a new local minimum appears, such that at $T=T_c$ both minima are degenerate (orange dashed line). $\phi\neq0$ is the true ground state for $T<T_c$ (blue solid line). Quantum or thermal fluctuations allow for tunneling between the vacua. }}
\label{fig:1stPT}
\end{figure}

 In these notes, we focus on symmetry-breaking phase transitions. Here, \emph{first-order phase transitions} are processes of symmetry breaking from a symmetric phase to a broken phase, in which the previous vacuum is no longer the true vacuum. The symmetry breaking can naturally occur if the properties of the system change according to external parameters, such as temperature, as shown in Fig.~\ref{fig:1stPT}. As a result, an enormous amount of energy can be released once one comes from the symmetric phase to the broken one. In this context, bubble collisions, magneto-hydrodynamics turbulence, and sound waves are phenomena sourcing GWs during first-order phase transitions \cite{Witten:1984rs,Hogan:1986qda,Kamionkowski:1993fg,Caprini:2007xq,Caprini:2009yp, Huber:2008hg,Hindmarsh:2013xza,Giblin:2013kea,Giblin:2014qia,Caprini:2015zlo,Hindmarsh:2015qta,Caprini:2019egz}.

\begin{figure}[h!]
\centering 
\hspace{-1cm}\includegraphics[width=0.85\linewidth]{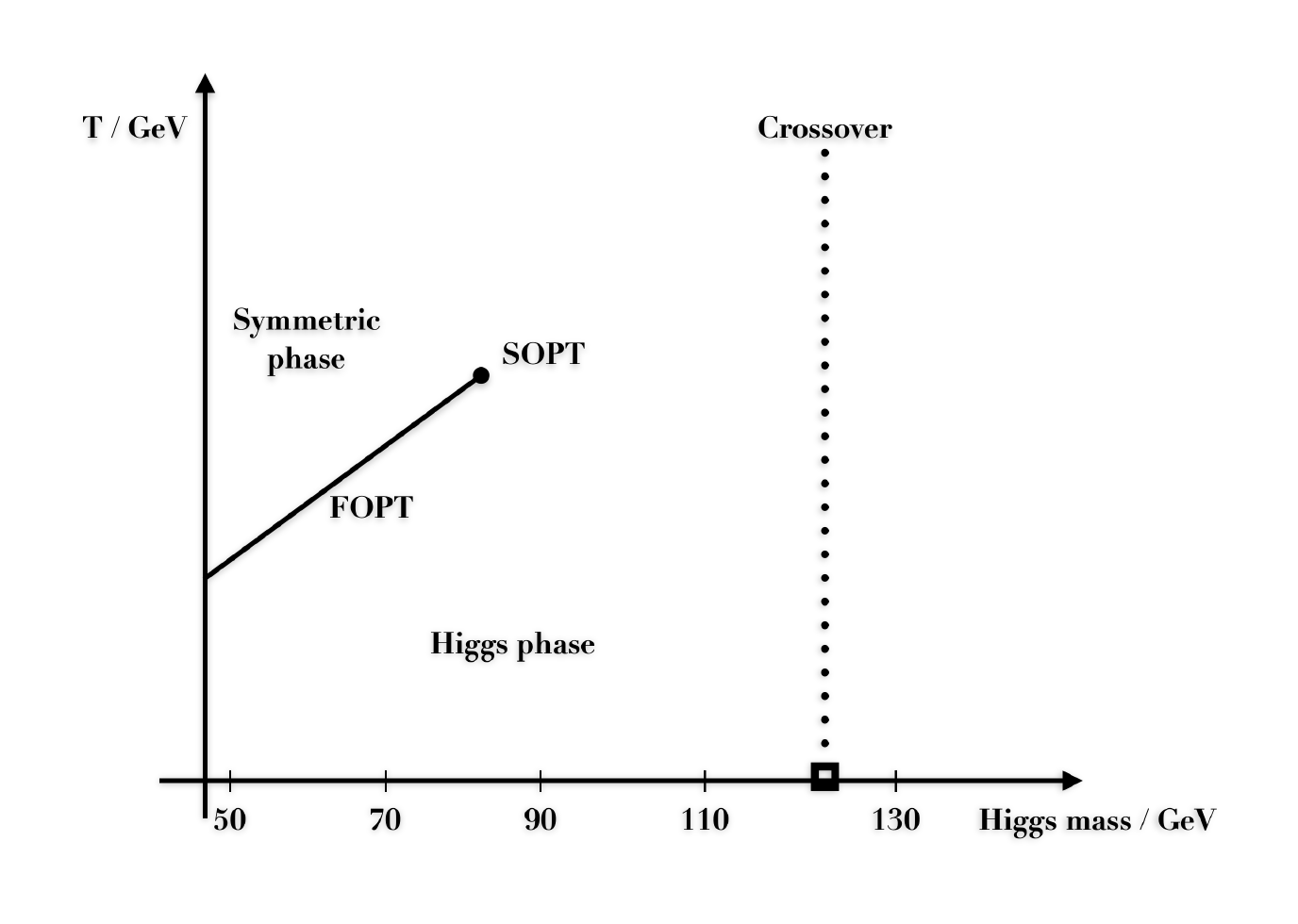} 
\caption{\textit{Representation of the phase diagram of the SM as a function of Higgs mass and temperature. For the observed Higgs mass (small square) at around 125 GeV, we have a crossover. The Higgs is too heavy to allow the SM to undergo a first-order phase transition (FOPT) between the symmetric and the broken (Higgs phase) phases; it is also too heavy for a second-order phase transition (SOPT).}}
\label{fig:SM_diagram}
\end{figure}

The peak frequency is around (using $\epsilon_*\sim 10^{-3}$ in Eq.~\ref{eq:peak_freq_char}),
\be 
f_{\text{peak}}\approx 10^{-3} \text{Hz} \left(\dfrac{T}{100 \text{GeV}}\right),
\ee
where $100$ GeV corresponds to the Standard Model (SM) electroweak phase transition temperature. No signal from the SM electroweak phase transition is expected because the SM does not have a first-order phase transition for the observed Higgs mass \cite{Kajantie:1996mn}, instead, it has a crossover at around 125 GeV, see Fig.~\ref{fig:SM_diagram}. However, several BSM models lead to a first-order electroweak phase transition that could be probed by LISA. From BBN, $ \Omega_{\rm GW} \lesssim 10^{-6} $ already constrained some phase transition models, depending on the strength of the first-order phase transitions \cite{NANOGrav:2020bcs}. See also constraints from LVK in \cite{Romero:2021kby} and PTA searches in \cite{NANOGrav:2023hvm,EPTA:2023xxk,Figueroa:2023zhu}. For more on cosmological phase transitions, see, for instance, \cite{Caprini:2015zlo} and the lecture notes \cite{Hindmarsh:2020hop}.

\subsection{Topological defects: cosmic strings}

\begin{figure}[t]\centering 
\includegraphics[width=\textwidth]{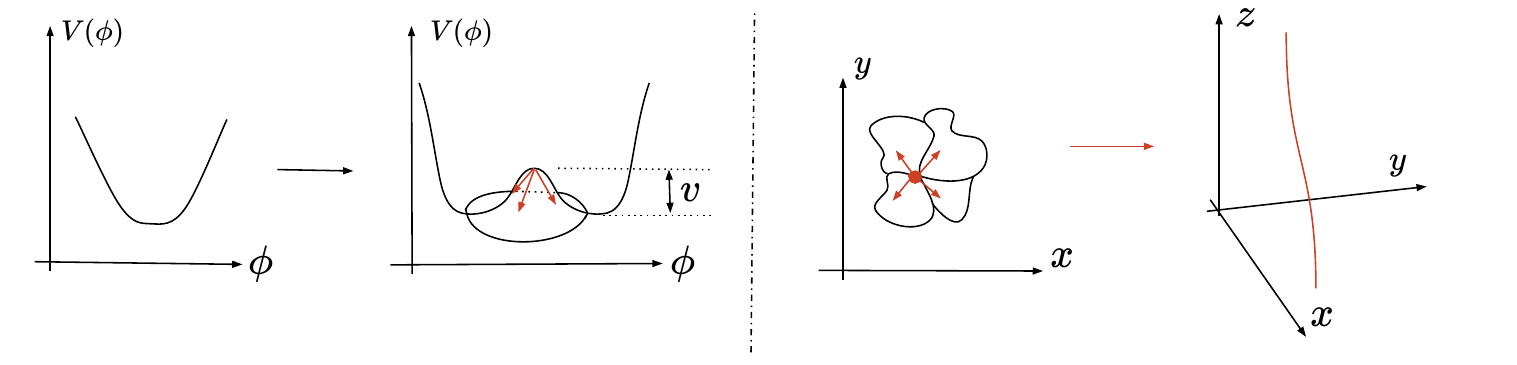} \caption{\textit{Representation of cosmic strings - one-dimensional topological defects. In the first two plots, we show the potential of a complex scalar field versus the scalar field configuration, in which some mechanism allows for a phase transition with a non-vanishing expectation value $v$. We obtain the last two plots by mapping the solution to real space (position). In the 2D plot (third, from left to right), we show the location of a local extremal point (false vacuum) by an orange dot and regions where the scalar field configuration assumes different values and are causally not connected \cite{Kibble:1980mv,Kibble:1981gv}. By continuity, these regions intersect each other where the vacuum expectation value $\expval{\phi} $ corresponds to the false vacuum. In the 3D plot (last plot, from left to right), we extrapolate the false vacuum region to three spatial dimensions; the reason for the name \emph{strings} becomes clearer.}}\label{fig:cosmicstrings}
\end{figure}

Topological defects are products of phase transitions associated with spontaneous symmetry breaking (SSB) of a gauge group. When the system has a topologically nontrivial vacuum manifold, fields in different regions in space can fall into different ground states. When SSB occurs, a network of topological defects can emerge. The formation of such topological defects is usually present in the SSB of grand unification theories (GUT) down to the SM gauge group \cite{Jeannerot:2003qv}. Mathematically, there can be topological defects when the vacuum manifold $M$ is topologically nontrivial, i.e. $\pi_n(M)\neq \mathit{I}$. Here, $\pi_n(M)$ stands for the $n$-th homotopy group in a manifold $M$ and counts the number of equivalence classes of loops in $M$.  They can be domain walls ($n=0$), strings ($n=1$), monopoles ($n=2$) or textures ($n=3$) \cite{Kibble:1976sj}.  See, for instance, \cite{Kibble:1976sj,Vilenkin:1984ib,Hindmarsh:1994re,Vilenkin:2000jqa,Mukhanov:2005sc}. 

\begin{figure}[t]\centering 
\includegraphics[width=0.85\textwidth]{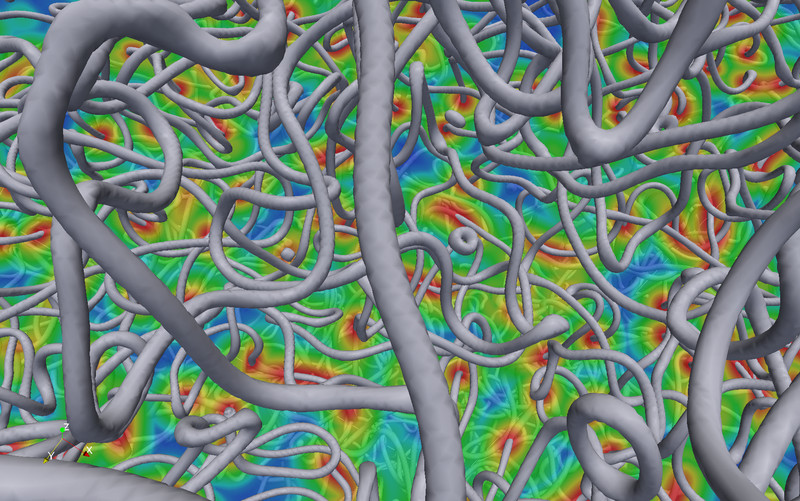} 
\caption{\textit{3D representation of cosmic strings (gray) from a simulation credited to David Daverio, from the group of Professor Martin Kunz, Université de Genève, using simulation data obtained at the Swiss National Supercomputer Centre.}}\label{fig:cosmicstringssimulation}
\end{figure}

 The mechanism is sketched in Fig.~\ref{fig:cosmicstrings}, where we can see how a non-trivial vacuum manifold can induce the formation of a topological defect once the symmetry is broken. The scalar field configuration can assume different ground states in regions of space that are not causally connected \cite{Kibble:1980mv,Kibble:1981gv}. Once the SSB happens, a new vacuum is formed, and a phase transition occurs inside these regions, while a topological defect is formed exactly where these regions intersect each other (they are in the false vacuum of the theory, preventing the phase transition). Because the symmetry has not been broken, the topological defects store a great amount of energy. If this topological defect is one-dimensional, we have a cosmic string. From now on, we focus on cosmic strings rather than other topological defects.\footnote{Note here that cosmic strings must not be understood as products of some string theory.}  See a simulation of a cosmic-string network in Fig.~\ref{fig:cosmicstringssimulation}. 
 
 Cosmic strings are hypothetical one-dimensional topological defects left after cosmic inflation. Cosmic strings arise in phase transitions if and only if, for $G\rightarrow H$, $\pi_1 (G/H) \neq \mathit{I} $. There is no cosmic string within the SM, but they can be present in extensions of the SM, for instance, in GUTs, in which almost any simple gauge group can lead to the formation of topological defects \cite{Jeannerot:2003qv}. Cosmic-string models are often associated with the SSB of a local U(1) symmetry in some BSM/GUT scenario \cite{Kibble:1976sj,Vilenkin:1984ib,Hindmarsh:1994re}. One example of such a BSM scenario is the breaking of a $B-L$ symmetry, accounting for the difference between baryon and lepton numbers \cite{Buchmuller:2012wn}. So, complementary to collider searches, GWs can also probe GUT physics \cite{Jeannerot:2003qv,Buchmuller:2019gfy,Buchmuller:2021mbb} and even be related to quantum gravity physics, see, for instance, \cite{Eichhorn:2023gat,King:2023ayw,King:2023ztb,LinoSantos:2023ylg}. 

 In the simplest models of stable cosmic strings, the string tension $\mu$ -- the energy per unit of length -- is the only free parameter that characterizes the string. It is related to the amplitude of the vacuum expectation value $v$ through \cite{Hindmarsh:1994re}
\begin{equation}
v \sim  \left(\frac{G\mu}{10^{-7}}\right)^{1/2} 10^{16}\,\,\text{GeV}. \label{eq:vev}
\end{equation}

\begin{figure}[t]
\centering 
\includegraphics[width=0.85\textwidth]{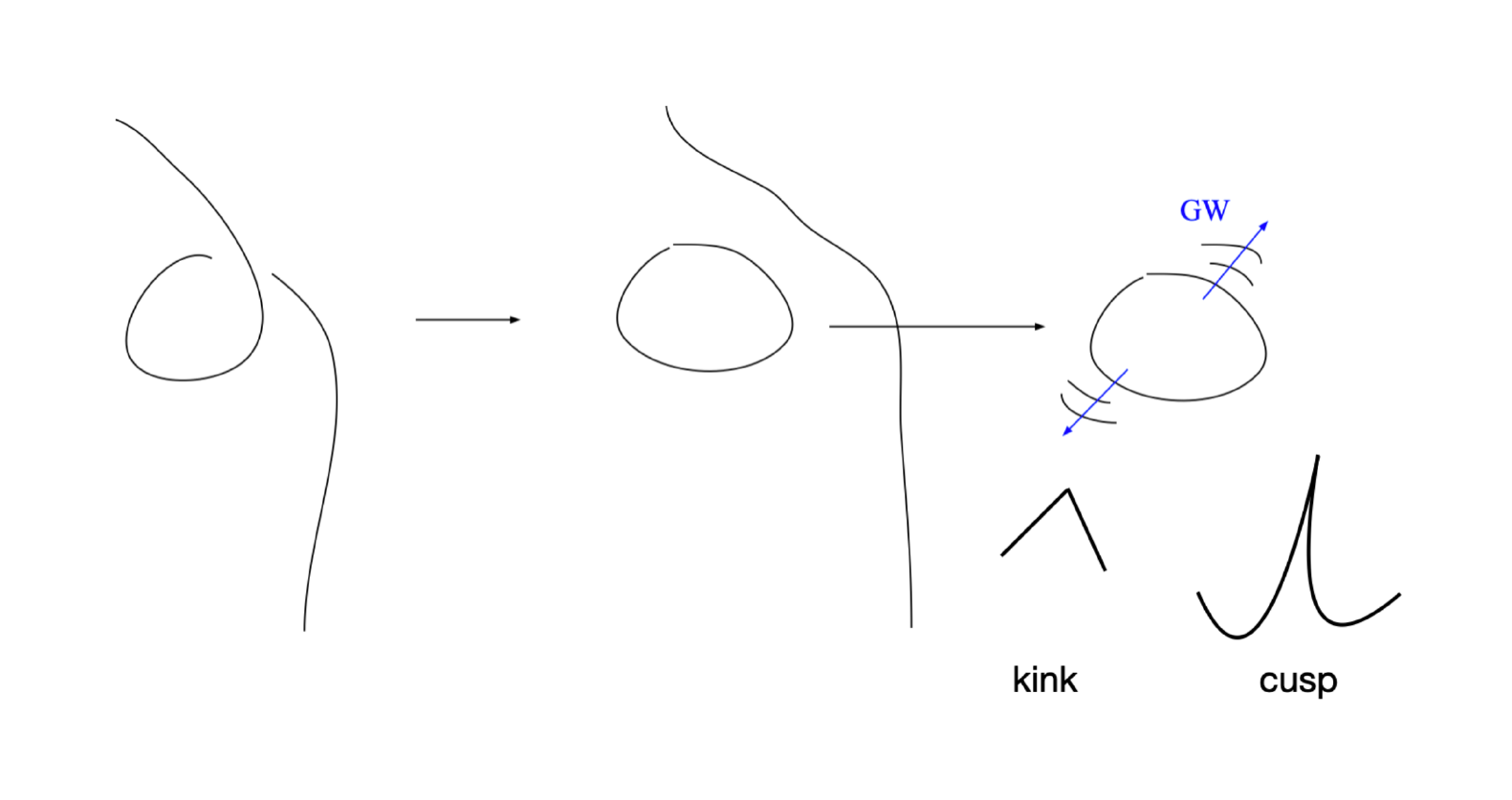} 
\caption{\textit{Representation of gravitational waves emitted by kinks and cusps in loops of cosmic-string networks. Taken from \cite{LinoSantos:2023ylg}. }}
\label{fig:cosmicstrings3}
\end{figure}

In the evolution of cosmic-string networks, (self-)intersection generates loops. Loops are more energetically favorable, so most searches focus on loops. During the string evolution, there is emission of particles and GWs by the excitations of the loops. Moreover, most of the GW radiation is emitted by cusps and kinks \cite{Vilenkin:1981bx,Vachaspati:1984gt,Caldwell:1991jj}, see Fig.~\ref{fig:cosmicstrings3}. To evolve the cosmic-string network, heavy numerical simulation is needed and, in most cases, the energy of strings is believed to be concentrated in a very thin region of space in such a way that their dynamics can be modeled according to the Nambu-Goto paradigm,\footnote{The Nambu-Goto action describes the dynamics of a particle traveling on a $1+1$ spacetime worldsheet.} see, for instance, \cite{Kibble:1984hp,Vilenkin:2000jqa,Ringeval:2005kr,Lorenz:2010sm,Blanco-Pillado:2011egf,Blanco-Pillado:2013qja,Blanco-Pillado:2015ana,Blanco-Pillado:2017oxo,Auclair:2019wcv,Blanco-Pillado:2023sap}.  Whether or not the Nambu-Goto strings agree with the dynamical evolution of field-theoretical cosmic strings \cite{Nielsen:1973cs,Hindmarsh:1994re,Vincent:1997cx,Moore:2001px,Bevis:2006mj,Bevis:2010gj,Correia:2019bdl,Hindmarsh:2022awe} is an open question \cite{Auclair:2019wcv,Hindmarsh:2022awe,Blanco-Pillado:2023sap}. Below, we report on a few properties found in the literature.

The scaling regime is a fixed point of the cosmic-string evolution with the property \cite{Hindmarsh:1994re}
\be 
\dfrac{\rho_{\text{CS}}}{\rho_{\text{total}}} \approx \text{constant},
\ee
with $\mathcal{O}(1)$ cosmic strings per Hubble volume. This property follows from the fact that in the scaling regime, the only physical scale is the Hubble radius $H^{-1}$. Therefore, the energy density of cosmic strings is $ \rho = \mu \times[M]^2 \propto \mu H^2 $, while the critical energy density is given by $\rho_{\text{total}}=\rho_{\text{crit}} = 3H^2/(8\pi G)$, so that the ratio $\rho_{\text{CS}}/\rho_{\text{total}}$ is constant.\footnote{ This property is essential for cosmic strings phenomenology, and it distinguishes strings from monopoles and domain walls. For these topological defects, the ratio $\rho_{\text{CS}}/\rho_{total}$ is not constant and their energy density is overproduced, above the experimental limits.}

The GW signal is characterized by
\be 
\rho_{\rm GW}(t,f) \propto \sum_{n=1}^{\infty} C_n(f) P_{gw,n}.
\ee
In this expression, $P_{gw,n}$ is the power of a single loop. The larger the string tension $\mu$, the larger $P_{gw}$. Furthermore, $n$ corresponds to the $n$-th harmonic, and $C_n(f)$ gives the number of loops emitting GWs that are observed today at frequency $f$ at time $t$, 
\be
C_n(f)=\dfrac{2n}{f^2}\int dz \dfrac{N(l(z),t(z))}{H(z)(1+z)^6}. \label{eq:cnloop}
\ee
Above, the denominator $H(z)(1+z)^6$ tells about the cosmological history. $N(l(z),t(z))$ is the number of loops of length $l$ at time $t$, where the length $l$ is given by $ l = 2n/ (f(1+z))$. Since it is a continuous process, the spectrum is broader. The computation is not straightforward and there are different methods to compute it. To solve the integral analytically, the usual assumption is that loops are sourced with $lH=\alpha=$ constant. Numerically, see, for instance, \cite{Blanco-Pillado:2013qja,Blanco-Pillado:2015ana,Blanco-Pillado:2017oxo}. 

\begin{figure}[t]
\centering 
\includegraphics[scale=0.5]{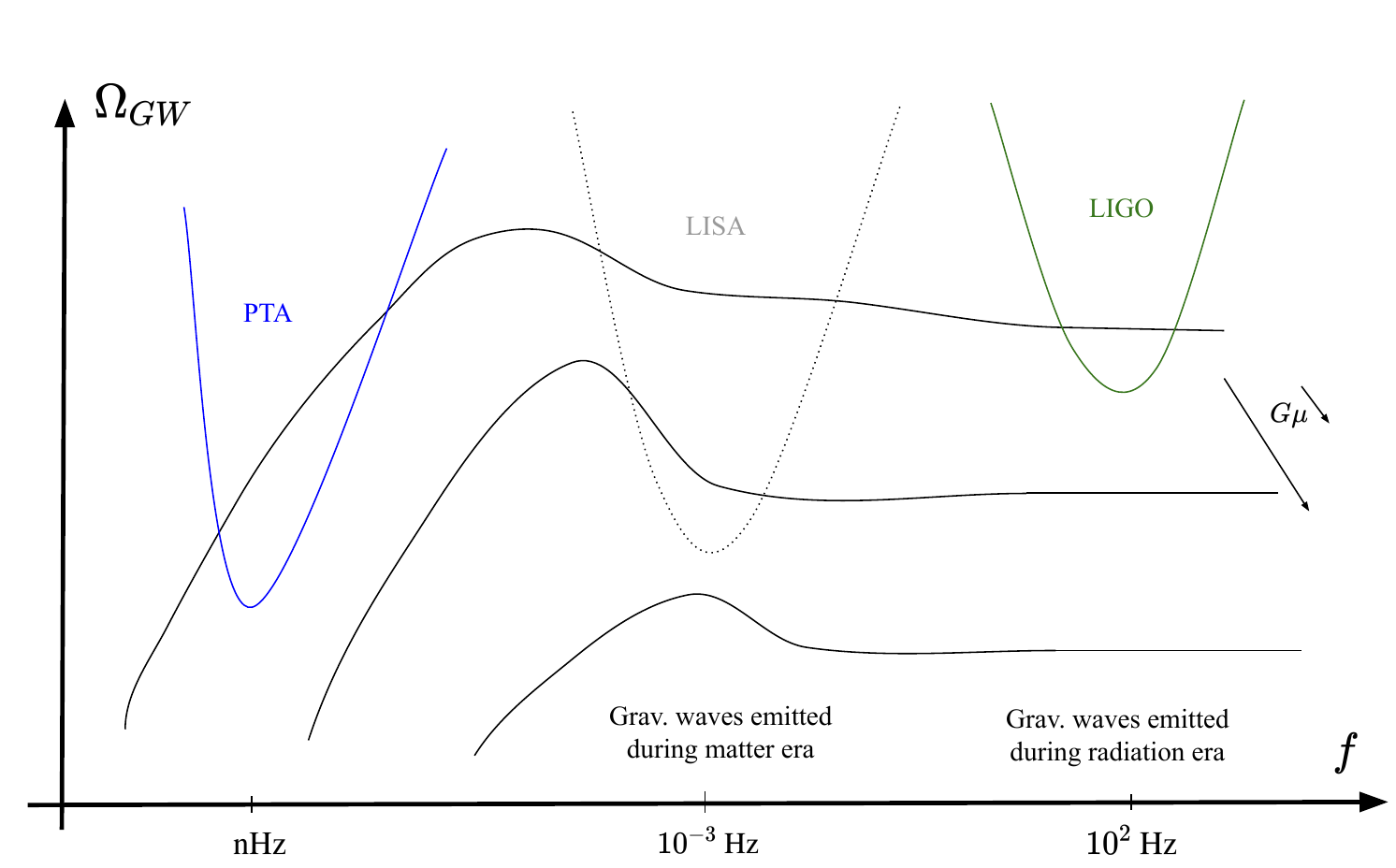} 
\caption{\textit{We sketch the amplitude of the 
 gravitational wave density generated by cosmic strings, with different string tensions $\mu$, as a function of frequency. The arrows on $G\mu$ mean that the smaller is $G\mu$, the smaller is the amplitude of $\Omega_{\rm GW}$. We also plot the frequency range probed, or expected to be probed, by LVK, LISA, and PTA collaborations. PTA  signals already constrain cosmic-string models with large $G\mu$, whose frequency peak is below or at the nHz range. LVK data can constrain models with small $G\mu$, whose frequency peak is around the Hz range. }}
\label{fig:cosmicstringsgw}
\end{figure}

Fig.~\ref{fig:cosmicstringsgw} summarizes a few properties of the GW spectrum from cosmic strings. First, the larger $G\mu$, the stronger the GW signal $\Omega_{\rm GW}$. Second, these signals are associated with a largely flat spectrum at high frequencies and with a mild peak at nHz-mHz frequencies. The flat spectrum is due to the decay of strings produced and emitted during radiation domination, while the peak corresponds to strings produced either during radiation or matter domination and decaying in matter domination. Third, the larger $G\mu$, the lower the frequency at which the spectrum peaks. Consequently, GW searches constrain the string tension $G\mu$ and bound the symmetry-breaking scale $v$ through Eq.~\ref{eq:vev}. For instance, large symmetry-breaking scales for topologically stable cosmic strings were already excluded by PTAs,
$$ G\mu \lesssim 10^{-10} \rightarrow v \lesssim 5\times10^{14} \text{GeV},$$ which excludes the originally GUT-scale motivated strings produced at $\nu\sim10^{16}\,$ GeV, with string tension in the range $G\mu \sim 10^{-8\cdots-6}$.\footnote{If one tries to work with GUTs whose SSB scales are lower than $10^{16}\,$ GeV, one can dangerously run into trouble because of proton decay. On the one hand, the gauge couplings tend to unify at the $10^{16}\,$ GeV scale, explaining why this scale is well-motivated and preferred. On the other hand, since GUTs can produce topological defects, these well-motivated GUT models can be strongly endangered and even excluded by the GW phenomenology of cosmic strings, at least if their production mechanism is simply the one of a stable cosmic string.}
See the search for four different models of stable cosmic strings with the latest NANOGrav 15-year dataset in \cite{NANOGrav:2023hvm}, the search with the latest EPTA dataset in \cite{EuropeanPulsarTimingArray:2023lqe,EPTA:2023xxk}, and a recent search with NANOGrav and EPTA combined data in \cite{Figueroa:2023zhu}. The string tension had also been probed before by the LVK collaboration in \cite{LIGOScientific:2021nrg}.
 
Finally, as a note on possible production mechanisms, the SM cannot produce cosmic strings, but BSM theories can. Although stable cosmic strings seem not to be favored by any GW signal, there can be emission through \emph{metastable defects}. For instance, in some GUT models, strings can decay via monopole pair production. A metastable defect relies on a sequence of phase transitions \cite{Preskill:1992ck} 
\be 
G \rightarrow G^{'} \rightarrow SM, \qquad \pi_n(G/SM)=\mathbb{I} \qquad \text{and}  \qquad \pi_n(G/G^{'})\neq\mathbb{I}, \pi_m(G^{'}/SM)\neq\mathbb{I},
\ee
i.e., the manifolds $(G/G^{'})$ and $(G^{'}/SM)$ have non-trivial homotopy groups, but the homotopy group of $(G/SM)$ is trivial, so that the defect is not topologically stable.

For example, if we have the symmetry groups $G=SO(10)$ and $G^{'}=U(1)/SM$, $G\rightarrow G^{'}$ generates monopoles, $G^{'}\rightarrow SM$ generates cosmic strings, but  $\pi_1(S0(10)/SM)=\mathbb{I}$. Therefore, strings are not topologically stable, thus metastable. There are at least two decaying mechanisms. In the first mechanism, there is an initial population of monopoles and strings; then, the \emph{string-monopole gas} decays fast. In a second mechanism, relevant if inflation dilutes away the initial monopole population, strings can only decay via spontaneous Schwinger monopole production with a decay rate $\propto e^{-m^2/\mu}$, where $m$ is the monopole mass; then, these \emph{metastable strings} can emit GWs \cite{Buchmuller:2019gfy}. At low frequencies, the spectrum $\Omega_{\rm GW}$ is suppressed at low $f$ and it cannot be excluded by the PTA bounds \cite{NANOGrav:2023hvm}, while allowing for larger spectra at larger frequencies, which opens a discovery space for the LISA and LVK collaborations \cite{Buchmuller:2021mbb}. See a recent review on metastable strings in \cite{Buchmuller:2023aus} and a search with the latest NANOGrav 15-year dataset in \cite{NANOGrav:2023hvm}.

\section{Summary}
\label{sec:conclusions}
In these lecture notes on gravitational waves from the early universe, we derived the main properties of GWs, introduced the stochastic gravitational wave background (SGWB), discussed ongoing and future detection efforts, and introduced some primordial sources of GWs.

We started with the basics in Sec.~\ref{sec:Linearized Einstein equations}, with the derivation of the Einstein equations for a linear perturbed metric $g_{\mn}=\eta_{\mn}+h_{\mn}$. We evaluated the degrees of freedom of the linearized metric tensor and fixed all the nonphysical degrees of freedom. The solution of the Einstein equation in vacuum is a wave equation. Hence, any test mass follows a sinusoidal geodesic while a GW passes by. This effect on test masses is the gist of GW measurement in interferometers.

In Sec.~\ref{sec:Emission of gravitational waves}, we derived the GW emitted by a source given by an energy-momentum tensor  $T_{\mn}$. We showed that GWs carry energy that curves the background. An important aspect of this section is the separation of scales and the relation between the GW radiation and the curved background. The power of gravitational radiation given by Einstein's quadrupole formula and the power spectrum of tensor perturbations are found in the last parts of this section.

In Sec.~\ref{sec:The stochastic gravitational wave background}, we introduced the SGWB. This background is defined as the composition of GWs with different wavelengths, amplitudes, and phases,  coming from all directions in the sky. The incoherently summed waves produce a stochastic background, which cannot be individually detected. Hence, these waves behave like noise, making it a challenge to detect them. SGWB signals are different than the transient signals coming from binary mergers, and more similar to the cosmic microwave background (CMB). However, the SGWB can possibly probe earlier stages of the universe, since GWs can travel freely through the hot plasma of the early universe, whereas photons could not. We have discussed spectral properties and sources of the SGWB, besides an overview of experimental efforts to probe it. 

Then, in Sec.~\ref{sec:searching}, we discussed the experimental settings  in more detail and derived the overlap reduction function for interferometers and PTA searches. In particular, we carefully derived the Hellings-Downs correlation for an isotropic, unpolarized, Gaussian, stationary SGWB.

The SGWB can have astrophysical and cosmological origins. The astrophysical sources are associated with supermassive black-hole binaries. A direct detection of waves produced by these objects would confirm again a prediction of General Relativity, now for a different range of masses that ground-based interferometers cannot probe. In a binary merger, the greater the black hole masses, the lower the frequencies of the emitted GWs. PTA collaborations expect to find evidence for astrophysical signals in the very near future.

On top of this astrophysical background, there are also cosmological sources from the early universe. These primordial sources produced GWs way before the emission of the CMB radiation. These waves traveled freely through the hot plasma of the early universe. Since BSM physics relies on mechanisms at energy scales beyond the ones current accelerators can probe, the phenomenology of the SGWB from the early universe is an essential step toward probing BSM physics. 

In this context, in Sec.~\ref{sec:Primordial gravitational waves}, we introduced GWs in an expanding universe and discussed how we can use data from GWs as complementary probes to the CMB and BBN bounds to constrain BSM physics and the earlier stages of the universe. 

Then, in Sec.~\ref{sec:Probing cosmology and BSM physics with the SGWB}, we discussed how very different early universe sources can give rise to a background of GWs. We introduced these sources chronologically according to at which stage the waves are produced. This has to do with the time the tensor perturbations re-enter the horizon. We focused on the spectrum produced by the following sources: the cosmic gravitational wave background, inflationary gravitational waves, axion inflation, scalar-induced gravitational waves, first-order phase transitions, and cosmic strings.\\

In summary, GW data opened a new window to the phenomenology of new physics and can give important insights into new physics. We have already experienced how data from the latest PTA searches can constrain new physics models. Much more is expected in the next decade, when we expect the development of detection techniques and prospects for the detection of cosmological sources from different collaborations, such as LVK, LISA, PTAs, SKA, the Einstein Telescope, and the Cosmic Explorer.  Better stay tuned!\\

Finally, we highlight that these notes are an introduction or an executive summary of the different lines of research and experiments described here. That is why we highlighted the main aspects of some cosmological sources and provided references for further reading. We encourage the readers to deepen their understanding of our presented material in the original works that are available in the literature!

\section*{Acknowledgements}
We thank all the organizers and students for the friendly and productive atmosphere in the 27th W.E. Heraeus Summer School ``Saalburg '' for Graduate Students on ``Foundations and New Methods in Theoretical Physics''. We thank Valerie Domcke for her encouragement to prepare these notes, for allowing us to use her lecture notes as the starting point of our work, and for reading previous versions of the manuscript. R.R.L.d.S. thanks the NANOGrav Collaboration; Richard von Eckardstein and Tobias Schröder, for feedback on the manuscript; Kai Schmitz and the Particle Cosmology Münster group at the University of Münster, for discussions and hospitality during the preparation of the first version of these notes. He also thanks the group LASSCO at the Universidade Federal do Ceará for hospitality during the last stages of preparation of the latest version of these notes.

% TODO: include author contributions
\paragraph{Author contributions}
Rafael R. Lino dos Santos and Linda M. van Manen wrote the notes. A very early version of the manuscript was based on lecture notes by Valerie Domcke, which the authors later enlarged. 

% TODO: include funding information
\paragraph{Funding information}
The work of R.R.L.d.S. was supported by a research grant (29405) from VILLUM FONDEN. R.R.L.d.S. was supported in part by the National Science Centre (Poland) under the research Grant No.\ 2020/38/E/ST2/00126. L.M.v.M is supported by the Volkswagen Foundation.

\bibliography{references}

\nolinenumbers

\end{document}